\documentclass[11pt]{article}

\usepackage[final]{acl}

\usepackage{times}
\usepackage{latexsym}
\usepackage{placeins}

\usepackage[T1]{fontenc}

\usepackage[utf8]{inputenc}

\usepackage{microtype}

\usepackage{inconsolata}

\usepackage{graphicx}

\usepackage{newfloat}
\usepackage{listings}
\usepackage{booktabs}
\usepackage{latexsym}
\usepackage{amsmath}
\usepackage{amssymb}
\usepackage{enumitem}
\usepackage{comment}
\usepackage{algorithm}
\usepackage{algpseudocode}
\usepackage{minted}
\usepackage[most]{tcolorbox}
\usepackage{subcaption}
\usepackage{siunitx}
\usepackage{multirow}
\usepackage[table]{xcolor}

\usepackage{xcolor}  
\lstdefinestyle{mypython}{
	language=Python,
	basicstyle=\ttfamily\small,
	keywordstyle=\color{blue},
	stringstyle=\color{orange},
	commentstyle=\color{gray},
	showstringspaces=false,
	breaklines=true,
	tabsize=4,
	frame=single,
	numbers=none,                   
	numberstyle=\tiny\color{gray},  
	numbersep=8pt,                  
	xleftmargin=2em,             
	moredelim=**[is][\color{red}]{@}{@},
}  

\usepackage{makecell}
\usepackage{ragged2e}
\usepackage{tabularx}
\usepackage{multirow}
\usepackage{eso-pic} 

\lstdefinestyle{mystyle}{
    basicstyle=\ttfamily\footnotesize,
numberstyle=\ttfamily\tiny\color{gray},
    breakatwhitespace=true,
    breaklines=true,
    breakautoindent=true,
    breakindent=0pt,
    captionpos=b,
    keepspaces=true,
    numbers=none,
    numbersep=2pt,
    showspaces=false,
    showstringspaces=false,
    showtabs=false,
    tabsize=2,
    frame=single,
    captionpos=b 
}

\usepackage{natbib}
\usepackage{caption}

\definecolor{myviolet}{RGB}{133, 96, 136}
\definecolor{myblue}{RGB}{0,0,0}
\definecolor{rowgray}{gray}{0.92}

\title{LLMs are Brittle to Simple Code Transformations: Introducing CETBench -- A Benchmark for Code-Equivalence Checking}

\author{
    Neeva Oza$^{1\dagger*}$ \quad
    Ishaan Govil$^{1\dagger}$\thanks{\hspace{0.3em} Equal Contribution, $^\dagger$Work done while at IIT Delhi. \\Corresponding Authors: \\
    \texttt{neevahoza@gmail.com}, \texttt{ishaangovil870@gmail.com}} \quad 
    Parul Gupta$^{1}$\quad
    Dinesh Khandelwal$^{2}$ \\
    {\bf Dinesh Garg}$^{2}$ \quad
    {\bf Parag Singla}$^{1}$\\
    $^{1}$Indian Institute of Technology Delhi \quad $^{2}$IBM Research\\
}

\def\dataset{CETBench}

\begin{document}

\AddToShipoutPictureFG*{%
  \AtPageLowerLeft{%
    \put(0,35){%
      \parbox{\paperwidth}{%
        \centering
        \footnotesize
        Findings of the Association for Computational Linguistics: ACL 2026,
        pages 41653--41685\\
        July 2--7, 2026 \textcopyright 2026 Association for Computational Linguistics
      }%
    }%
  }%
}
\maketitle

\begin{abstract}

We study how well LLMs can determine whether two programs are functionally equivalent. This is an important problem because benchmarking code equivalence helps assess LLM capability in tasks such as code rewriting and translation. 
To this end, we introduce \dataset — Code Equivalence with Transformations Benchmark — built from a repository of programs that may solve the same or different tasks. Each dataset instance is created by sampling a program pair and applying a random sequence of predefined code transformations, yielding either equivalent or non-equivalent pairs. Our analysis shows that even simple transformations cause a significant performance drop in state-of-the-art LLMs on code-equivalence checking. These challenges are further amplified in the cross-lingual setting when comparing programs written in different languages. To remedy this, we present a simple fine-tuning-based approach to boost LLM performance on the transformed pairs of programs.
Our approach for dataset generation is generic, supporting cross-lingual equivalence checking, generation of program pairs with varying difficulty levels, and the application of diverse transformations. In our experiments, we perform ablations over the difficulty level of original programs, as well as the kind of transformations used in generating pairs for equivalence checking. Our analysis presents deep insights into the working of LLMs for the task of code-equivalence, and points to the fact that they may still be far from what could be termed as a semantic understanding of the underlying code.
\end{abstract}

\section{Introduction}
The field of AI-assisted code generation and understanding has witnessed a rapid rise in adoption and capability over the past few years, with models increasingly integrated into real-world software development workflows ~\cite{jin2025llmsllmbasedagentssoftware}. The reasons behind this trend include (i) the emergence of powerful, general-purpose large language models and (ii) the development of standardised benchmarks like HumanEval \cite{Humaneval} and MBPP \cite{MBPP}, which evaluate text-to-code generation.
In this work, our focus is on the related task of checking {\em code-equivalence}: 
given two pieces of code, $p1$ and $p2$, how effective are LLMs in determining whether the two pieces of code are equivalent or not.
While there are many definitions of equivalence in the literature, we focus on the most general one, which is {\em functional equivalence}, i.e., two programs are said to be equivalent if for any given input, they produce exactly the same output. Doing this task successfully can help with multiple auxiliary tasks, such as verifying the correctness of code rewrites, code translation, etc. While there has been a good amount of work in the classical literature on this task, using AST and program analysis ~\cite{baxter1998clone, gabel2008scalable}, this is very tedious, does not scale, and often has multiple challenges. Of late, there have been some attempts to examine the capabilities of LLMs for the task of code-equivalence checking, with some of the approaches being parallel to our work~\cite{maveli-etal-2025-large, wei2025equibench}. We highlight the differences from these works, as well as contrast with other related work in the related work section.  

In this work, we provide a general-purpose dataset 
to evaluate the robustness of LLMs for the undecidable task of code equivalence. The dataset is generated using the following steps: (a) we start with a repository of programs written to solve a variety of programming tasks. (b) We construct a set of pairs of programs by randomly picking programs which (i) solve the same task (either correctly or incorrectly) (ii) solve different tasks. (c) each pair in the set is taken through a series of (randomly picked) pre-defined transformations, such as {\it variable renaming}, {\it swap of if-else blocks}, etc., to generate the corresponding perturbed pair. We note that some of the transformations are semantic preserving (sp), and others are not (np). These steps result in the dataset of perturbed pairs, and are referred to as \dataset. 
We perform a detailed analysis of relative performance among different pair categories, before and after introducing the perturbations. Our analysis shows that for a variety of SOTA LLMs, including both open source and closed source, the performance drops significantly on the perturbed dataset. Of specific interest is the drop in performance when the starting pair represents a single program.
We present additional insights on LLM performance, we vary the difficulty level~\cite{mikemirzayanov2018codeforces, codeforces_ratings_blog} of the original program repository. Additional ablations present the variation in performance across a variety of LLMs, with varying kinds of transformations applied. Furthermore, we extend our analysis to the cross-lingual setting, revealing that these challenges are exacerbated when comparing code from different programming languages.

Finally, we present a simple fine-tuning-based approach to boost LLM performance on the unperturbed as well as perturbed datasets. Best results are obtained when the fine-tuning is done using a combination of pairs from the perturbed as well as unperturbed datasets. We also demonstrate that the knowledge gained from this fine-tuning successfully transfers to unseen, more difficult tasks in a zero-shot setting.
Overall, our experiments present deep insights into the working of LLMs for the task of code-equivalence checking for a variety of categories. Further, our findings clearly point to the fact that LLMs are still quite far from semantic understanding of underlying programs, and likely rely on statistical patterns for the task of code equivalence.

The contributions of this paper can be summarised as follows: (a) We present a novel dataset of perturbed program pairs generated via a series of code transformations applied randomly to pairs of programs in a repository. (b) We present a detailed analysis demonstrating a significant drop in LLM performance over the perturbed program pairs. We also present additional ablations across varying program difficulty, cross-lingual settings, and the types of transformations applied. 
(c) We propose a fine-tuning approach that not only boosts performance on the perturbed dataset (proving LLMs can learn with explicit supervision) but also demonstrates strong generalizability, as the learned skills successfully transfer to more difficult and cross-lingual tasks. {\it We release our dataset and code here\footnote{\url{https://github.com/dair-iitd/CETBench}}}. 
\section{Related Work}\label{sec:related}

Determining code equivalence—whether two programs yield identical outputs for all possible inputs—is a foundational challenge in computer science, relevant to program verification~\cite{chakraborty2023ranking}, refactoring~\cite{liu2025exploring}, translation~\cite{roziere2020unsupervised}, and optimization~\cite{rosas2024should}. While undecidable in the general case~\cite{goldblatt2012well}, several heuristic and formal methods have been developed.

\paragraph{Traditional Approaches:}
Static program analysis techniques (e.g., AST comparison~\cite{baxter1998clone}, data-flow analysis~\cite{gabel2008scalable}) struggle with semantics-preserving syntactic variations. Formal methods like symbolic execution~\cite{kim2011mecc} and SMT solving~\cite{stolee2014solving} offer stronger guarantees but often lack scalability. Test-case-based execution is more practical 
in real-world settings; however, it has fundamental limitations in establishing true functional equivalence. This limitation arises because test suites inherently cover only a finite subset of potentially infinite program input domains and thus passing these tests merely indicates identical behaviour for those specific inputs rather than universal functional equivalence. Additionally, developing comprehensive test suites remains a significant practical challenge for complex, real-world software.

\paragraph{LLM-as-a-Judge for Code Evaluation:}
The \emph{LLM-as-a-Judge} paradigm~\cite{gu2024survey}—where LLMs are used to assess task outputs—has gained traction for its scalability and flexibility in both NLP~\cite{zheng2023judging,dong2024can} and code-related tasks~\cite{tong2024codejudge, zhuo2024ice, he2025code}. However, functional equivalence checking demands deeper semantic reasoning, making LLM reliability in this setting an open question. It requires more than understanding surface-level syntax or problem intent; it demands robust semantic reasoning to judge whether two programs, even if structurally dissimilar, will produce identical behaviour across all possible inputs.

\paragraph{Benchmarks for Code Equivalence:}
Prior datasets like BigCloneBench~\cite{svajlenko2014towards} suffer from false positives and class imbalance~\cite{krinke2022bigclonebench}. GPTCloneBench~\cite{alam2023gptclonebench} relies heavily on LLM-generated code, limiting its diversity. EqBench~\cite{badihi2021eqbench} is limited in scale and diversity, with only 147 equivalent and 125 non-equivalent C/Java pairs, making it less suitable for robust LLM evaluation.

\paragraph{Recent Parallel Work:}
Two recent benchmarks are closely related to our work: (More detailed experimental comprision in Appendix~\ref{sec:appendix-sota-cmp})

\paragraph{SeqCoBench:}~\citet{maveli-etal-2025-large} apply over 20 single-step transformations to Python programs from MBPP~\cite{MBPP}, focusing on subtle functional changes. In contrast, \dataset \ applies composed sequences of up to seven transformations, often across distinct but algorithmically similar programs. Additionally, our work evaluates \dataset \  on frontier models such as GPT-4o, extending beyond the models used in SeqCoBench.

\paragraph{EquiBench:}~\citet{wei2025equibench} present a multi-language benchmark. While three of EquiBench's six categories (OJ\_A, OJ\_V, OJ\_VA) are Python-based and sourced from online judge submissions like \dataset , key differences remain:

\noindent$\bullet$ \textbf{Transformation Depth:} \dataset \  applies composed sequences of up to seven diverse transformations. In contrast, EquiBench’s OJ\_A and OJ\_V use single transformation types (algorithmic difference or variable renaming), while OJ\_VA combines only two. Our compositional complexity provides a more rigorous test of LLM robustness.

\noindent$\bullet$ \textbf{Nature of Negative Pairs:} \dataset \  generates negative pairs via subtle, semantic-breaking changes between similar or identical programs, unlike EquiBench’s use of syntactically distinct wrong-answer submissions, which are often easier to classify.

\noindent$\bullet$ \textbf{Prompting Context:} EquiBench includes problem descriptions in prompts; \dataset \  omits them to focus solely on LLMs' standalone code reasoning abilities. 

\noindent$\bullet$ \textbf{Cross-Lingual:} \dataset \ is the first benchmark to include comparing two programs written in two different languages and applying a transformation on one.

\section{\dataset: Constructing a Perturbation-Based Benchmark}
\label{sec:dataset-curation}

To evaluate the ability of LLMs to determine program equivalence, we construct a perturbation-based benchmark that systematically generates \emph{both} semantically equivalent and non-equivalent pairs of code.  Our design mirrors practical scenarios in code translation~\cite{roziere2020unsupervised}, repair~\cite{le2019automated}, and synthesis~\cite{devlin2017robustfill}, where seemingly minor changes in the code can either leave the intended behaviour unchanged or silently change it. 

\subsection{Source Corpora}

Our benchmark is constructed by extending the CodeContests dataset\footnote{\url{https://github.com/google-deepmind/code_contests}}, originally introduced by DeepMind to train AlphaCode~\cite{CodeContests}. It consists of approximately $10000$ competitive programming problems curated from platforms such as Codeforces\footnote{\url{https://codeforces.com/}}. Each problem includes complex problem descriptions, test cases (input–output pairs) for validation, correct/incorrect human solutions in multiple programming languages, and metadata including difficulty ratings, time/memory constraints, and competition source tags. For this benchmark implementation, we use Python, Java and C++ language solutions from the CodeContests dataset and perturb only the Python language solution. However, our perturbation-based methodology is general and can be applied to other programming languages and datasets, as long as they include correct solutions to the problems.

For each problem, we group the solutions that solve it into a \emph{cluster}. We denote a cluster of solutions for a given problem $i$ as $P_i$. Such a cluster $P_i$ may contain multiple solutions, including those that are correct, incorrect, syntactically different but functionally equivalent, or even identical. The entire collection of these clusters, across all $N$ problems, is denoted $\mathcal{P}=\{P_1,\dots,P_{N}\}$.

\subsection{Perturbation Pipeline}
\label{subsec:perturbation-pipeline}

Our pipeline begins by sampling an initial pair of human-written programs, $(s_{1}, s_{2})$, from $\mathcal{P}$ using the $\textsc{SamplePair}$ function (see Algorithm~\ref{alg:perturb}). Please note that for the benchmark dataset, $\mathcal{P}$ is filtered to contain only Python programs. And for the cross-lingual setting described in Section~\ref{sec:cross-lingual}, programs from multiple languages are filtered accordingly.  This function is designed to select pairs that reflect diverse initial relationships—such as being identical, functionally equivalent (but syntactically different), one correct and one incorrect solution to the same problem, or correct solutions to entirely different problems (these initial relationships are further detailed in the dataset categorization Section~\ref{sec:dataset-categorization}). If the pair obtained after sampling from $\mathcal{P}$ doesn't fall under any one of these four types, it is discarded.

Once an initial pair of programs is selected, the algorithm applies a sequence of $k$ perturbations ($1 \leq k \leq K_{\max}$). Before applying any perturbations, the algorithm randomly decides whether the full sequence should preserve the original semantic relationship between the program pair or introduce a semantic break. To preserve semantics, all $k$ perturbations must be semantic-preserving. To break semantics, at least one of the $k$ perturbations must be semantic-non-preserving, with the remainder (if any) drawn from the semantic-preserving variants. At each step, one of the two programs ($s_1$ or $s_2$) is randomly selected for modification. A perturbation type is then randomly selected from the set of available types described in the types of perturbation section~\ref{subsec:perturbation-categories}, and the corresponding variant—semantic-preserving or semantic-non-preserving—is applied based on the overall goal for the sequence. Algorithm~\ref{alg:perturb} formalizes the procedure for generating perturbed code pairs via controlled semantic modifications.

\begin{algorithm}[ht!]
	\caption{Perturbed Program Pair Generation}
	\label{alg:perturb}
	\begin{algorithmic}[1]
		\Require A set of solution clusters $\mathcal{P} = \{P_1,\dots,P_N\}$ with $P_i=\{s_{i1},\dots,s_{i|P_i|}\}$; perturbation budget $K_{\max}$, \texttt{max\_tries}
		\State $\{s_1, s_2\} \gets \textsc{SamplePair}(\mathcal{P})$ \Comment{With probability ~$p_{\text{same}}$ both from one $P_i$, else from different clusters}
		\State $k \sim \text{Uniform}(1, K_{\max})$ \Comment{Number of perturbations to apply.}
		\State $\text{preserve} \sim \text{Bernoulli}(0.5)$ \Comment{Decide if the $k$-step perturbation sequence should be net semantic-preserving.}
		\If{\textbf{not} preserve}
		\State $k_{\text{neg}} \sim \text{Uniform}(1, k)$ \Comment{Number of semantic-non-preserving perturbations.}
		\Else
		\State $k_{\text{neg}} \gets 0$
		\EndIf
		\For{$j = 1$ to $k - k_{\text{neg}}$}
		\State $\hat{s} \gets \textsc{RandomChoice}(\{s_1, s_2\})$
		\State Apply to $\hat{s}$: the \textit{semantic-preserving version} of a randomly chosen perturbation type. Retry up to \texttt{max\_tries} on failure. 
		\EndFor
		
		\State $\hat{s} \gets \textsc{RandomChoice}(\{s_1, s_2\})$
        \For{$j = 1$ to $k_{\text{neg}}$}
		    \State Apply to $\hat{s}$: the \textit{semantic-non-preserving version} of a randomly chosen perturbation type. Retry up to \texttt{max\_tries} on failure.
		\EndFor
		\State \Return {Perturbed solution pair}
	\end{algorithmic}
\end{algorithm}

\subsection{Types of Perturbations}
\label{subsec:perturbation-categories}

We design seven distinct perturbation types to generate controlled code variations, each with \textit{semantic-preserving} and \textit{semantic-non-preserving} variants that simulate both benign refactorings and subtle logic-breaking edits. The semantic-preserving variants are theoretically behaviour-preserving transformations, empirically validated by executing the perturbed code against all test cases from the original problem; instances that fail validation are discarded. Similarly, semantic non-preserving transformations are validated and retained only if they cause at least one test case to fail for the program they are applied to, while all test cases continue to pass for the other program.
 This perturbation set thus enables structured control over code semantics, facilitating fine-grained evaluation of LLMs' ability to distinguish syntactic from semantic changes. Table~\ref{tab:perturb-summary} summarizes all perturbation types, with detailed explanations and examples in Appendix~\ref{sec:appendix-per-types}.

\begin{table}[h!]
	\centering
	
	{%
		\scriptsize
		\setlength{\tabcolsep}{5pt}
		\renewcommand{\arraystretch}{1.08}
		\setlength{\extrarowheight}{2.5pt} 
		
		\begin{tabular}{|
				>{\centering\arraybackslash}p{0.17\columnwidth}
				|>{\RaggedRight\arraybackslash}p{0.35\columnwidth}
				|>{\RaggedRight\arraybackslash}p{0.35\columnwidth}
				|}
			\hline
			\parbox[t]{0.17\columnwidth}{\centering\textbf{Perturbation\\Type}\strut} &
			\parbox[t]{0.35\columnwidth}{\centering\textbf{Semantic-Preserving Variant}\strut} &
			\parbox[t]{0.35\columnwidth}{\centering\textbf{Semantic-Non-Preserving Variant}\strut} \\ \hline
			
			\parbox[t]{0.17\columnwidth}{\rule{0pt}{3.0ex}If-Else\\Swapping\strut} &
			\parbox[t]{0.35\columnwidth}{\rule{0pt}{3.0ex}Negates the \texttt{if} condition and swaps the \texttt{if}/\texttt{else} blocks.\strut} &
			\parbox[t]{0.35\columnwidth}{\rule{0pt}{3.0ex}Negates the \texttt{if} condition but does not swap the blocks.\strut} \\ \hline
			
			\parbox[t]{0.17\columnwidth}{\rule{0pt}{3.0ex}For-While\\Swapping\strut} &
			\parbox[t]{0.35\columnwidth}{\rule{0pt}{3.0ex}Correctly converts a \texttt{for} loop to an equivalent \texttt{while} loop.\strut} &
			\parbox[t]{0.35\columnwidth}{\rule{0pt}{3.0ex}Converts to a \texttt{while} loop with an intentional error (e.g., faulty iterator update).\strut} \\ \hline
			
			\parbox[t]{0.17\columnwidth}{\rule{0pt}{3.0ex}If Condition\\Flipping\strut} &
			\parbox[t]{0.35\columnwidth}{\rule{0pt}{3.0ex}Rewrites the \texttt{if} condition to a logically equivalent form (e.g., $a<b$ to \texttt{not} $(a\ge b)$).\strut} &
			\parbox[t]{0.35\columnwidth}{\rule{0pt}{3.0ex}Inverts the relational operator in the condition (e.g., $a<b$ to $a\ge b$).\strut} \\ \hline
			
			\parbox[t]{0.17\columnwidth}{\rule{0pt}{3.0ex}Variable\\Renaming\strut} &
			\parbox[t]{0.35\columnwidth}{\rule{0pt}{3.0ex}Consistently renames a variable across its entire scope.\strut} &
			\parbox[t]{0.35\columnwidth}{\rule{0pt}{3.0ex}Renames a variable inconsistently, breaking its semantic link in one or more places.\strut} \\ \hline
			
			\parbox[t]{0.17\columnwidth}{\rule{0pt}{3.0ex}Boolean\\Inversion\strut} &
			\parbox[t]{0.35\columnwidth}{\rule{0pt}{3.0ex}Inverts a boolean variable's declaration and all its subsequent uses.\strut} &
			\parbox[t]{0.35\columnwidth}{\rule{0pt}{3.0ex}Inverts the variable's declaration but leaves its subsequent uses unchanged.\strut} \\ \hline
			
			\parbox[t]{0.17\columnwidth}{\rule{0pt}{3.0ex}Statement\\Reordering\strut} &
			\parbox[t]{0.35\columnwidth}{\rule{0pt}{3.0ex}Swaps two statements that are data-independent.\strut} &
			\parbox[t]{0.35\columnwidth}{\rule{0pt}{3.0ex}Swaps two data-dependent statements, altering the execution logic.\strut} \\ \hline
			
			\parbox[t]{0.17\columnwidth}{\rule{0pt}{3.0ex}Expression\\Reformatting\strut} &
			\parbox[t]{0.35\columnwidth}{\rule{0pt}{3.0ex}Refactors a sub-expression into a temporary variable without changing the final result.\strut} &
			\parbox[t]{0.35\columnwidth}{\rule{0pt}{3.0ex}Incorrectly refactors an expression, altering the final computational result.\strut} \\ \hline
			
		\end{tabular}
	}
    \caption{Summary of Perturbation Types. Each type has a semantic-preserving (SP) variant that maintains program logic and a semantic-non-preserving (NP) variant that intentionally breaks it.}
	\label{tab:perturb-summary}
\end{table}

\subsection{Dataset Categorization:}
\label{sec:dataset-categorization}

Each resulting pair from Algorithm~\ref{alg:perturb} assigned a structured label \mbox{$\langle\textit{prefix},k,{\textit{flag}}\rangle$}. The \emph{prefix} encodes the initial relationship of the pair before any perturbations were applied. The value \emph{$k$} is the total number of perturbations that were actually applied. The \emph{flag} indicates the net semantic effect of the perturbations. The following are the four possible values of the \emph{prefix}:

\vspace{0.25em}
\noindent
{
\centering
\setlength{\tabcolsep}{4pt}
\renewcommand{\arraystretch}{1.1}
\begin{tabular}{|l|p{0.845\columnwidth}|}
	\hline
	\texttt{id} & identical correct submissions \\ \hline
	\texttt{fe} & functionally equivalent but syntactically different solutions \\ \hline
	\texttt{ne} & one correct, one incorrect submission \\ \hline
	\texttt{di} & correct submissions of \emph{different} problems \\ \hline
\end{tabular}
\par
}
\vspace{0.5em}

\noindent\textbf{Example:}  
A pair starting from correct solutions to \emph{different problems} (\texttt{di}), then perturbed twice ($k=2$), with one of those perturbations being a semantic-non-preserving ($k_{\text{neg}}=1$), would be labeled $\langle\textit{di},2, \text{np}\rangle$.

Our benchmark results in program pairs classified into \emph{eight categories} by their structured labels (reflecting four initial relationship types combined with two overall semantic impact flags). The dataset consists of $id$ and $di$ categories each containing 400 original program pairs, evenly split into 200 $np$ and 200 $sp$ perturbed samples, while the $fe$ and $ne$ categories each contain 2000 original pairs, with 1000 $np$ and 1000 $sp$ samples per category. 
For the benchmark test set, we chose $K_{max}$ = 5. Hence, $k$ ranges from 1 through 5. Please note that here, $k$ is the count of perturbation applied to a given sample and each perturbation can be any of the 7 types of perturbations in Table~\ref{tab:perturb-summary}. For more discussion on the choice of sample count, please refer to Table~\ref{tab:benchmark-distribution} in Appendix \ref{sec:appendix-exp} and Appendix \ref{sec:appendix-var-bounds}. 
\dataset \ enables a comprehensive and fine-grained assessment of large language models under controlled semantic variation, facilitating systematic analysis of model robustness and failure modes across different solution types and perturbation impacts.

\section{Experiments}
\subsection{Methodology}
To observe the effect of perturbation on LLMs' performance in equivalence checking, we curate two parallel sets of code pairs. One set contains the categorised perturbed code pairs generated using our dataset generation method, while the other contains the corresponding original pairs prior to perturbation. For benchmarking, we use problems with a difficulty level of $900$ from the CodeContests dataset. We prompt various LLMs to check whether the codes in a given pair are semantically equivalent for both sets. Detailed description of the prompts is present in the Appendix \ref{sec:appendix-exp}. 
The LLMs used in the experiments are Qwen: Qwen2.5-Coder-7B-Instruct, Mistral: Mistral-7B-Instruct-v0.3, DSC: Deepseek-coder-6.7b-instruct, GPT-4o: GPT-4o-2024-08-06. We additionally report results for newer models (GPT-5-nano: GPT-5-nano-2025-08-07 and GPT-5.1: GPT-5.1-2025-11-13). We report accuracies and F1 scores.
\begin{table*}[t]
    \resizebox{\textwidth}{!}{
    \setlength{\tabcolsep}{1.7mm}
     \begin{tabular}{lcccccccccc}
     
        \toprule
       \textbf{Model} & \textbf{id} & \textbf{fe} & \textbf{ne} & \textbf{di} & \textbf{Avg}  &\textbf{W-Avg} & \textbf{Mac-F1} & \textbf{Mic-F1} & \textbf{PosF1} & \textbf{NegF1}  \\ 
       
       & \makecell{\begin{tabular}{@{}r  r@{}} 
                    \textbf{np} & \textbf{sp}
                \end{tabular}}
        & \makecell{\begin{tabular}{@{}r r@{}} 
                    \textbf{np} & \textbf{sp}
                \end{tabular}}
        & \makecell{\begin{tabular}{@{}r r@{}} 
                    \textbf{np} & \textbf{sp} 
                \end{tabular}}
        & \makecell{\begin{tabular}{@{}r r@{}} 
                    \textbf{np} & \textbf{sp}
                \end{tabular}} & & & & & & \\  \midrule

\rowcolor{rowgray}        
\multirow{3}{*}{Qwen}
    & \makecell{\begin{tabular}{@{}r r@{}}  100.0 & 99.0  \end{tabular}} 
    & \makecell{\begin{tabular}{@{}r r@{}}   80.7 & 78.7 \end{tabular} }
    & \makecell{ \begin{tabular}{@{}r r@{}} 44.0 & 42.9  \end{tabular} }
    & \makecell{ \begin{tabular}{@{}r r@{}} 100.0 & 100.0 \end{tabular} } 
    & \makecell{80.6} & \makecell{80.6} & \makecell{81.2}
    & \makecell{81.2} & \makecell{72.5} & \makecell{66.8} \\
\cline{2-11} \noalign{\vskip 2pt} 
    & \makecell{ \begin{tabular}{@{}r r@{}} 30.5 & 98.0  \end{tabular}}
    & \makecell{ \begin{tabular}{@{}r r@{}}  39.7 & 76.9  \end{tabular}}
    & \makecell{ \begin{tabular}{@{}r r@{}}  61.4 & 47.6  \end{tabular}} 
    & \makecell{ \begin{tabular}{@{}r r@{}} 99.5 & 100.0 \end{tabular}}
    & \makecell{69.2} & \makecell{69.2} & \makecell{78.7}
    & \makecell{78.7} & \makecell{69.2} & \makecell{78.7}
  \\ \hline \noalign{\vskip 3pt}       

\rowcolor{rowgray}
\multirow{3}{*}{Mistral} 
    & \makecell{\begin{tabular}{@{}r r@{}}   100.0 & 100.0  \end{tabular}} 
    & \makecell{\begin{tabular}{@{}r r@{}}   90.3 & 90.3 \end{tabular} }
    & \makecell{ \begin{tabular}{@{}r r@{}} 25.2 & 24.7   \end{tabular} }
    & \makecell{ \begin{tabular}{@{}r r@{}}  86.5 & 89.5 \end{tabular} } 
    & \makecell{76.3} & \makecell{76.3} & \makecell{82.6}
    & \makecell{65.1} & \makecell{71.7} & \makecell{58.4}\\
\cline{2-11} \noalign{\vskip 2pt} 
    & \makecell{ \begin{tabular}{@{}r r@{}} 0.2 & 98.0   \end{tabular}}
    & \makecell{ \begin{tabular}{@{}r r@{}}  16.5 & 86.5   \end{tabular}}
    & \makecell{ \begin{tabular}{@{}r r@{}}  33.0 & 27.8  \end{tabular}} 
    & \makecell{ \begin{tabular}{@{}r r@{}} 90.0 & 89.0 \end{tabular}}
    & \makecell{56.6} & \makecell{68.3} & \makecell{64.4}
    & \makecell{49.9} & \makecell{44.8} & \makecell{54.9}
  \\ \hline \noalign{\vskip 3pt}        
 
\rowcolor{rowgray}
\multirow{3}{*}{DSC}
    & \makecell{\begin{tabular}{@{}r r@{}}  100.0 & 100.0   \end{tabular}} 
    & \makecell{\begin{tabular}{@{}r r@{}}  100.0 & 99.9 \end{tabular} }
    & \makecell{ \begin{tabular}{@{}r r@{}} 0.1 & 0.3   \end{tabular} }
    & \makecell{ \begin{tabular}{@{}r r@{}} 2.5 & 2.0  \end{tabular} } 
    & \makecell{50.6} & \makecell{50.6} & \makecell{51.1}
    & \makecell{32.5} & \makecell{63.1} & \makecell{1.9}\\
\cline{2-11} \noalign{\vskip 2pt} 
    & \makecell{ \begin{tabular}{@{}r r@{}} 0.0 & 99.5   \end{tabular}}
    & \makecell{ \begin{tabular}{@{}r r@{}}  0.0 & 99.9    \end{tabular}}
    & \makecell{ \begin{tabular}{@{}r r@{}}  1.2 & 0.3  \end{tabular}} 
    & \makecell{ \begin{tabular}{@{}r r@{}} 2.5 & 2.5 \end{tabular}}
    & \makecell{23.6} & \makecell{50.6} & \makecell{27.8}
    & \makecell{20.2} & \makecell{37.6} & \makecell{2.8}
  \\  \hline \noalign{\vskip 3pt} 
  
  \rowcolor{rowgray} 
  \multirow{3}{*}{GPT-4o}
    & \makecell{\begin{tabular}{@{}r r@{}} 99.5 & 100.0  \end{tabular}} 
    & \makecell{\begin{tabular}{@{}r r@{}}  72.5 & 73.0  \end{tabular} }
    & \makecell{ \begin{tabular}{@{}r r@{}} 67.7 & 66.7  \end{tabular} }
    & \makecell{ \begin{tabular}{@{}r r@{}} 100.0 & 100.0 \end{tabular} } 
    & \makecell{84.9} & \makecell{84.9} & \makecell{91.1}
    & \makecell{76.8} & \makecell{75.5} & \makecell{78.1}\\
\cline{2-11} \noalign{\vskip 2pt} 
    & \makecell{ \begin{tabular}{@{}r r@{}} 81.0 & 97.0  \end{tabular}}
    & \makecell{ \begin{tabular}{@{}r r@{}}  77.4 & 71.8  \end{tabular}}
    & \makecell{ \begin{tabular}{@{}r r@{}}  91.8 & 72.2  \end{tabular}} 
    & \makecell{ \begin{tabular}{@{}r r@{}} 100.0 & 100.0 \end{tabular}}
    & \makecell{86.4} & \makecell{85.7} & \makecell{92.3}
    & \makecell{77.3} & \makecell{66.7} & \makecell{88.1}
  \\ \hline \noalign{\vskip 3pt}
    
    \rowcolor{rowgray}
    \multirow{3}{*}{GPT-5-nano}
    & \makecell{\begin{tabular}{@{}r r@{}}  100.0 & 100.0  \end{tabular} }
    & \makecell{ \begin{tabular}{@{}r r@{}} 69.8 & 71.8  \end{tabular} }
    & \makecell{ \begin{tabular}{@{}r r@{}} 87.1 & 86.6 \end{tabular} } 
     & \makecell{ \begin{tabular}{@{}r r@{}} 100.0 & 100.0 \end{tabular} } 
    & \makecell{89.4} & \makecell{89.4} & \makecell{94.0}
    & \makecell{83.4} & \makecell{81.1} & \makecell{85.7}\\
\cline{2-11} \noalign{\vskip 2pt}
    & \makecell{ \begin{tabular}{@{}r r@{}} 90.5 & 98.5 \end{tabular}}
    & \makecell{ \begin{tabular}{@{}r r@{}} 78.4 & 70.7  \end{tabular}} 
    & \makecell{ \begin{tabular}{@{}r r@{}} 96.1 & 89.8 \end{tabular}}
    & \makecell{ \begin{tabular}{@{}r r@{}} 100.0 & 100.0 \end{tabular}}
    & \makecell{90.5} & \makecell{88.5} & \makecell{94.7}
    & \makecell{82.2} & \makecell{72.9} & \makecell{91.5}
  \\ \hline \noalign{\vskip 3pt}
  
    \rowcolor{rowgray}
  \multirow{3}{*}{GPT-5.1} 
    & \makecell{\begin{tabular}{@{}r r@{}}  100.0 & 100.0  \end{tabular} }
    & \makecell{ \begin{tabular}{@{}r r@{}} 50.4 & 57.1  \end{tabular} }
    & \makecell{ \begin{tabular}{@{}r r@{}} 85.2 & 85.2 \end{tabular} } 
     & \makecell{ \begin{tabular}{@{}r r@{}} 100.0 & 100.0 \end{tabular} } 
    & \makecell{84.7} & \makecell{84.7} & \makecell{90.5}
    & \makecell{75.6} & \makecell{70.7} & \makecell{80.4}\\
\cline{2-11} \noalign{\vskip 2pt}
    & \makecell{ \begin{tabular}{@{}r r@{}} 93.0 & 90.5 \end{tabular}}
    & \makecell{ \begin{tabular}{@{}r r@{}} 92.6 & 54.8  \end{tabular}} 
    & \makecell{ \begin{tabular}{@{}r r@{}} 96.9 & 88.2 \end{tabular}}
    & \makecell{ \begin{tabular}{@{}r r@{}} 100.0 & 100.0 \end{tabular}}
    & \makecell{89.5} & \makecell{83.9} & \makecell{93.8}
    & \makecell{79.4} & \makecell{67.3} & \makecell{91.4}
  \\ 
        
               \bottomrule
    \end{tabular}
    }
    \captionof{table}{Results comparing performance of various models before and after perturbations. First 10 columns show accuracy values. W-Avg is average accuracy after re-weighing examples to have equal weight on positively and negatively labelled examples. Macro F1, Micro F1, and F1 scores of each individual labels are also reported.} 
    \label{tab:stricter_prompt}
\end{table*}

\subsection{Main Results}
Table~\ref{tab:stricter_prompt} shows the effect of perturbations on various models for the task of code similarity detection. There are four values corresponding to each category (id, fe, ne, and di). Within each row, the left column corresponds to non-semantic-preserving perturbations, and the right column to semantic-preserving perturbations. Overall accuracy and F1 scores are computed by aggregating results across all four categories, for both the original (top row) and perturbed (bottom row) code pairs. We observe a significant decrease in overall performance across 4 different LLMs. 
We observe that for the \textit{id} category - i.e. pairs of identical code, all models have a significant drop in performance after perturbation, while having nearly 100\% accuracy on the corresponding unperturbed set.
Moreover, a similar trend is observed for \textit{fe} category - i.e. functionally equivalent original pairs. This suggests that models severely struggle to tell similar-looking code as inequivalent on adding simple perturbations. 
This drop is surprising, since humans can easily spot such semantic differences. We observe proprietary model GPT-4o being able to gain an advantage in identifying inequivalent pairs, especially for $ne$ category, by being able to spot cases of semantic inequivalence. However, it still remains vulnerable in the positive class. Both these effects tend to counterbalance each other, keeping the weighted accuracy nearly stable, though there is a significant decrease in F1 for the positive class. We observe similar trends for GPT-5-nano and GPT-5.1 as well. We also observe low performance for \textit{fe} category for GPT-5.1 before perturbation. 

\subsection{Comparison to Existing Benchmarks}
We present performance on the Python split of the EquiBench dataset \cite{wei2025equibench} in Table \ref{tab:cmp_equibench}. EquiBench contains three categories: OJA, OJV, and OJVA. The positive-label pairs in OJA are analogous to our $fe$ category before perturbation, while the negative-label pairs in OJA correspond to $ne$ before perturbation. The positive and negative pairs in OJV are similar to $\langle id, k, sp \rangle$ and $\langle id, k, np \rangle$, respectively. Likewise, OJVA is similar to the perturbed $fe$ category in our dataset. For SeqCoBench \cite{maveli-etal-2025-large}, the dataset contains samples that are similar to the $\langle fe, 1, np \rangle$ and $\langle fe, 1, sp \rangle$ categories in CETBench, and the comparison results are reported in Table \ref{tab:cmp_secobench}. We observe that the perturbed pairs in our dataset are more challenging: performance decreases consistently after perturbation, whereas this decrease is not observed for the positive-label pairs when moving from OJA to OJVA. For results comparing against additional SOTA benchmarks, please see Appendix \ref{sec:appendix-sota-cmp}.

\subsection{Comparison with a Non-LLM Baseline}
To evaluate the complexity of our dataset, we compare against ASTNN~\cite{Zhang2019ANN}, a specialized code clone detection model that does not rely on LLMs. ASTNN decomposes the abstract syntax tree into statement-level subtrees, encodes them as vectors using an RNN, aggregates these representations with a GRU to obtain a program-level embedding, and applies a classifier to predict code equivalence. From Table~\ref{tab:astnn}, we observe that: (a) when trained on unperturbed data, ASTNN collapses to predicting equivalence and fails to learn discriminative patterns; (b) training on perturbed data increases sensitivity to transformations, but the improvement remains marginal and close to random performance; and (c) perturbed pairs remain challenging for ASTNN, with performance significantly lower than that of LLM-based models.

\begin{table*}[!hbt]
\centering 
\begin{subtable}[t]{0.5\textwidth}
    \setlength{\tabcolsep}{1mm}
    \begin{tabular}{lccc}
        \toprule
       \textbf{Model} & \textbf{OJA} & \textbf{OJV} & \textbf{OJVA}  \\ 
        & \makecell{\begin{tabular}{@{}r  r@{}}  
                     \textbf{neg} & \textbf{pos}
                 \end{tabular}}
         & \makecell{\begin{tabular}{@{}r r@{}}  
                     \textbf{neg} & \textbf{pos}
                 \end{tabular}}
         
         & \makecell{\begin{tabular}{@{}r  r@{}}  
                     \textbf{neg} & \textbf{pos}
                 \end{tabular}} \\ \midrule
Qwen  
    
     & \makecell{ \begin{tabular}{@{}r r@{}} 39.0 & 83.5 \end{tabular}}
    & \makecell{ \begin{tabular}{@{}r r@{}} 36.0 & 99.5 \end{tabular}}
    & \makecell{ \begin{tabular}{@{}r r@{}} 41.0 & 83.5 \end{tabular}} 
     
  \\ \hline \noalign{\vskip 3pt}
  
Mistral
    
      & \makecell{ \begin{tabular}{@{}r r@{}} 17.5 & 91.5   \end{tabular}}
    & \makecell{ \begin{tabular}{@{}r r@{}} 16.0 & 99.0 \end{tabular}}
    & \makecell{ \begin{tabular}{@{}r r@{}} 19.0 & 90.5 \end{tabular}} \\ \hline 
       \noalign{\vskip 3pt}
  
DSC  
    
    & \makecell{ \begin{tabular}{@{}r r@{}} 99.5 & 1.0 \end{tabular}}
    & \makecell{ \begin{tabular}{@{}r r@{}}  99.5 & 2.0 \end{tabular}}
    & \makecell{ \begin{tabular}{@{}r r@{}} 99.5 & 2.0 \end{tabular}} 
    \\

        \bottomrule
    \end{tabular}
    \caption{Performance on EquiBench}
    \label{tab:cmp_equibench}
\end{subtable}
\begin{subtable}[t]{0.33\textwidth} 
    \setlength{\tabcolsep}{1mm}
\begin{tabular}{p{1.2cm}lcc}
        \toprule
       \textbf{Model} & \textbf{neg} & \textbf{pos}    \\ \midrule
        Qwen  & 92.2 & 92.9    \\ \hline
        Mistral & 81.1 & 98.1   \\ \hline
        DSC & 3.5 & 99.4   \\
        \bottomrule
       \end{tabular}
    \caption{Performance on SeqCoBench}
    \label{tab:cmp_secobench}
\end{subtable}

    \caption{Comparing the performance of our Benchmark to existing Benchmarks}
\label{fig:cmp}
\end{table*}

\begin{table*}[t]
    \resizebox{\textwidth}{!}{
    \setlength{\tabcolsep}{1.7mm}
     \begin{tabular}{lcccccccccc}
     
        \toprule
       \textbf{Model} & \textbf{id} & \textbf{fe} & \textbf{ne} & \textbf{di} & \textbf{Avg}  &\textbf{W-Avg} & \textbf{Mac-F1} & \textbf{Mic-F1} & \textbf{PosF1} & \textbf{NegF1}  \\ 
       
       & \makecell{\begin{tabular}{@{}r  r@{}} 
                    \textbf{np} & \textbf{sp}
                \end{tabular}}
        & \makecell{\begin{tabular}{@{}r r@{}} 
                    \textbf{np} & \textbf{sp}
                \end{tabular}}
        & \makecell{\begin{tabular}{@{}r r@{}} 
                    \textbf{np} & \textbf{sp} 
                \end{tabular}}
        & \makecell{\begin{tabular}{@{}r r@{}} 
                    \textbf{np} & \textbf{sp}
                \end{tabular}} & & & & & & \\  \midrule
 
    \rowcolor{rowgray}
\multirow{3}{*}{ASTNN}
    & \makecell{\begin{tabular}{@{}r r@{}} 100.0 & 100.0  \end{tabular} }
    & \makecell{ \begin{tabular}{@{}r r@{}} 100.0 & 100.0 \end{tabular} }
    & \makecell{ \begin{tabular}{@{}r r@{}} 0.0 & 0.0 \end{tabular} } 
    & \makecell{ \begin{tabular}{@{}r r@{}} 0.0 & 0.0 \end{tabular} }
    & \makecell{50.0} & \makecell{50.0} & \makecell{50.0}
    & \makecell{31.6} & \makecell{63.2} & \makecell{0.0}\\
\cline{2-11} \noalign{\vskip 2pt}
   
    & \makecell{ \begin{tabular}{@{}r r@{}} 47.5 & 59.5 \end{tabular}}
    & \makecell{ \begin{tabular}{@{}r r@{}} 43.3 & 56.7 \end{tabular}} 
    & \makecell{ \begin{tabular}{@{}r r@{}} 43.8 & 43.1 \end{tabular}}
    & \makecell{ \begin{tabular}{@{}r r@{}} 43.0 & 39.0 \end{tabular}}
    & \makecell{47.0} & \makecell{50.7} & \makecell{63.6}
    & \makecell{44.1} & \makecell{33.0} & \makecell{55.3}
  \\

               \bottomrule
    \end{tabular}
    }
    \captionof{table}{Results comparing performance of ASTNN baseline.} 
    \label{tab:astnn}
\end{table*}

\subsection{Ablation Analysis}
\label{sec:ablation}
We conduct a comprehensive analysis to understand how various models respond to three key factors: the presence of Chain-of-Thought (CoT) reasoning, different difficulty levels of the problem, and the type of perturbation applied. 

\subsubsection{Effect of Chain of Thought Prompt}
\label{subsec:cot}
From table \ref{tab:s_cot2}, we observe that asking the model to first generate a CoT of reasoning before concluding to the final answers helps weaker models like Mistral, DeepSeekCoder significantly. We observe a significant boost in their performance in both original pairs as well as the perturbed pairs, as compared to prompting without chain of thought. However, we observe a drop in the performance of GPT-4o in perturbed pairs. This signifies the brittleness of reasoning for equivalence on the generated perturbations. We present the results for newer models for this and the following sections in Appendix \ref{sec:newer-models}. We also experiment with self-consistent decoding and observe the performance to be consistent with that in Table~\ref{tab:stricter_prompt}. Details and results are present in Appendix \ref{sec:appendix_self_consist}.

\subsubsection{Error Analysis}
On human evaluation of the Chain of thought generations, it is observed that for original pairs of programs, LLMs struggle to correctly identify that two different algorithms are solving the same problem and hence are equivalent. For many of the \mbox{$\langle\textit{fe},k,{\textit{np}}\rangle$} samples where the model correctly generates the final answer as no, its reasoning is flawed - incorrectly stating that the difference in the algorithm leads to inequivalent behaviour. The model fails to focus on the perturbation and the error caused by it in the intended algorithm. Apart from this, a significant number of correct final generations have missing reasoning details despite explicit instructions in the prompt to add them. For detailed results and sample generation examples, please refer to the Appendix \ref{sec:appendix-error-ana}.

\begin{table*}
    \centering
    \resizebox{\textwidth}{!}{
    \setlength{\tabcolsep}{1.7mm}
    \begin{tabular}{lcccccccccc}
        \toprule
       \textbf{Model} & \textbf{id} & \textbf{fe} & \textbf{ne} & \textbf{di} & \textbf{Avg} &\textbf{W-Avg} & \textbf{Mac-F1}  & \textbf{Mic-F1} & \textbf{PosF1} & \textbf{NegF1}  \\ 
       & \makecell{\begin{tabular}{@{}r  r@{}} 
                    \textbf{np} & \textbf{sp}
                \end{tabular}}
        & \makecell{\begin{tabular}{@{}r r@{}} 
                    \textbf{np} & \textbf{sp}
                \end{tabular}}
        & \makecell{\begin{tabular}{@{}r r@{}} 
                    \textbf{np} & \textbf{sp} 
                \end{tabular}}
        & \makecell{\begin{tabular}{@{}r r@{}} 
                    \textbf{np} & \textbf{sp}
                \end{tabular}} & & & & & & \\ \midrule

\rowcolor{rowgray}
\multirow{3}{*}{Qwen}
    & \makecell{\begin{tabular}{@{}r r@{}} 100.0 & 98.5  \end{tabular}} 
    & \makecell{\begin{tabular}{@{}r r@{}} 56.8 & 57.7  \end{tabular} }
    & \makecell{ \begin{tabular}{@{}r r@{}} 54.5 & 57.5 \end{tabular} }
    & \makecell{ \begin{tabular}{@{}r r@{}} 99.0 & 100.0 \end{tabular} } 
    & \makecell{79.6} & \makecell{79.6} & \makecell{87.3}
    & \makecell{69.0} & \makecell{67.6} & \makecell{70.4}\\
\cline{2-11} \noalign{\vskip 2pt} 
    & \makecell{ \begin{tabular}{@{}r r@{}} 24.1 & 85.7 \end{tabular}}
    & \makecell{ \begin{tabular}{@{}r r@{}} 43.3 & 58.9 \end{tabular}}
    & \makecell{ \begin{tabular}{@{}r r@{}} 63.5 & 59.6 \end{tabular}} 
    & \makecell{ \begin{tabular}{@{}r r@{}}  99.5 & 99.5  \end{tabular}}
    & \makecell{66.7} & \makecell{68.6} & \makecell{77.2}
    & \makecell{58.0} & \makecell{77.2} & \makecell{43.9}
  \\ \hline \noalign{\vskip 3pt}

    \rowcolor{rowgray}
\multirow{3}{*}{Mistral}
    & \makecell{\begin{tabular}{@{}r r@{}} 100.0 & 98.9  \end{tabular}} 
    & \makecell{\begin{tabular}{@{}r r@{}} 56.8 & 57.7 \end{tabular} }
    & \makecell{ \begin{tabular}{@{}r r@{}} 63.7 & 64.5 \end{tabular} }
    & \makecell{ \begin{tabular}{@{}r r@{}} 96.0 & 93.0 \end{tabular} } 
    & \makecell{78.8} & \makecell{78.8} & \makecell{86.9}
    & \makecell{68.4} & \makecell{65.5} & \makecell{71.5}\\
\cline{2-11} \noalign{\vskip 2pt} 
    & \makecell{ \begin{tabular}{@{}r r@{}} 1.57 & 94.9   \end{tabular}}
    & \makecell{ \begin{tabular}{@{}r r@{}} 39.6 & 59.6 \end{tabular}}
    & \makecell{ \begin{tabular}{@{}r r@{}} 59.0 & 60.2 \end{tabular}} 
    & \makecell{ \begin{tabular}{@{}r r@{}}  93.0 & 89.5 \end{tabular}}
    & \makecell{62.1} & \makecell{67.2} & \makecell{71.5}
    & \makecell{55.9} & \makecell{42.8} & \makecell{68.9}
  \\ \hline \noalign{\vskip 3pt}
       
    \rowcolor{rowgray}
\multirow{3}{*}{DSC}
    & \makecell{\begin{tabular}{@{}r r@{}} 98.9 & 98.7 \end{tabular}} 
    & \makecell{\begin{tabular}{@{}r r@{}} 87.2 & 87.2 \end{tabular} }
    & \makecell{ \begin{tabular}{@{}r r@{}} 23.8 & 24.0 \end{tabular} }
    & \makecell{ \begin{tabular}{@{}r r@{}} 89.0 & 86.0 \end{tabular} } 
    & \makecell{74.3} & \makecell{74.3} & \makecell{81.1}
    & \makecell{62.4} & \makecell{69.4} & \makecell{55.5}\\
\cline{2-11} \noalign{\vskip 2pt} 
    & \makecell{ \begin{tabular}{@{}r r@{}} 16.7 & 92.3 \end{tabular}}
    & \makecell{ \begin{tabular}{@{}r r@{}}  18.8 & 85.2 \end{tabular}}
    & \makecell{ \begin{tabular}{@{}r r@{}} 30.4 & 26.0 \end{tabular}} 
    & \makecell{ \begin{tabular}{@{}r r@{}} 92.0 & 89.5 \end{tabular}}
    & \makecell{56.3} & \makecell{48.5} & \makecell{65.8}
    & \makecell{67.1} & \makecell{43.8} & \makecell{53.2}
  \\ \hline \noalign{\vskip 3pt}
   
     \rowcolor{rowgray}
     \multirow{3}{*}{GPT-4o}
    & \makecell{\begin{tabular}{@{}r r@{}} 100.0 & 100.0 \end{tabular}} 
    & \makecell{\begin{tabular}{@{}r r@{}} 76.2 & 76.3 \end{tabular} }
    & \makecell{ \begin{tabular}{@{}r r@{}} 60.5 & 61.1 \end{tabular} }
    & \makecell{ \begin{tabular}{@{}r r@{}} 100.0 & 100.0 \end{tabular} } 
    & \makecell{84.3} & \makecell{84.3} & \makecell{90.5}
    & \makecell{75.7} & \makecell{75.3} & \makecell{76.2}\\
\cline{2-11} \noalign{\vskip 2pt} 
    & \makecell{ \begin{tabular}{@{}r r@{}} 65.5 & 99.0 \end{tabular}}
    & \makecell{ \begin{tabular}{@{}r r@{}} 47.6 & 74.6 \end{tabular}}
    & \makecell{ \begin{tabular}{@{}r r@{}} 74.0 & 66.4 \end{tabular}} 
    & \makecell{ \begin{tabular}{@{}r r@{}} 100.0 & 100.0 \end{tabular}}
    & \makecell{78.3} & \makecell{81.1} & \makecell{86.6}
    & \makecell{68.0} & \makecell{56.6} & \makecell{79.5}
  \\ 
        
               \bottomrule
    \end{tabular}
    }
    \caption{Results comparing various models using COT prompt}
    \label{tab:s_cot2}
\end{table*}

\subsubsection{Failure-Mode–Based Benchmarking} 
\label{subsec:failure-mode}  
We also analysed how different types of perturbations impact the models by comparing the performance drop (delta) before and after applying individual perturbation; the detailed results are provided in Appendix~\ref{sec:appendix_abalation_perturb}. On observing GPT-4o's delta plot, we observe that it struggles on various perturbations except the Variable Renaming type of perturbation. We hypothesised that the presence of a relatively easy-to-spot perturbation is helping the model to label inequivalent pairs correctly. To test this hypothesis, we generated a separate test set consisting of all the types of perturbations except Variable Renaming. In table \ref{tab:except_var}, the results of GPT-4o on this set show a significant decrease in the $\langle\textit{fe},k,\text{np}\rangle$  category and a higher decrease in the $\langle\textit{id},k,\text{np}\rangle$  category, unlike in table \ref{tab:stricter_prompt}. And we now observe a drop in GPT-4o's average accuracy for a combination of different types of perturbations, similar to individual perturbations, pinning down its weakness. We also performed analysis of perturbing code pairs of different difficulty levels and found a consistent gap in perturbed pairs for each of the difficulty levels. Detailed results are presented in Appendix~\ref{sec:appendix_ablation_diffi}.

\begin{table*}[t]
    \centering
    \resizebox{\textwidth}{!}{
    \setlength{\tabcolsep}{1.7mm}
    \begin{tabular}{lcccccccccc}
        \toprule
       \textbf{Model} & \textbf{id} & \textbf{fe} & \textbf{ne} & \textbf{di} & \textbf{Avg} &\textbf{W-Avg} & \textbf{Mac-F1}  & \textbf{Mic-F1} & \textbf{PosF1} & \textbf{NegF1}  \\ 
       & \makecell{\begin{tabular}{@{}r  r@{}} 
                    \textbf{np} & \textbf{sp}
                \end{tabular}}
        & \makecell{\begin{tabular}{@{}r r@{}} 
                    \textbf{np} & \textbf{sp}
                \end{tabular}}
        & \makecell{\begin{tabular}{@{}r r@{}} 
                    \textbf{np} & \textbf{sp} 
                \end{tabular}}
        & \makecell{\begin{tabular}{@{}r r@{}} 
                    \textbf{np} & \textbf{sp}
                \end{tabular}} & & & & & & \\ \midrule

    \rowcolor{rowgray}
    \multirow{3}{*}{GPT-4o} 
    & \makecell{\begin{tabular}{@{}r r@{}} 100.0 & 100.0 \end{tabular}} 
    & \makecell{\begin{tabular}{@{}r r@{}} 81.0 & 78.0 \end{tabular} }
    & \makecell{ \begin{tabular}{@{}r r@{}} 67.0 & 67.0 \end{tabular} }
    & \makecell{ \begin{tabular}{@{}r r@{}} 100.0 & 100.0 \end{tabular} } 
    & \makecell{86.6} & \makecell{86.6} & \makecell{92.2}
    & \makecell{89.0} & \makecell{87.0} & \makecell{90.1}\\
\cline{2-11} \noalign{\vskip 3pt} 
    & \makecell{ \begin{tabular}{@{}r r@{}} 75.5 & 94.0 \end{tabular}}
    & \makecell{ \begin{tabular}{@{}r r@{}} 65.5 & 80.5 \end{tabular}}
    & \makecell{ \begin{tabular}{@{}r r@{}} 90.5 & 71.0 \end{tabular}} 
    & \makecell{ \begin{tabular}{@{}r r@{}} 100.0 & 100.0 \end{tabular}}
    & \makecell{84.7} & \makecell{85.6} & \makecell{91.2}
    & \makecell{83.1} & \makecell{74.1} & \makecell{92.0}
  \\  

               \bottomrule
    \end{tabular}
    }
    \caption{Results on GPT-4o on combinations of all other perturbations except variable-renaming}
    \label{tab:except_var}
\end{table*}

\subsection{Finetuning}

We fine-tuned multiple large language models (LLMs) using three distinct data configurations: the original clean dataset, a perturbed version of the dataset, and a balanced 50:50 mixture of both. To assess performance, we evaluated each model against two test sets: one comprising unseen, unperturbed programs and another comprising perturbed programs. The models trained on the combined 50:50 dataset consistently achieved the best results (shown in table \ref{tab:stricter_prompt_ft}) for each LLM. (Other configurations are in Appendix \ref{sec:FT-ratios}.) These optimal models not only demonstrated improved performance on both the unperturbed and perturbed test sets but also reached a level of performance comparable to GPT-4o. Further experimental details are available in Appendix \ref{sec:appendix-exp}.

\begin{table*}[!t]
    \centering
     \resizebox{\textwidth}{!}{    
   \setlength{\tabcolsep}{1.7mm}
    \begin{tabular}{lcccccccccc}
        \toprule
       \textbf{Model} & \textbf{id} & \textbf{fe} & \textbf{ne} & \textbf{di} & \textbf{Avg} &\textbf{W-Avg} & \textbf{Mac-F1}  & \textbf{Mic-F1} & \textbf{PosF1} & \textbf{NegF1}  \\ 
       & \makecell{\begin{tabular}{@{}r  r@{}} 
                    \textbf{np} & \textbf{sp}
                \end{tabular}}
        & \makecell{\begin{tabular}{@{}r r@{}} 
                    \textbf{np} & \textbf{sp}
                \end{tabular}}
        & \makecell{\begin{tabular}{@{}r r@{}} 
                    \textbf{np} & \textbf{sp} 
                \end{tabular}}
        & \makecell{\begin{tabular}{@{}r r@{}} 
                    \textbf{np} & \textbf{sp}
                \end{tabular}} & & & & & & \\ \midrule

    \rowcolor{rowgray}
\multirow{3}{*}{Qwen} 
    & \makecell{\begin{tabular}{@{}r r@{}}  98.0 & 97.0 \end{tabular} }
    & \makecell{ \begin{tabular}{@{}r r@{}} 81.0 & 81.5  \end{tabular} }
    & \makecell{ \begin{tabular}{@{}r r@{}} 46.3 & 45.1 \end{tabular} } 
     & \makecell{ \begin{tabular}{@{}r r@{}} 99.5 & 99.0 \end{tabular} } 
    & \makecell{80.9} & \makecell{80.9} & \makecell{87.6}
    & \makecell{71.5} & \makecell{73.1} & \makecell{69.8}\\
\cline{2-11} \noalign{\vskip 2pt} 
    & \makecell{ \begin{tabular}{@{}r r@{}} 82.7 & 96.5 \end{tabular}}
    & \makecell{ \begin{tabular}{@{}r r@{}} 82.2 & 90.1 \end{tabular}} 
    & \makecell{ \begin{tabular}{@{}r r@{}} 85.3 & 39.4 \end{tabular}}
    & \makecell{ \begin{tabular}{@{}r r@{}}  99.5 & 97.5 \end{tabular}}
    & \makecell{84.2} & \makecell{87.3} & \makecell{90.1}
    & \makecell{75.8} & \makecell{66.9} & \makecell{84.8}
  \\ \hline \noalign{\vskip 3pt}
 
    \rowcolor{rowgray}  
 \multirow{3}{*}{Mistral}
    & \makecell{\begin{tabular}{@{}r r@{}}  100.0 & 99.5  \end{tabular} }
    & \makecell{ \begin{tabular}{@{}r r@{}} 63.6 & 63.7  \end{tabular} }
    & \makecell{ \begin{tabular}{@{}r r@{}} 61.5 & 59.3 \end{tabular} } 
     & \makecell{ \begin{tabular}{@{}r r@{}} 99.0 & 98.5 \end{tabular} } 
    & \makecell{80.6} & \makecell{80.6} & \makecell{88.0}
    & \makecell{65.7} & \makecell{87.7} & \makecell{64.7}\\
\cline{2-11} \noalign{\vskip 2pt} 
    & \makecell{ \begin{tabular}{@{}r r@{}} 77.0 & 87.5 \end{tabular}}
    & \makecell{ \begin{tabular}{@{}r r@{}} 76.7 & 67.6  \end{tabular}} 
    & \makecell{ \begin{tabular}{@{}r r@{}} 87.5 & 55.6 \end{tabular}}
    & \makecell{ \begin{tabular}{@{}r r@{}} 100.0 & 98.0 \end{tabular}}
    & \makecell{81.2} & \makecell{80.0} & \makecell{88.9}
    & \makecell{71.2} & \makecell{58.5} & \makecell{83.9}
  \\ \hline \noalign{\vskip 3pt} 
   
    \rowcolor{rowgray}
 \multirow{3}{*}{DSC}
    & \makecell{\begin{tabular}{@{}r r@{}} 100.0 & 100.0   \end{tabular} }
    & \makecell{ \begin{tabular}{@{}r r@{}} 84.1 & 83.9   \end{tabular} }
    & \makecell{ \begin{tabular}{@{}r r@{}} 42.6 & 41.0 \end{tabular} } 
     & \makecell{ \begin{tabular}{@{}r r@{}} 99.5 & 100.0 \end{tabular} } 
    & \makecell{81.3} & \makecell{81.3} & \makecell{87.5}
    & \makecell{71.2} & \makecell{73.6} & \makecell{68.7}\\
\cline{2-11} \noalign{\vskip 2pt} 
    & \makecell{ \begin{tabular}{@{}r r@{}} 89.0 & 91.41 \end{tabular}}
    & \makecell{ \begin{tabular}{@{}r r@{}} 84.2 & 88.9  \end{tabular}} 
    & \makecell{ \begin{tabular}{@{}r r@{}} 87.4 & 39.0  \end{tabular}}
    & \makecell{ \begin{tabular}{@{}r r@{}} 100.0 & 98.5 \end{tabular}}
    & \makecell{84.8} & \makecell{86.5} & \makecell{90.4}
    & \makecell{76.2} & \makecell{67.0} & \makecell{85.4}
  \\ 

               \bottomrule
    \end{tabular}
    }
    \captionof{table}{Effect of finetuning}
    \label{tab:stricter_prompt_ft}
\end{table*}
\subsection{Cross-lingual Equivalence}
\label{sec:cross-lingual}

We extend our benchmark to evaluate cross-lingual program equivalence. For this, we created a test set of Java-Python code pairs of the same size as our monolingual sets. We generate two different variants of the cross-lingual dataset: (a) Single code perturbed, and (b) Both codes perturbed. 
In the single code perturbed cross-lingual setting, we apply perturbations only to the Python programs. We then compare the performance of LLMs on perturbed pairs and their corresponding original versions (unperturbed pairs). As shown in Table~\ref{tab:java}, the LLMs find it more difficult to verify equivalence for unperturbed cross-lingual pairs than for their monolingual counterparts. Furthermore, perturbing the Python program leads to an additional drop in performance, mirroring the trend observed in the monolingual setting. In the Both codes perturbed setting, we randomly select either the Java or the Python code to be perturbed. The results are shown in Table~\ref{tab:java_both}. They are quantitatively similar to those in Table~\ref{tab:java}, pointing to the finding that our analysis is agnostic to which language(s) the perturbation is applied to in the cross-lingual setting. Further details for cross-lingual datasets generation are present in \ref{sec:appendix_crossling_dataset}. Similar results were observed for C++ and Python pairs, with detailed results available in Appendix \ref{sec:appendix-analysis}.

\begin{table*}[!t]
    \centering
    \resizebox{\textwidth}{!}{
    \begin{tabular}{lcccccccccc}
        \toprule
       \textbf{Model} &  \textbf{fe} & \textbf{ne} & \textbf{di} & \textbf{Avg} &\textbf{W-Avg} & \textbf{Mac-F1}  & \textbf{Mic-F1} & \textbf{PosF1} & \textbf{NegF1}  \\ 
       & \makecell{\begin{tabular}{@{}r  r@{}} 
                    \textbf{np} & \textbf{sp}
                \end{tabular}}
        & \makecell{\begin{tabular}{@{}r r@{}} 
                    \textbf{np} & \textbf{sp}
                \end{tabular}}
        & \makecell{\begin{tabular}{@{}r r@{}} 
                    \textbf{np} & \textbf{sp}
                \end{tabular}} & & & & & & & \\ \midrule

\rowcolor{rowgray}
\multirow{3}{*}{Qwen} 
    & \makecell{\begin{tabular}{@{}r r@{}} 87.5 & 87.8  \end{tabular} }
    & \makecell{ \begin{tabular}{@{}r r@{}} 33.4 & 33.9 \end{tabular} }
    & \makecell{ \begin{tabular}{@{}r r@{}} 100.0 & 99.5 \end{tabular} } 
    & \makecell{73.7} & \makecell{77.2} & \makecell{81.2}
    & \makecell{68.3} & \makecell{72.1} & \makecell{64.4}\\
\cline{2-11} \noalign{\vskip 2pt} 
   
    & \makecell{ \begin{tabular}{@{}r r@{}} 46.7 & 82.8 \end{tabular}}
    & \makecell{ \begin{tabular}{@{}r r@{}} 65.7 & 41.3 \end{tabular}} 
    & \makecell{ \begin{tabular}{@{}r r@{}}  100.0 & 100.0  \end{tabular}}
    & \makecell{72.8} & \makecell{75.0} & \makecell{82.0}
    & \makecell{63.0} & \makecell{52.8} & \makecell{73.2}
  \\ \hline \noalign{\vskip 3pt}  
 
    \rowcolor{rowgray}
\multirow{3}{*}{Mistral}
    & \makecell{\begin{tabular}{@{}r r@{}} 46.7 & 82.8 \end{tabular} }
    & \makecell{ \begin{tabular}{@{}r r@{}} 63.7 & 64.5 \end{tabular} }
    & \makecell{ \begin{tabular}{@{}r r@{}} 96.0 & 93.0 \end{tabular} } 
    & \makecell{78.8} & \makecell{78.8} & \makecell{86.9}
    & \makecell{68.4} & \makecell{65.5} & \makecell{71.5}\\
\cline{2-11} \noalign{\vskip 2pt} 
   
    & \makecell{ \begin{tabular}{@{}r r@{}} 39.6 & 59.6 \end{tabular}}
    & \makecell{ \begin{tabular}{@{}r r@{}} 59.0 & 60.2 \end{tabular}} 
    & \makecell{ \begin{tabular}{@{}r r@{}}  93.0 & 89.5 \end{tabular}}
    & \makecell{62.1} & \makecell{67.8} & \makecell{71.5}
    & \makecell{55.9} & \makecell{64.0} & \makecell{68.9}
  \\ \hline \noalign{\vskip 3pt}
  
    \rowcolor{rowgray}      
\multirow{3}{*}{DSC}
    & \makecell{\begin{tabular}{@{}r r@{}} 99.7 & 99.3 \end{tabular} }
    & \makecell{ \begin{tabular}{@{}r r@{}} 0.9 & 0.8 \end{tabular} }
    & \makecell{ \begin{tabular}{@{}r r@{}} 3.5 & 2.5 \end{tabular} } 
    & \makecell{34.5} & \makecell{50.7} & \makecell{35.8}
    & \makecell{33.1} & \makecell{63.3} & \makecell{2.9}\\
\cline{2-11} \noalign{\vskip 2pt} 
    
    & \makecell{ \begin{tabular}{@{}r r@{}}  2.2 & 99.5 \end{tabular}}
    & \makecell{ \begin{tabular}{@{}r r@{}} 2.6 & 1.0 \end{tabular}} 
    & \makecell{ \begin{tabular}{@{}r r@{}} 10.5 & 2.5 \end{tabular}}
    & \makecell{19.7} & \makecell{51.6} & \makecell{22.5}
    & \makecell{21.8} & \makecell{38.0} & \makecell{5.6}
  \\ \hline \noalign{\vskip 3pt}
  
    \rowcolor{rowgray}
  \multirow{3}{*}{GPT-4o} 
    & \makecell{\begin{tabular}{@{}r r@{}} 68.0 & 64.7 \end{tabular} }
    & \makecell{ \begin{tabular}{@{}r r@{}} 74.8 & 76.6 \end{tabular} }
    & \makecell{ \begin{tabular}{@{}r r@{}} 100.0 & 100.0 \end{tabular} } 
    & \makecell{80.7} & \makecell{76.9} & \makecell{88.6}
    & \makecell{74.6} & \makecell{71.1} & \makecell{78.1}\\
\cline{2-11} \noalign{\vskip 2pt} 
   
    & \makecell{ \begin{tabular}{@{}r r@{}} 79.7 & 64.8 \end{tabular}}
    & \makecell{ \begin{tabular}{@{}r r@{}} 93.4 & 79.5 \end{tabular}} 
    & \makecell{ \begin{tabular}{@{}r r@{}} 100.0 & 100.0 \end{tabular}}
    & \makecell{86.2} & \makecell{75.9} & \makecell{92.1}
    & \makecell{75.3} & \makecell{62.4} & \makecell{88.2}
  \\ 
        
               \bottomrule
    \end{tabular}
    }
    \caption{Results comparing various models for pairs with one Java code and one Python code  }
    \label{tab:java}
\end{table*}

\begin{table*}[!t]
    \centering
    \resizebox{\textwidth}{!}{
    \begin{tabular}{lcccccccccc}
        \toprule
       \textbf{Model} &  \textbf{fe} & \textbf{ne} & \textbf{di} & \textbf{Avg} &\textbf{W-Avg} & \textbf{Mac-F1}  & \textbf{Mic-F1} & \textbf{PosF1} & \textbf{NegF1}  \\ 
       & \makecell{\begin{tabular}{@{}r  r@{}} 
                    \textbf{np} & \textbf{sp}
                \end{tabular}}
        & \makecell{\begin{tabular}{@{}r r@{}} 
                    \textbf{np} & \textbf{sp}
                \end{tabular}}
       
        & \makecell{\begin{tabular}{@{}r r@{}} 
                    \textbf{np} & \textbf{sp}
                \end{tabular}} & & & & & & & \\ \midrule

    \rowcolor{rowgray}
\multirow{3}{*}{Qwen}
    & \makecell{\begin{tabular}{@{}r r@{}} 81.6 & 84.4  \end{tabular} }
    & \makecell{ \begin{tabular}{@{}r r@{}} 37.9 & 40.9 \end{tabular} }
    & \makecell{ \begin{tabular}{@{}r r@{}} 100.0 & 100.0 \end{tabular} } 
    & \makecell{73.9} & \makecell{76.0} & \makecell{82.3}
    & \makecell{67.7} & \makecell{68.3} & \makecell{67.2}\\
\cline{2-11} \noalign{\vskip 2pt}
   
    & \makecell{ \begin{tabular}{@{}r r@{}} 39.6 & 82.1 \end{tabular}}
    & \makecell{ \begin{tabular}{@{}r r@{}} 60.9 & 42.1 \end{tabular}} 
    & \makecell{ \begin{tabular}{@{}r r@{}} 100.0 & 100.0 \end{tabular}}
    & \makecell{70.8} & \makecell{75.3} & \makecell{80.3}
    & \makecell{60.0} & \makecell{48.3} & \makecell{71.7}
  \\ \hline \noalign{\vskip 3pt}  
 
    \rowcolor{rowgray}
\multirow{3}{*}{Mistral}
    & \makecell{\begin{tabular}{@{}r r@{}} 93.6 & 93.4 \end{tabular} }
    & \makecell{ \begin{tabular}{@{}r r@{}} 12.6 & 12.9 \end{tabular} }
    & \makecell{ \begin{tabular}{@{}r r@{}} 72.0 & 73.0 \end{tabular} } 
    & \makecell{59.6} & \makecell{68.1} & \makecell{67.8}
    & \makecell{54.7} & \makecell{64.7} & \makecell{44.8}\\
\cline{2-11} \noalign{\vskip 2pt}
   
    & \makecell{ \begin{tabular}{@{}r r@{}} 13.3 & 91.7 \end{tabular}}
    & \makecell{ \begin{tabular}{@{}r r@{}} 22.2 & 15.3 \end{tabular}} 
    & \makecell{ \begin{tabular}{@{}r r@{}} 78.0 & 73.5 \end{tabular}}
    & \makecell{49.0} & \makecell{66.1} & \makecell{59.1}
    & \makecell{42.2} & \makecell{39.8} & \makecell{44.6}
  \\ \hline \noalign{\vskip 3pt}
 
    \rowcolor{rowgray}       
\multirow{3}{*}{DSC}
    & \makecell{\begin{tabular}{@{}r r@{}} 99.3 & 99.2 \end{tabular} }
    & \makecell{ \begin{tabular}{@{}r r@{}} 1.4 & 0.6 \end{tabular} }
    & \makecell{ \begin{tabular}{@{}r r@{}} 8.0 & 19.0 \end{tabular} } 
    & \makecell{37.9} & \makecell{53.3} & \makecell{41.8}
    & \makecell{34.5} & \makecell{60.2} & \makecell{8.9}\\
\cline{2-11} \noalign{\vskip 2pt}
    
    & \makecell{ \begin{tabular}{@{}r r@{}} 2.0 & 98.9 \end{tabular}}
    & \makecell{ \begin{tabular}{@{}r r@{}} 2.4 & 1.2 \end{tabular}} 
    & \makecell{ \begin{tabular}{@{}r r@{}} 16.5 & 13.5 \end{tabular}}
    & \makecell{22.4} & \makecell{53.0} & \makecell{27.1}
    & \makecell{22.0} & \makecell{35.2} & \makecell{8.8}
  \\ \hline \noalign{\vskip 3pt}
 
    \rowcolor{rowgray}
\multirow{3}{*}{GPT-4o}
    & \makecell{\begin{tabular}{@{}r r@{}} 61.9 & 62.9 \end{tabular} }
    & \makecell{ \begin{tabular}{@{}r r@{}} 78.1 & 76.8 \end{tabular} }
    & \makecell{ \begin{tabular}{@{}r r@{}} 100.0 & 100.0 \end{tabular} } 
    & \makecell{79.9} & \makecell{75.6} & \makecell{88.0}
    & \makecell{73.5} & \makecell{67.5} & \makecell{79.6}\\
\cline{2-11} \noalign{\vskip 2pt}
   
    & \makecell{ \begin{tabular}{@{}r r@{}} 83.5 & 61.9 \end{tabular}}
    & \makecell{ \begin{tabular}{@{}r r@{}} 95.7 & 77.4 \end{tabular}} 
    & \makecell{ \begin{tabular}{@{}r r@{}} 100.0 & 100.0 \end{tabular}}
    & \makecell{86.4} & \makecell{76.6} & \makecell{92.1}
    & \makecell{74.7} & \makecell{60.3} & \makecell{89.2}
  \\

               \bottomrule
    \end{tabular}
    }
    \caption{Results comparing various models for pairs with one Java code and one Python code (Both Perturbed) }
    \label{tab:java_both}
\end{table*}

\subsection{Transferability of Finetuned Models}

\setcounter{dbltopnumber}{10}  
To determine if the skills learned during fine-tuning generalize, we evaluated models in a zero-shot transfer setting. Models fine-tuned on 900-level Python-Python pairs were tested without further training on: (1) Python programs of higher difficulty levels, and (2) cross-lingual code pairs. Table~\ref{tab:transfer_difficulty} shows that finetuned models exhibit stable improvements across difficulty levels, indicating effective knowledge transfer.
Similarly, Table~\ref{tab:transfer_language} demonstrates consistent accuracy gains across Python-Python, Python-Java, and Python-C++ pairs, where the perturbation is always applied to a single Python program. Notably, models with initially poor performance improved the most, highlighting the benefits of our fine-tuning in challenging or noisy settings.

\begin{table*}[!hbt]
\centering %
\begin{subtable}{0.5\textwidth}
\resizebox{\textwidth}{!}{
    \setlength{\tabcolsep}{1mm}
    \begin{tabular}{lccccc}
        \toprule
       \textbf{Model} & \textbf{900} & \textbf{1200} & \textbf{1600} & \textbf{1900} & \textbf{2200}  \\ 
        & \makecell{\begin{tabular}{@{}r  r@{}}  
                     \textbf{B} & \textbf{A}
                 \end{tabular}}
         & \makecell{\begin{tabular}{@{}r r@{}}  
                     \textbf{B} & \textbf{A}
                 \end{tabular}}
         & \makecell{\begin{tabular}{@{}r r@{}}  
                     \textbf{B} & \textbf{A} 
                 \end{tabular}}
         & \makecell{\begin{tabular}{@{}r r@{}}  
                     \textbf{B} & \textbf{A}
                 \end{tabular}}  
         & \makecell{\begin{tabular}{@{}r  r@{}}  
                     \textbf{B} & \textbf{A}
                 \end{tabular}} \\ \midrule
  
    \rowcolor{rowgray}
    \multirow{3}{*}{Qwen}
     & \makecell{\begin{tabular}{@{}r r@{}}  80.6 & 80.9 \end{tabular} }
     & \makecell{ \begin{tabular}{@{}r r@{}} 76.1 & 74.9  \end{tabular} }
     & \makecell{ \begin{tabular}{@{}r r@{}} 78.1 & 77.5 \end{tabular} } 
      & \makecell{ \begin{tabular}{@{}r r@{}} 74.8 & 76.8 \end{tabular} }  
       & \makecell{ \begin{tabular}{@{}r r@{}} 80.6 & 80.4 \end{tabular} } \\
\cline{2-6} \noalign{\vskip 2pt} 
     & \makecell{ \begin{tabular}{@{}r r@{}} 69.2 & 84.2 \end{tabular}}
     & \makecell{ \begin{tabular}{@{}r r@{}} 62.7 & 80.3 \end{tabular}} 
     & \makecell{ \begin{tabular}{@{}r r@{}} 68.1 & 81.4 \end{tabular}}
     & \makecell{ \begin{tabular}{@{}r r@{}} 62.1 & 81.6 \end{tabular}}
     & \makecell{ \begin{tabular}{@{}r r@{}} 66.6 & 75.6 \end{tabular} } 
     
  \\ \hline \noalign{\vskip 3pt}
   
    \rowcolor{rowgray} 
 \multirow{3}{*}{Mistral}
     & \makecell{\begin{tabular}{@{}r r@{}}  76.3 & 80.6 \end{tabular} }
     & \makecell{ \begin{tabular}{@{}r r@{}} 67.0 & 84.3  \end{tabular} }
     & \makecell{ \begin{tabular}{@{}r r@{}} 65.5 & 79.4 \end{tabular} } 
      & \makecell{ \begin{tabular}{@{}r r@{}} 66.7 & 77.5 \end{tabular} }  
       & \makecell{ \begin{tabular}{@{}r r@{}} 58.2 & 77.3 \end{tabular} } \\
\cline{2-6} \noalign{\vskip 2pt} 
     & \makecell{ \begin{tabular}{@{}r r@{}} 56.6 & 81.2 \end{tabular}}
     & \makecell{ \begin{tabular}{@{}r r@{}} 46.3 & 77.4 \end{tabular}} 
     & \makecell{ \begin{tabular}{@{}r r@{}} 44.3 & 75.5 \end{tabular}}
     & \makecell{ \begin{tabular}{@{}r r@{}} 46.2 & 72.7 \end{tabular}}
     & \makecell{ \begin{tabular}{@{}r r@{}} 40.4 & 72.5 \end{tabular} } 
     
  \\ \hline \noalign{\vskip 3pt}
   
    \rowcolor{rowgray} 
 \multirow{3}{*}{DSC}
     & \makecell{\begin{tabular}{@{}r r@{}}  50.6 & 81.3 \end{tabular} }
     & \makecell{ \begin{tabular}{@{}r r@{}} 52.3 & 82.9  \end{tabular} }
     & \makecell{ \begin{tabular}{@{}r r@{}} 50.7 & 86.1 \end{tabular} } 
      & \makecell{ \begin{tabular}{@{}r r@{}} 53.0 & 78.1 \end{tabular} }  
       & \makecell{ \begin{tabular}{@{}r r@{}} 49.9 & 81.4 \end{tabular} } \\
\cline{2-6} \noalign{\vskip 2pt} 
     & \makecell{ \begin{tabular}{@{}r r@{}} 23.6 & 84.8 \end{tabular}}
     & \makecell{ \begin{tabular}{@{}r r@{}} 28.3 & 84.1 \end{tabular}} 
     & \makecell{ \begin{tabular}{@{}r r@{}} 27.3 & 82.7 \end{tabular}}
     & \makecell{ \begin{tabular}{@{}r r@{}}  29.0 & 81.6 \end{tabular}}
     & \makecell{ \begin{tabular}{@{}r r@{}} 25.9 & 79.6 \end{tabular} } 
     
  \\ 
        \bottomrule
    \end{tabular}
    }
    \caption{Transfer across problem difficulty levels}
    \label{tab:transfer_difficulty}
\end{subtable}
\begin{subtable}{0.33\textwidth}
\resizebox{\textwidth}{!}{
    \setlength{\tabcolsep}{1mm}
    \begin{tabular}{lccc}
        \toprule
       \textbf{Model} & \textbf{Py-Py} & \textbf{Py-Java} & \textbf{Py-C++}   \\ 
        & \makecell{\begin{tabular}{@{}r  r@{}}  
                     \textbf{B} & \textbf{A}
                 \end{tabular}}
         & \makecell{\begin{tabular}{@{}r r@{}}  
                     \textbf{B} & \textbf{A}
                 \end{tabular}}
         & \makecell{\begin{tabular}{@{}r r@{}}  
                     \textbf{B} & \textbf{A} 
                 \end{tabular}}
         \\ \midrule  
    \rowcolor{rowgray}
\multirow{3}{*}{Qwen}
     & \makecell{\begin{tabular}{@{}r r@{}}  81.4 & 82.7 \end{tabular} }
     & \makecell{ \begin{tabular}{@{}r r@{}} 73.7 & 75.6 \end{tabular} }
     & \makecell{ \begin{tabular}{@{}r r@{}} 73.3 & 67.1 \end{tabular} } 
\\
\cline{2-4} \noalign{\vskip 2pt} 
     & \makecell{ \begin{tabular}{@{}r r@{}} 74.2 & 83.6 \end{tabular}}
     & \makecell{ \begin{tabular}{@{}r r@{}} 72.8 & 77.0 \end{tabular}} 
     & \makecell{ \begin{tabular}{@{}r r@{}} 74.3 & 78.6 \end{tabular}}

  \\ \hline \noalign{\vskip 3pt}
    
    \rowcolor{rowgray}
 \multirow{3}{*}{Mistral}
     & \makecell{\begin{tabular}{@{}r r@{}}  77.2 & 81.3 \end{tabular} }
     & \makecell{ \begin{tabular}{@{}r r@{}} 59.8 & 72.5  \end{tabular} }
     & \makecell{ \begin{tabular}{@{}r r@{}} 58.9 & 72.0 \end{tabular} } 
      \\
\cline{2-4} \noalign{\vskip 2pt} 
     & \makecell{ \begin{tabular}{@{}r r@{}} 57.4 & 79.7 \end{tabular}}
     & \makecell{ \begin{tabular}{@{}r r@{}} 51.5 & 79.6 \end{tabular}} 
     & \makecell{ \begin{tabular}{@{}r r@{}} 48.2 & 77.2 \end{tabular}}

  \\ \hline \noalign{\vskip 3pt}
    
    \rowcolor{rowgray}
 \multirow{3}{*}{DSC}
     & \makecell{\begin{tabular}{@{}r r@{}}  50.6 & 81.8 \end{tabular} }
     & \makecell{ \begin{tabular}{@{}r r@{}} 34.5 & 75.3  \end{tabular} }
     & \makecell{ \begin{tabular}{@{}r r@{}} 34.7 & 75.5 \end{tabular} } 
     \\
\cline{2-4} \noalign{\vskip 2pt} 
     & \makecell{ \begin{tabular}{@{}r r@{}} 26.8 & 83.2 \end{tabular}}
     & \makecell{ \begin{tabular}{@{}r r@{}} 19.7 & 80.9 \end{tabular}} 
     & \makecell{ \begin{tabular}{@{}r r@{}} 19.0 & 81.3 \end{tabular}}

  \\ 

        \bottomrule
        
    \end{tabular}
    }
    \caption{Transfer across languages}
    \label{tab:transfer_language}
\end{subtable}

\caption{Testing finetuned model across difficulty levels and cross-lingual setting. \textbf{B}: Before, \textbf{A}: After Finetuning}
\label{fig:tranfer_learning_ft}
\end{table*}

\section{Conclusion}
We have introduced \dataset, a comprehensive benchmark for evaluating LLMs on the task of code-equivalence checking via controlled program transformations. Our experiments show that even simple perturbations applied to program code can significantly degrade the performance of state-of-the-art LLMs, highlighting their limitations in understanding functional semantics. \dataset’s methodology is generalizable, supporting extension to multiple programming languages and transformation types, and enables deeper analysis of LLM capabilities. Fine-tuning with perturbed examples improves performance, suggesting room for enhancement through targeted supervision. These findings underscore that current LLMs rely heavily on surface patterns rather than true semantic understanding of code. 

\newpage
\clearpage
\section*{Limitations}

While our analysis presents deep insights into the performance of LLMs on the task of code equivalence under various perturbations, several limitations remain. First, although we have performed experiments on a diverse set of models, the findings may not hold true for other models or future versions of the current models.

The dataset uses a binary Yes/No labelling scheme for functional equivalence, which oversimplifies nuanced cases where equivalence may be partial, conditional, or input-dependent. Furthermore, while the transformation set covers a wide range of syntactic and semantic changes, it is not exhaustive. Real-world coding variations—such as advanced refactoring patterns, concurrency-related modifications, or differences in library usage—are not fully captured.

Finally, although fine-tuning leads to performance gains within the distribution of the dataset, it remains unclear how well these models generalise to unseen perturbation types or real-world scenarios that lack the synthetic structure of our benchmarks.

\section*{Acknowledgements}
This work was supported by an IBM AI Horizons Network (AIHN) grant and IBM SUR Awards. Parag is supported by the Shanthi and K Ananth Krishnan Young Faculty Chair Professorship in AI, apart from the IBM AI Horizon Networks (AIHN) grant. We thank the IIT Delhi HPC facility\footnote{\url{https://supercomputing.iitd.ac.in/}} for computational resources, and Google Gemini Credits grant.  We thank anonymous reviewers for their insightful comments that helped in further improving our paper. Any opinions, findings, conclusions or recommendations expressed in this paper are those of the authors and do not necessarily reflect the views or official policies, either expressed or implied, of the funding agencies.
\bibliography{custom}

\appendix
\newpage
\section{Experimental Details}
\label{sec:appendix-exp}

\newtcolorbox{violetlisting}[1][]{%
  listing only,
  listing options={
    basicstyle=\ttfamily\small,
    breaklines=true,      
    breakatwhitespace=false,
    columns=fullflexible,
    keepspaces=true,
    language=Python      
  },
  colback=myviolet!5,
  colframe=myviolet,
  title=Code Example,
  fonttitle=\bfseries,
  sharp corners,
  enhanced,
  breakable,
  #1
}

\begin{figure*}[htbp]  
\centering
\begin{minipage}{0.95\textwidth}
\begin{violetlisting}[title={Basic Prompt}]
Given two programs, you must check if they solve the same problem - that is, for any input, they must produce identical outputs. \\
Your task is to answer only with one word: Yes or No. \\
No explanation, no extra words, no formatting.

Program 1: \{prog1\} \\
Program 2: \{prog2\}
\end{violetlisting}

\begin{violetlisting}[title={COT Prompt}]
You are given two programs. Your task is to analyze whether they solve the same problem — meaning, for any given input, both programs must always produce identical outputs. \\
\\
Start by reasoning step-by-step through the logic and structure of both programs. You must write your reasoning steps. \\
\\
After your analysis, you must end your response with exactly one of the following: \\
FINAL\_ANSWER\_IS\_YES \\
FINAL\_ANSWER\_IS\_NO \\
\\
Only use this exact format for the final answer, and do not write anything after it. \\

Program 1: \{prog1\} \\
Program 2: \{prog2\}
\end{violetlisting}

\end{minipage}
\end{figure*}

To keep our results reproducible, we use these hyperparameters of LLMs in our experiments: We set sampling to False, temperature to 0. Hence, we report the final results using a single run of the inference experiments.  \\
For finetuning, we used LoRA finetuning using A100 GPU. We finetuned for nearly 48 GPU hours and selected the model with the least validation loss. The training and validation datasets were generated by the same algorithm \ref{alg:perturb}. We initially split the 900-level problems in the CodeContests dataset into non-overlapping sets of 8, 8 and 40 problems, respectively, for test, validation and train sets. We then use the correct and incorrect submissions to these sets of problems to get the solution clusters for sampling. We used 2000 examples of each of the 8 categories and their original pairs for training. And we used 200 examples of each of the categories in the validation set.
We experimented with learning rates from $1e-3$ to $1e-6$, and experimented with different combinations of $[\ q\_proj,\ k\_proj,\ v\_proj,\ o\_proj \ ]$ layers for LoRA. \\ 
We made use of the \textbf{trl} \footnote{\url{https://github.com/huggingface/trl},  \url{  https://huggingface.co/docs/trl/en/sft_trainer}} library for SFTTrainer for finetuning and \textbf{ast} \footnote{\url{https://docs.python.org/3/library/ast.html}} and \textbf{astor} \footnote{\url{https://astor.readthedocs.io/en/latest/}} libraries for dataset perturbation. We would be publicly releasing the code, which includes the setting of seeds for any randomness used, along with setting the final seeds and hyperparameters used for reproducing the experiments.

The zero-shot prompt used in our experiments is shared in the following page. We also experimented with various few-shot prompting methods, but the performance of models degraded. We hypothesize that this is caused by a larger context that involves multiple long pieces of code pairs. In Table~\ref{tab:stricter_prompt_fs}, we present the results of few shot prompt with 4 examples each of positive and negative labels given along with the instruction of zero-shot prompt. It is observed that models get bias towards positive label.


\subsection{Cross-Lingual Dataset Generation}
\label{sec:appendix_crossling_dataset}

The cross-lingual setting differs from the procedure in Section~\ref{sec:dataset-curation} mainly in how we filter the original program clusters. Our source corpus, CodeContests, contains submissions to problems in multiple languages, along with a language label for each code snippet. Using these labels, we filter the solutions to retain only those written in the selected languages. This yields a collection $\mathcal{P}$ containing the two languages used in our experiments, e.g., Java and Python for Tables~\ref{tab:java} and~\ref{tab:java_both}.

In Algorithm~\ref{alg:perturb}, we additionally ensure at step 1 that the sampled code pair from $\mathcal{P}$ contains programs written in different languages, i.e., the languages of $s_1$ and $s_2$ are distinct. For the single-code perturbation setting, we skip steps 10 and 13 in Algorithm~\ref{alg:perturb}, and designate the Python program in the selected pair as $\hat{s}$, i.e., the code marked for perturbation. For the both-code perturbation setting, we keep the randomization unchanged.

\subsection{Data Statistics}

\begin{figure*}[htbp]
  \centering
  \begin{subfigure}{0.32\textwidth}
    \centering
    \includegraphics[width=\linewidth]{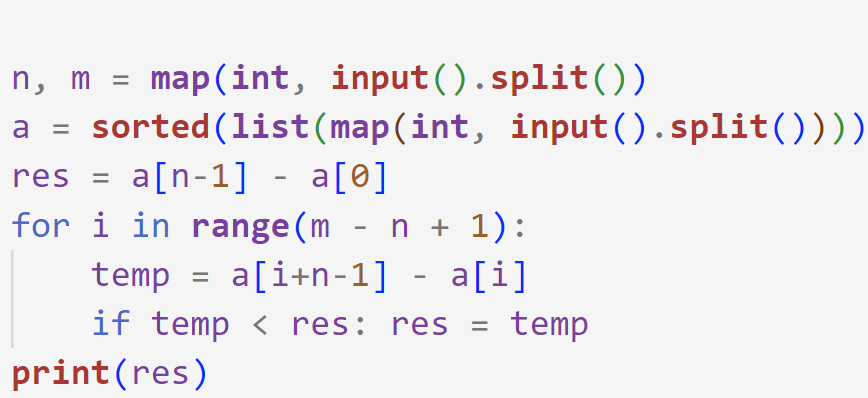}
    \caption{Original Code}
  \end{subfigure}
  \hfill
  \begin{subfigure}{0.32\textwidth}
    \centering
    \includegraphics[width=\linewidth]{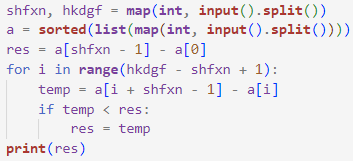}
    \caption{Perturbed Pair program -1}
  \end{subfigure}
  \hfill
  \begin{subfigure}{0.32\textwidth}
    \centering
    \includegraphics[width=\linewidth]{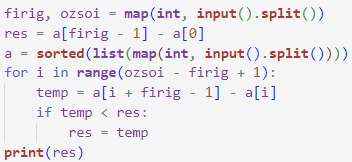}
    \caption{Perturbed Pair program -2}
  \end{subfigure}
  \caption{Example of perturbation on programs. }
  \label{fig:prog-ex}
\end{figure*}

\begin{table}[htbp]
	\centering
	\footnotesize
	\renewcommand{\arraystretch}{1.2}
	\setlength{\tabcolsep}{4pt}
	\begin{tabular}{|c|c|c|c|}
		\hline
		\makecell{\textbf{Original} \\ \textbf{Category}} & 
		\makecell{\textbf{Sample} \\ \textbf{Count}} & 
		\makecell{\textbf{Perturbed} \\ \textbf{Category}} & 
		\makecell{\textbf{Sample} \\ \textbf{Count}} \\
		\hline
		\multirow{2}{*}{\texttt{id}} & \multirow{2}{*}{400}   & $\langle\textit{id},k,\text{np}\rangle$   & 200 \\
		&                        & $\langle\textit{id},k,\text{sp}\rangle$   & 200 \\ \hline
		\multirow{2}{*}{\texttt{fe}} & \multirow{2}{*}{2000}  & $\langle\textit{fe},k,\text{np}\rangle$   & 1000 \\
		&                        & $\langle\textit{fe},k,\text{sp}\rangle$   & 1000 \\ \hline
		\multirow{2}{*}{\texttt{ne}} & \multirow{2}{*}{2000}  & $\langle\textit{ne},k,\text{np}\rangle$   & 1000 \\
		&                        & $\langle\textit{ne},k,\text{sp}\rangle$   & 1000 \\ \hline
		\multirow{2}{*}{\texttt{di}} & \multirow{2}{*}{400}   & $\langle\textit{di},k,\text{np}\rangle$   & 200 \\
		&                        & $\langle\textit{di},k,\text{sp}\rangle$   & 200 \\
		\hline
	\end{tabular}
	\caption{Sample Counts of original and perturbed program-pairs}
	\label{tab:benchmark-distribution}
\end{table}

\newpage
\subsection{Detailed explanation of types of perturbation}
\label{sec:appendix-per-types}

Given below are the implementation details and examples of the various types of perturbations:

\paragraph{1. \textbf{If-Else Swapping:}}
\textit{Semantic-preserving:} The \texttt{if} condition is logically negated, and the \texttt{if}/\texttt{else} blocks are interchanged, maintaining the original logic.
\textit{Non-preserving:} The \texttt{if} condition is logically negated, but blocks are \textit{not} swapped.

\begin{lstlisting}[style=mypython,caption=If-else swapping examples]
# Original
if a < b:
    print("A")
else:
    print("B")

# --- Semantic-preserving ---
if not (a < b):
    print("B")
else:
    print("A")

# --- Semantic-non-preserving ---
if not (a < b):
    print("A")
else:
    print("B")
\end{lstlisting}

\paragraph{2. \textbf{For-While Swapping}}  
\textit{Semantic-preserving:} Convert \texttt{for} loop to equivalent \texttt{while}.  
\textit{Non-preserving:} Converted to a \texttt{while} loop with an intentional error, such as incorrect iterator initialization 
a flawed loop termination condition, or an improper iterator update. 

\begin{lstlisting}[style=mypython, caption=For-while swapping examples]
# Original
for i in range(10):
    print(i)

# --- Semantic-preserving ---
i = 0
while i < 10:
    print(i)
    i += 1

# --- Semantic-non-preserving ---
i = 1
while i < 10:
    print(i)
    i += 1
\end{lstlisting}

\paragraph{3. \textbf{If Condition Flipping:}}  
\textit{Semantic-preserving:} The condition of an \texttt{if} statement is transformed to an equivalent form (e.g., $a<b$ to $not(a>=b)$ ), leaving the body unchanged.
\textit{Non-preserving:} The \texttt{if} condition's relational operator is inverted (e.g., $a<b$ to $a>=b$) or the condition negated.

\begin{lstlisting}[style=mypython, caption=If flipping examples]
# Original
if a < b:
    print("bigger")

# --- Semantic-preserving ---
if not (a >= b):
    print("bigger")

# --- Semantic-non-preserving ---
if a >= b:
    print("bigger")
\end{lstlisting}

\paragraph{4. \textbf{Variable Renaming:}}  
\textit{Semantic-preserving:} One or more randomly selected variables are consistently renamed to new, unique identifiers (e.g., random five-letter strings) across all their occurrences within their scope.
\textit{Non-preserving:} Similar to above, but for at least one selected variable, one of its occurrences is renamed to a different identifier than its other occurrences, effectively breaking its semantic link. 

\begin{lstlisting}[style=mypython, caption=Variable renaming examples]
# Original
x = 10
y = 15
if x > y:
    print("bigger")

# --- Semantic-preserving ---
abcde = 10
hogav = 15
if abcde > hogav:
    print("bigger")

# --- Semantic-non-preserving ---
abcde = 10
hogav = 15
if noagd > hogav:  # inconsistent renaming
    print("bigger")
\end{lstlisting}

\paragraph{5. \textbf{Boolean Variable Inversion:}}  
\textit{Semantic-preserving:} The initialization of a selected boolean variable is inverted (e.g., \texttt{True} to \texttt{False}). All subsequent usages of this variable in conditions are then modified to use its negation (e.g., \texttt{if(Flag)} becomes \texttt{if(not Flag)}).
\textit{Non-preserving:} The initialization of a boolean variable is inverted, but its subsequent usages are not altered to reflect this change.

\begin{lstlisting}[style=mypython, caption=Boolean Variable Inversion examples]
# Original
Flag = True
if Flag:
    print("Go")

# --- Semantic-preserving ---
Flag = False
if not Flag:
    print("Go")

# --- Semantic-non-preserving ---
Flag = False
if Flag:
    print("Go")
\end{lstlisting}

\paragraph{6. \textbf{Statement Reordering:}}  
\textit{Semantic-preserving:} Two data-independent statements (whose relative execution order doesn't affect program outcome) are swapped.
\textit{Non-preserving:} Two data-dependent statements are swapped, potentially causing errors like using variables before definition or with stale values.

\begin{lstlisting}[style=mypython, caption=Statement Swap examples]
# Original
a = 5
b = 0
print(2 * a)

# --- Semantic-preserving ---
a = 5
print(2 * a)
b = 0

# --- Semantic-non-preserving ---
print(2 * a)
b = 0
a = 5
\end{lstlisting}

\paragraph{7. \textbf{Expression Reformatting:}}  
\textit{Semantic-preserving:} A sub-expression is computed and stored in a new temporary variable, which then replaces the sub-expression in the main expression, maintaining the original computation.
\textit{Non-preserving:} A temporary variable may be used, but the main expression is reconstructed incorrectly (e.g., wrong operands/operators, re-introducing original terms), altering the result.

\begin{lstlisting}[style=mypython, caption=Expression reformatting examples]
# Original
ans = a + b - c

# --- Semantic-preserving ---
temp = a + b
ans = temp - c

# --- Semantic-non-preserving ---
temp = a + b
ans = temp + b - c
\end{lstlisting}

\section{Extended Analysis}\label{sec:appendix-analysis}
As we perturb the dataset to generate equal halves of the four original categories into $sp$ and $np$, the previously positive samples from $id$ and $fe$ categories now have half of the samples as positive and the remaining half as negative. Due to this, in the final perturbed dataset, we have 75\% negative samples and the remaining 25\% as positive samples. We want the model to be equally good on both positives and negatives. Hence, we evaluate the performance on weighted accuracy, which takes the average of the average accuracy on positive samples and the same on negative samples.

\subsection{Effect of Self-Consistency}
\label{sec:appendix_self_consist}

To study the effect of inference-time decoding strategies, we evaluate self-consistency decoding~\cite{wangself}, as shown in Table~\ref{tab:appendix_self-consistency}. For each sample, we generate 5 responses (temperature = 1) and select the final prediction via majority voting.  Self-consistency does not yield a significant improvement in accuracy. We observe that model predictions are highly consistent across runs; for most samples, the same answer is produced in all 5 iterations. This suggests that the errors stem from inherent limitations of the model rather than variance in decoding, and are therefore not mitigated by repeated sampling.

\subsection{Increasing Problem Difficulty Level}
\label{sec:appendix_ablation_diffi}
\begin{figure*}[ht!]
\begin{subfigure}{.45\textwidth}
    \includegraphics[width=\linewidth]{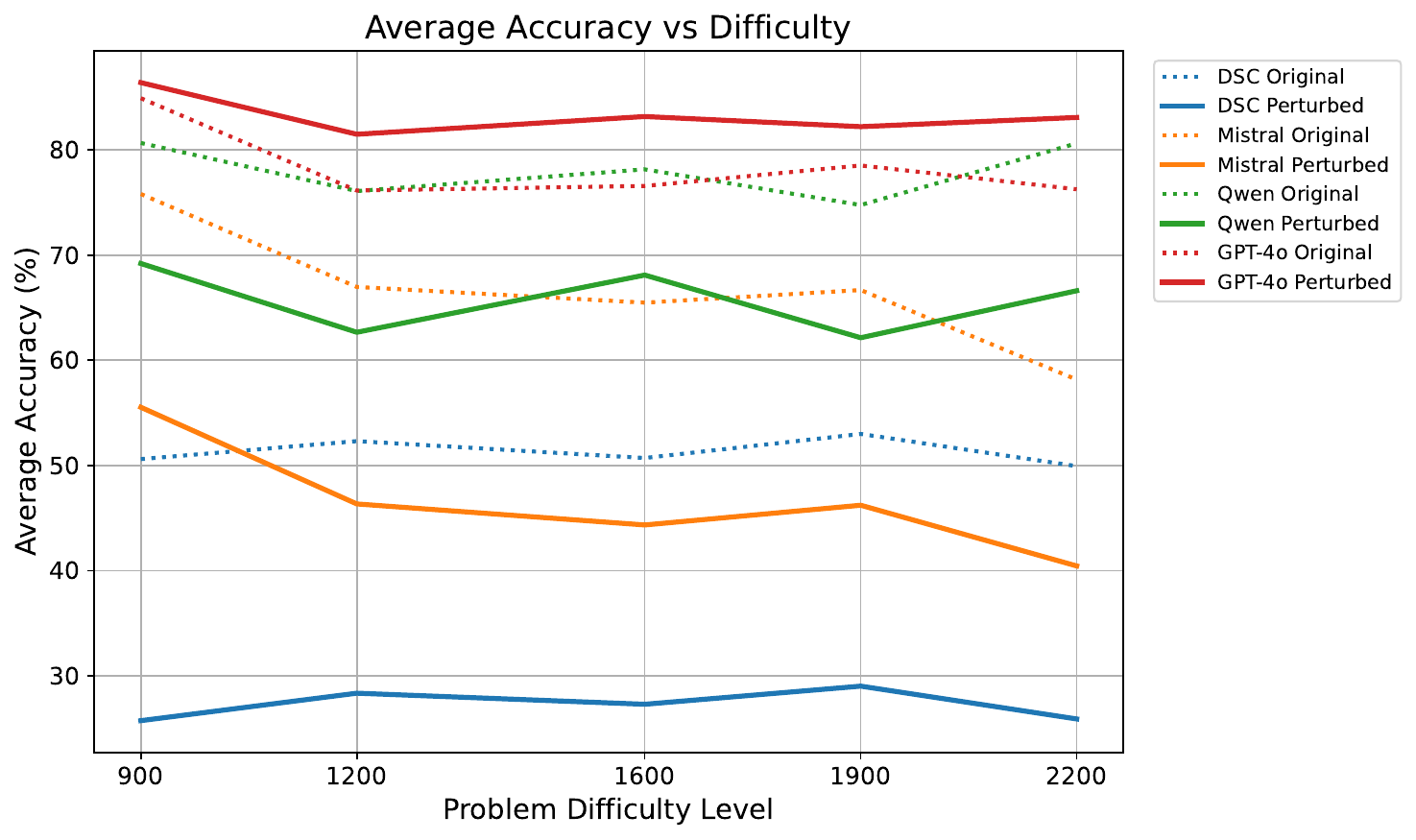}
    \caption{Average Accuracy vs Problem Difficulty}
    \label{fig:avg_vs_difficulty}
\end{subfigure}
\hfill
\begin{subfigure}{.45\textwidth}
    \includegraphics[width=\linewidth]{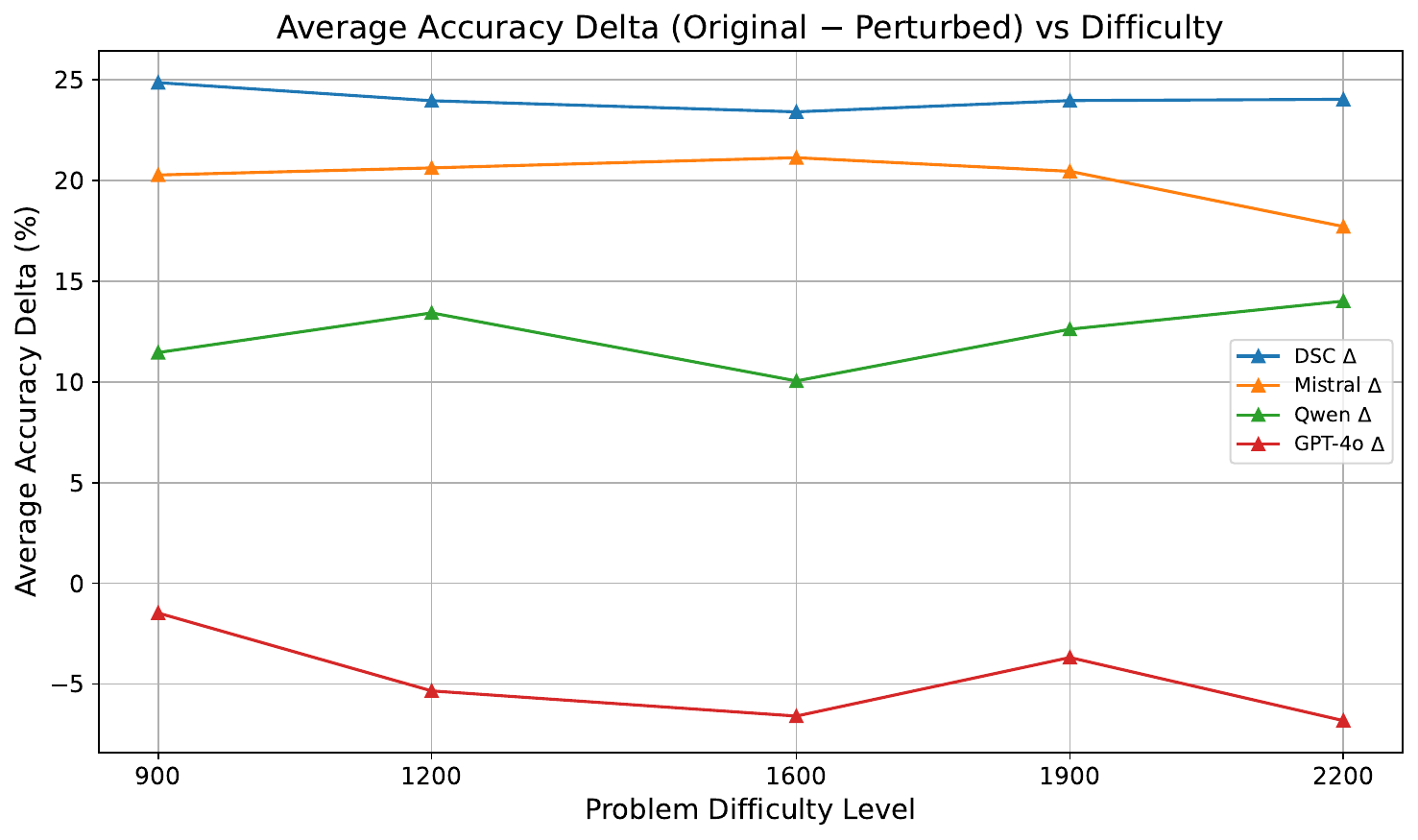}
    \caption{Average Accuracy Delta vs Problem Difficulty}
    \label{fig:avg_vs_difficulty_delta}
    
\end{subfigure}
\caption{Change in Average Accuracy with varying Difficulty Level for different Models}
\end{figure*}

\begin{figure*}[ht!]
\begin{subfigure}{.45\textwidth}
    \includegraphics[width=\linewidth]{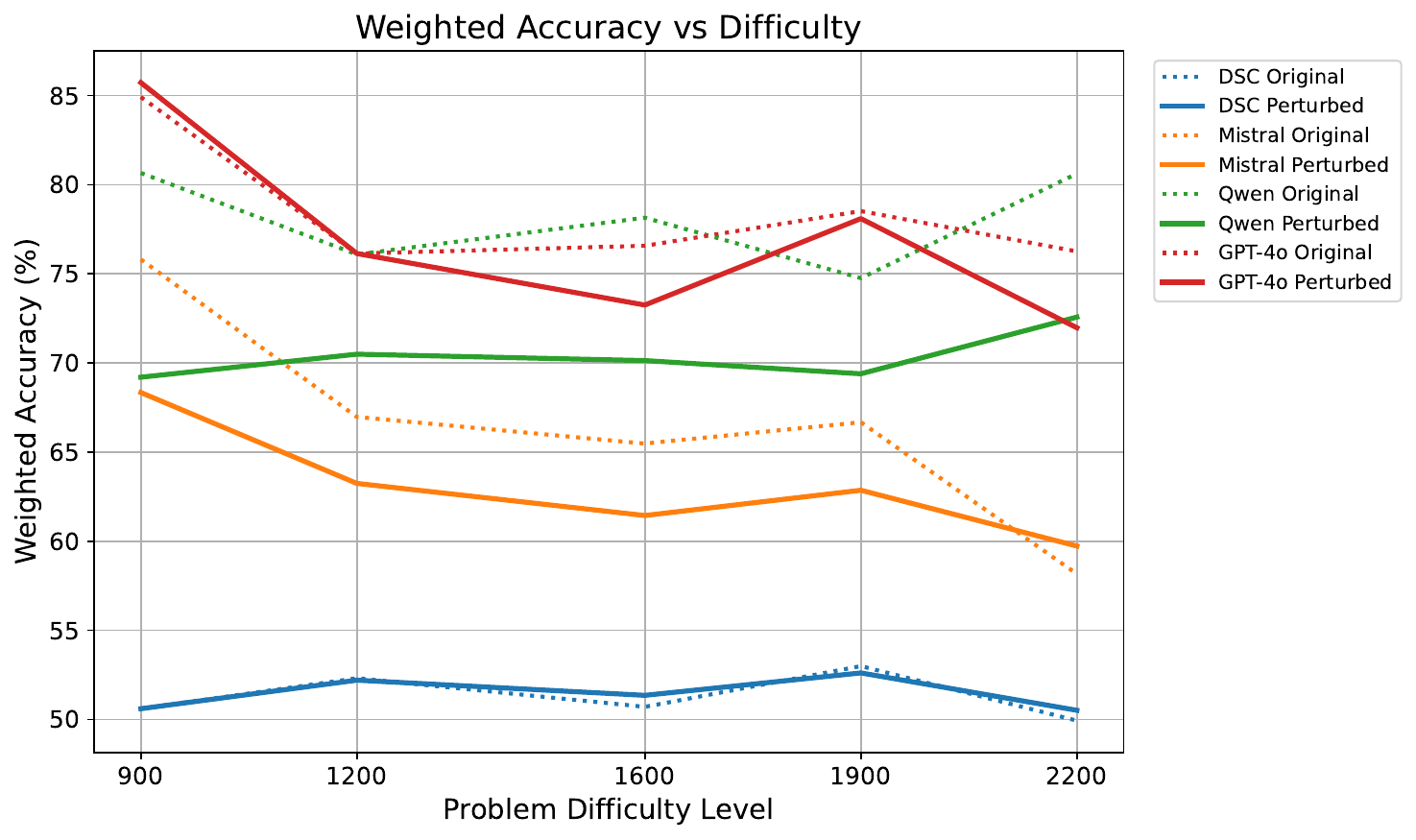}
    \caption{Weighted Accuracy vs Problem Difficulty}
    \label{fig:wacc_vs_difficulty}
\end{subfigure}
\hfill
\begin{subfigure}{.45\textwidth}
    \includegraphics[width=\linewidth]{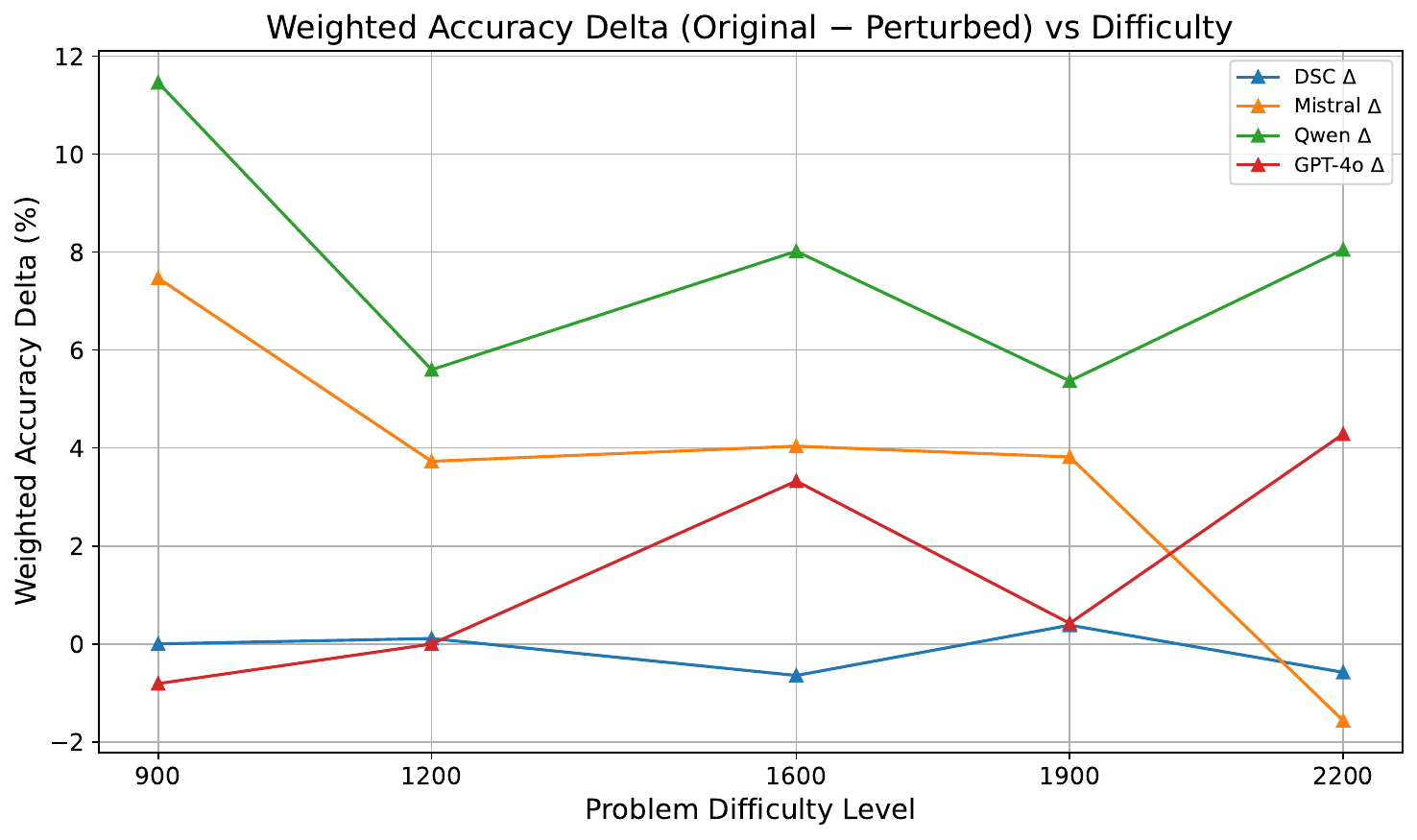}
    \caption{Weighted Accuracy Delta vs Problem Difficulty}
    \label{fig:wacc_vs_difficulty_delta}
    
\end{subfigure}
\caption{Change in Weighted Accuracy with varying Difficulty Level for different Models}
\end{figure*}

\begin{figure*}[ht!]
\begin{subfigure}{.45\textwidth}
    \includegraphics[width=\linewidth]{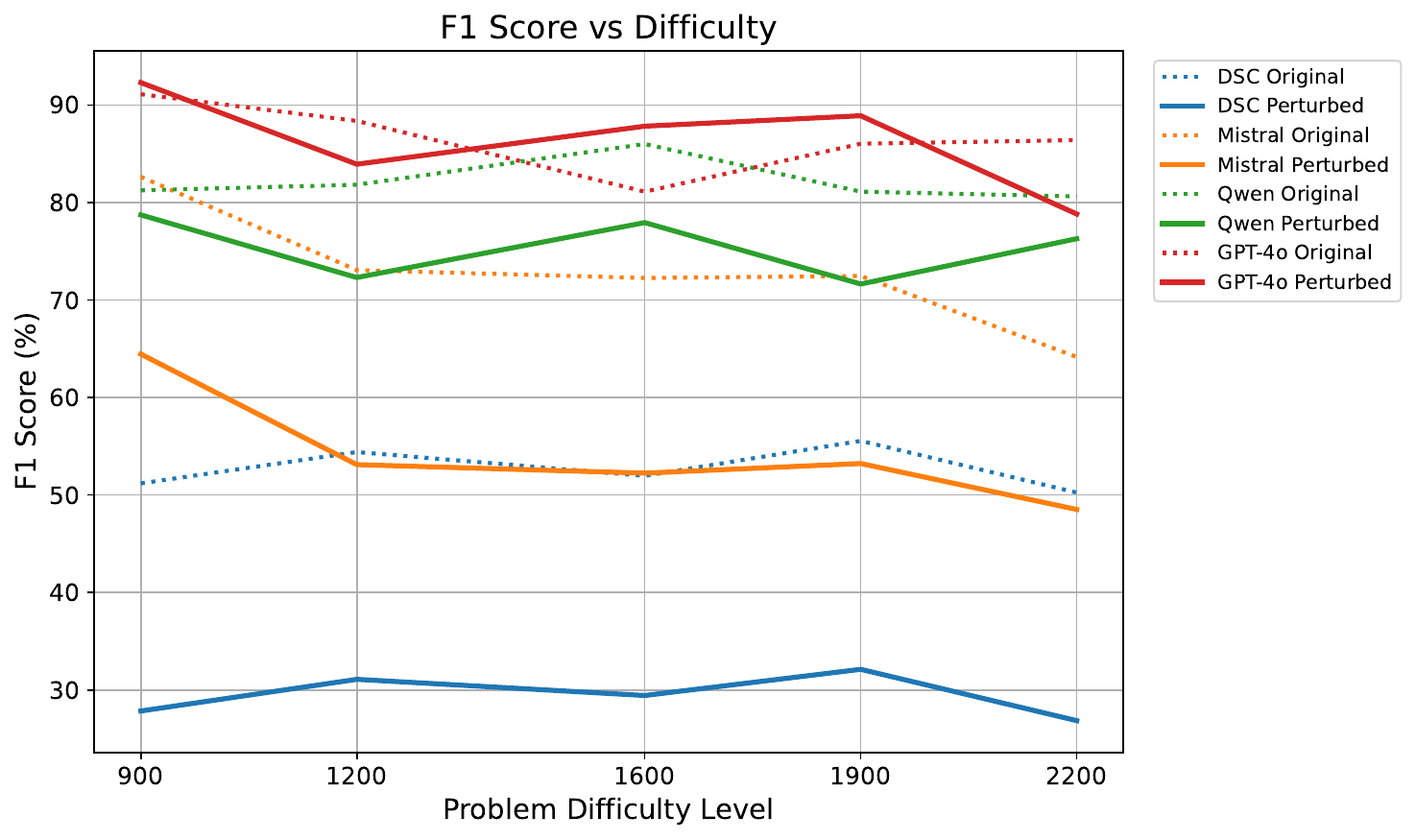}
    \caption{F1 Score vs Problem Difficulty}
    \label{fig:f1_vs_difficulty}
\end{subfigure}
\hfill
\begin{subfigure}{.45\textwidth}
    \includegraphics[width=\linewidth]{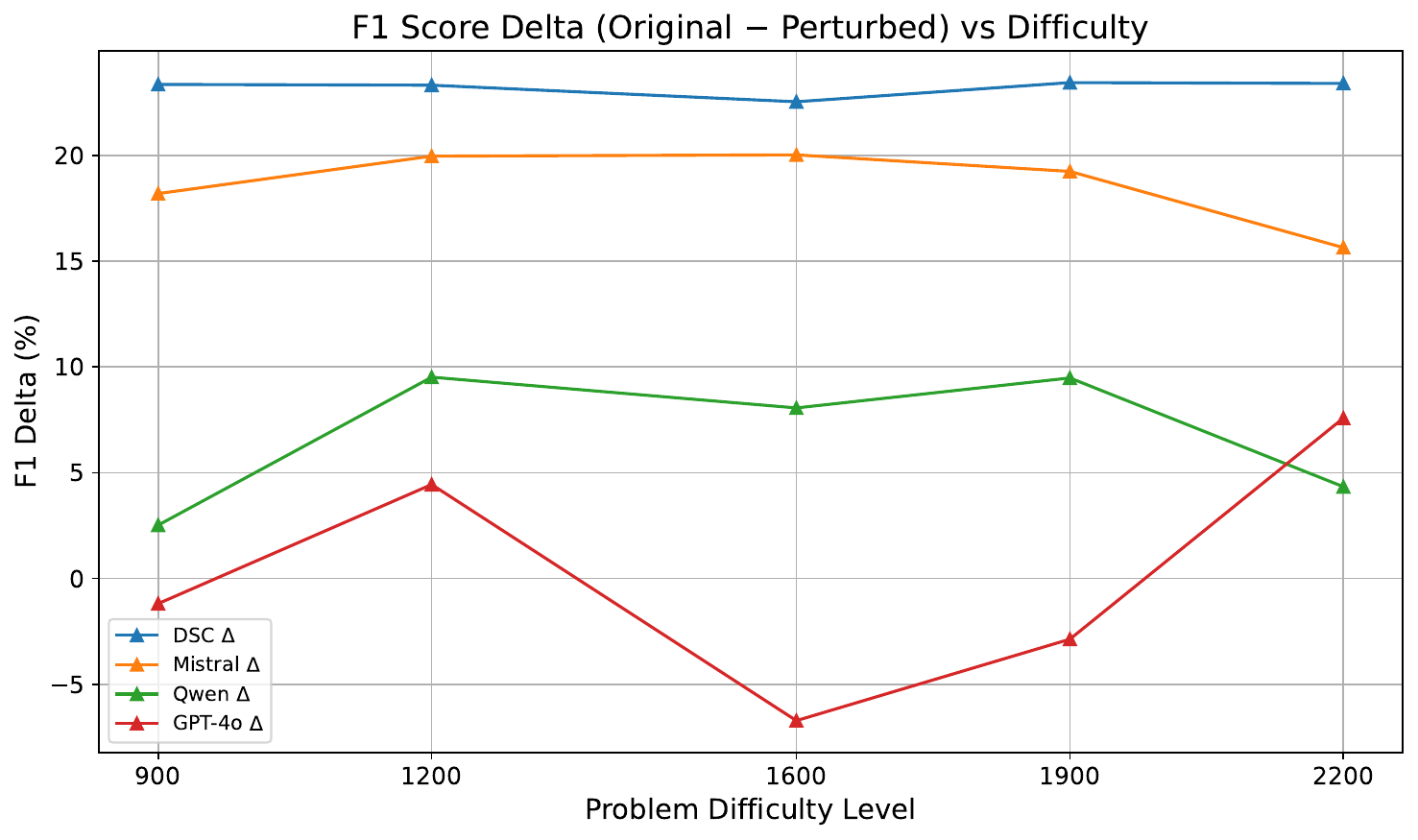}
    \caption{F1 Score Delta vs Problem Difficulty}
    \label{fig:f1_vs_difficulty_delta}
    
\end{subfigure}
\caption{Change in F1 score with varying Difficulty Level for different Models}
\end{figure*}

We also investigate how LLM model performance varies with increasing problem difficulty, using levels 900, 1200, 1600, 1900, and 2200, which represent a progression from simple, beginner-level problems to complex tasks requiring advanced algorithmic reasoning. These levels correspond to Codeforces problem ratings, a widely used metric in competitive programming. As shown in Fig~\ref{fig:avg_vs_difficulty} and Fig~\ref{fig:avg_vs_difficulty_delta}, there is a drop in accuracy—both before and after-perturbation across difficulty levels. The drop, however, remains constant across difficulty levels, which suggests that whether the problem is basic or complex, introducing perturbations impacts the model's reasoning in a similar way. For GPT-4o, we observe fluctuations in performance. Further, we note that while the drop for GPT-4o in terms of accuracy is negative, it is primarily due to high performance on $ne$ and $di$ categories. Similarly, Fig~\ref{fig:f1_vs_difficulty} and Fig~\ref{fig:f1_vs_difficulty_delta} show that F1 score metric too, fluctuations are observed for GPT-4o. For closer analysis, we refer to the corresponding graph in Fig~\ref{fig:wacc_vs_difficulty} and Fig~\ref{fig:wacc_vs_difficulty_delta}, plotting weighted Acc vs difficulty level, where we see that the drop is always positive across difficulty levels, pointing to nuanced performance behaviour of highly complex models.

\subsection{Effect of Individual Perturbations} 
\label{sec:appendix_abalation_perturb}
\begin{figure*}[ht!]
\begin{subfigure}{.45\textwidth}
    \includegraphics[width=\linewidth]{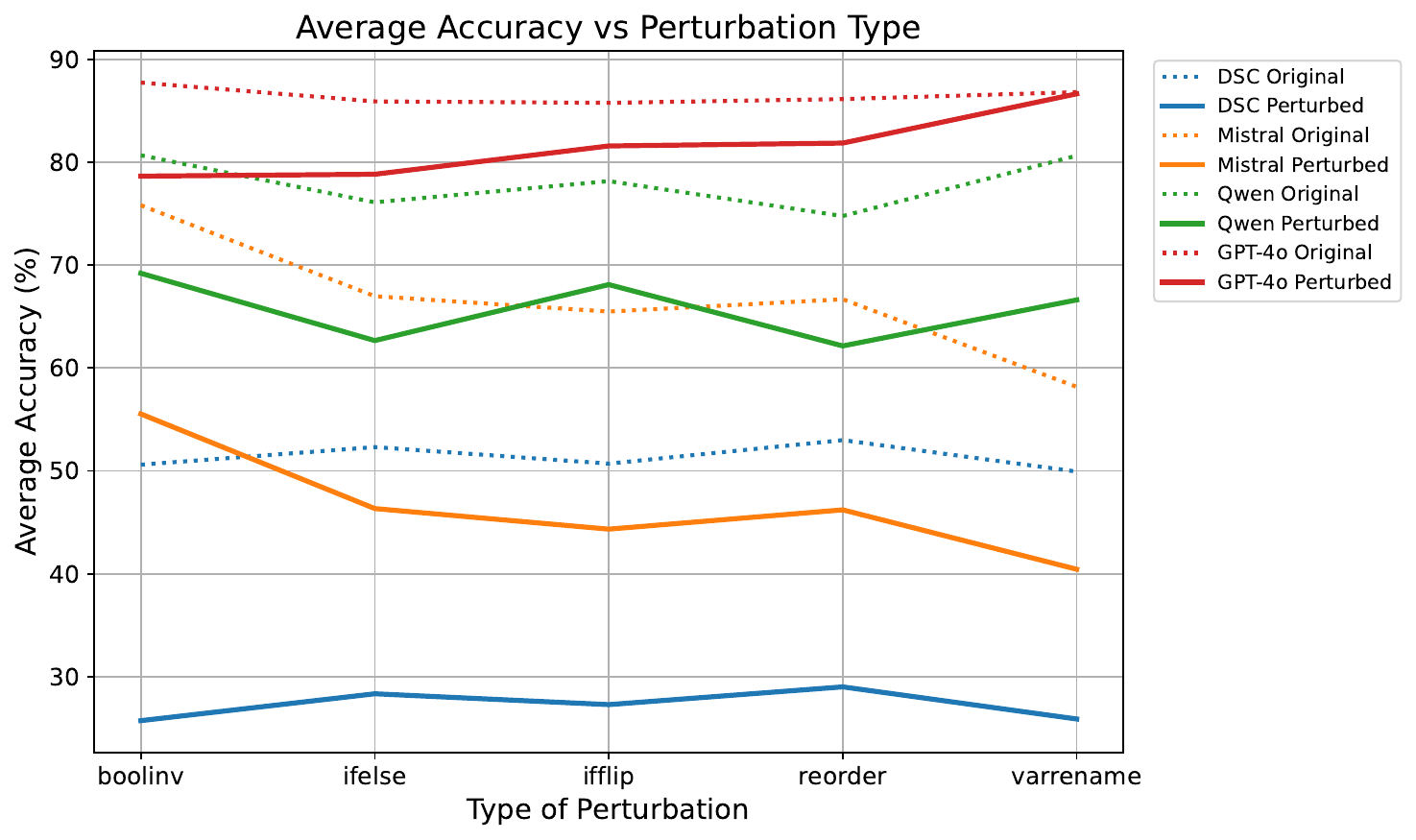}
    \caption{Average Accuracy vs Perturbation Type}
    \label{fig:avg_vs_perturbation}
\end{subfigure}
\hfill
\begin{subfigure}{.45\textwidth}
    \includegraphics[width=\linewidth]{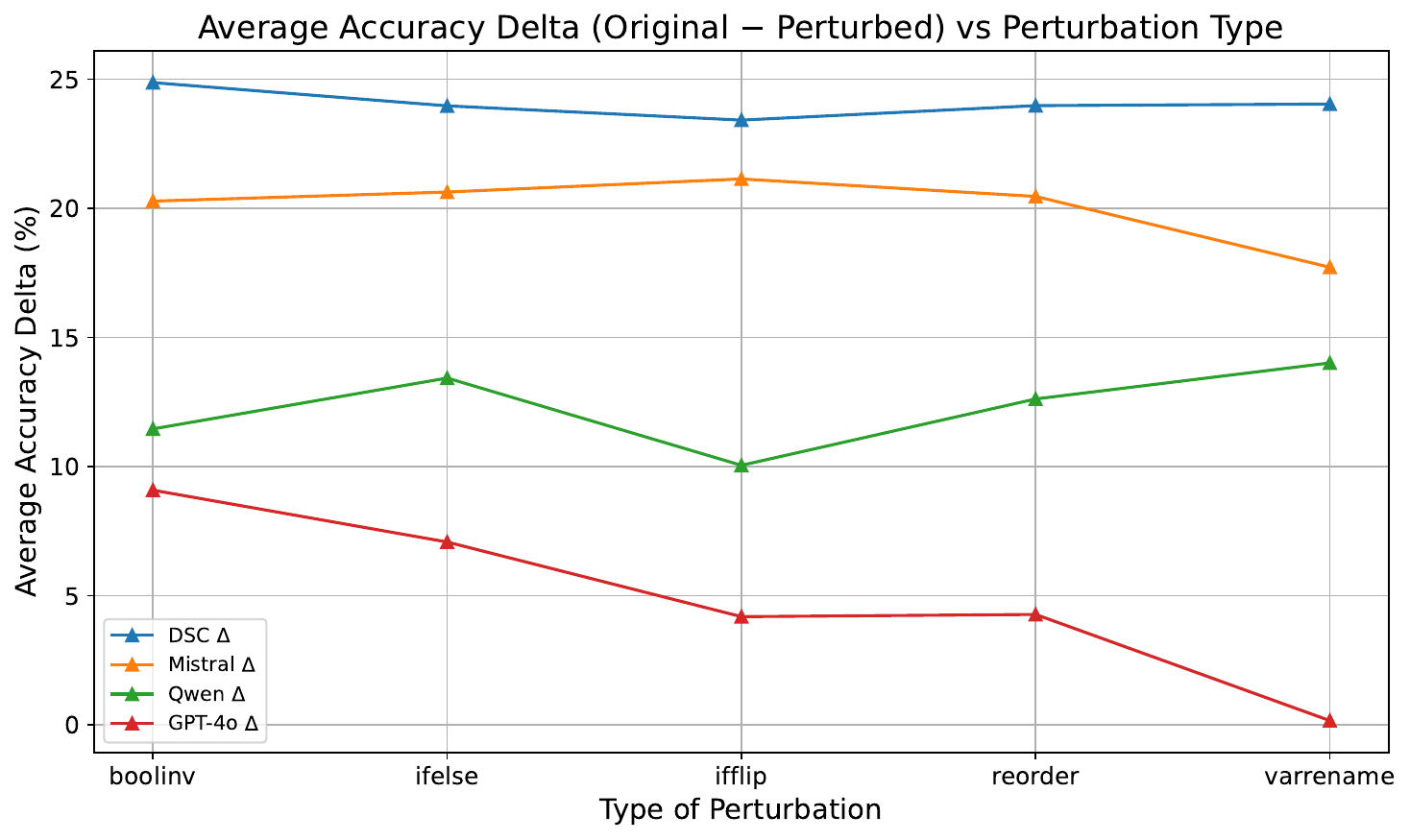}
    \caption{Average Accuracy Delta vs Perturbation Type}
    \label{fig:avg_vs_perturbation_delta}
    
\end{subfigure}
\caption{Change in Average Accuracy with applying individual Perturbation Types for different Models}
\end{figure*}

We also analysed how different types of perturbations impact the model by comparing the performance drop (delta) before and after applying each perturbation. We chose 5 perturbation types for which dataset generation was the quickest. Fig~\ref{fig:avg_vs_perturbation} and Fig~\ref{fig:avg_vs_perturbation_delta} show that logic-altering perturbations like boolean inversion, if condition flipping, and if-else swapping have a slightly higher delta, indicating that the model struggles to understand the transformed logic. In contrast, surface-level changes such as variable renaming and statement reordering are comparatively easier for the models to interpret, resulting in comparatively slightly lower delta values.

\begin{figure*}[ht!]
\begin{subfigure}{.45\textwidth}
    \includegraphics[width=\linewidth]{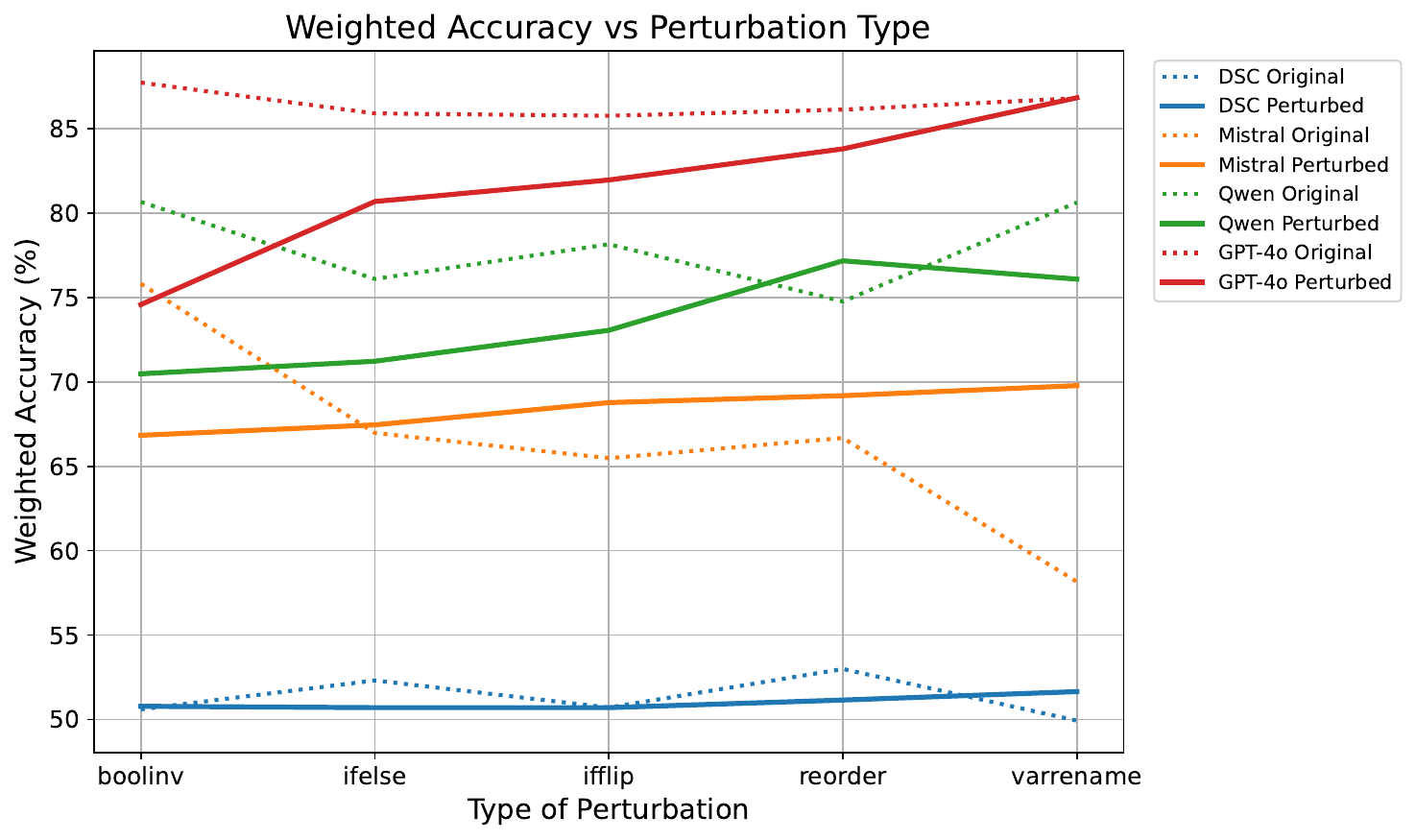}
    \caption{Weighted Accuracy vs Perturbation Type}
    \label{fig:wacc_vs_perturbation}
\end{subfigure}
\hfill
\begin{subfigure}{.45\textwidth}
    \includegraphics[width=\linewidth]{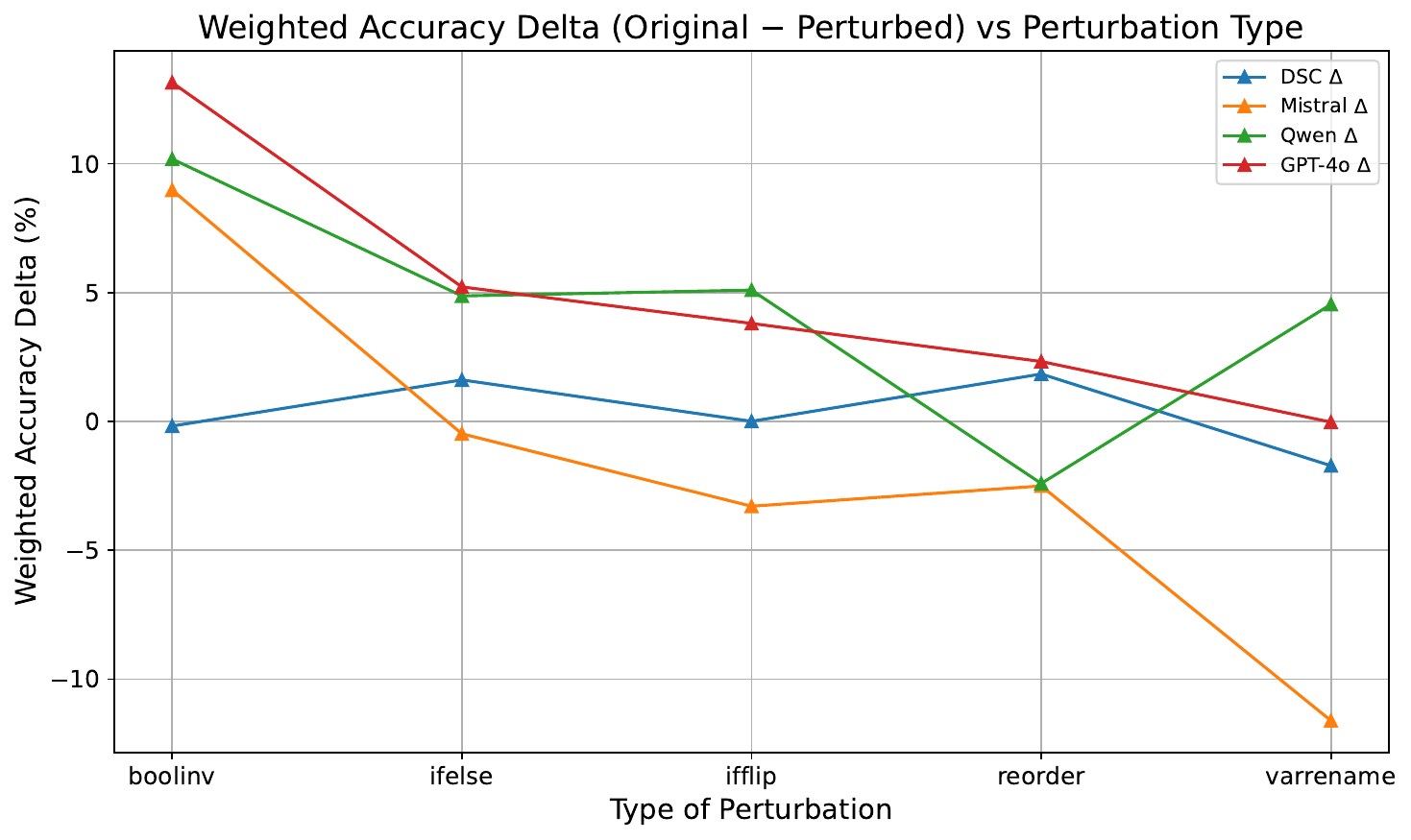}
    \caption{Weighted Accuracy Delta vs Perturbation Type}
    \label{fig:wacc_vs_perturbation_delta}
    
\end{subfigure}
\caption{Change in Weighted Accuracy with applying individual Perturbation Types for different Models}
\end{figure*}

\begin{figure*}[ht!]
\begin{subfigure}{.45\textwidth}
    \includegraphics[width=\linewidth]{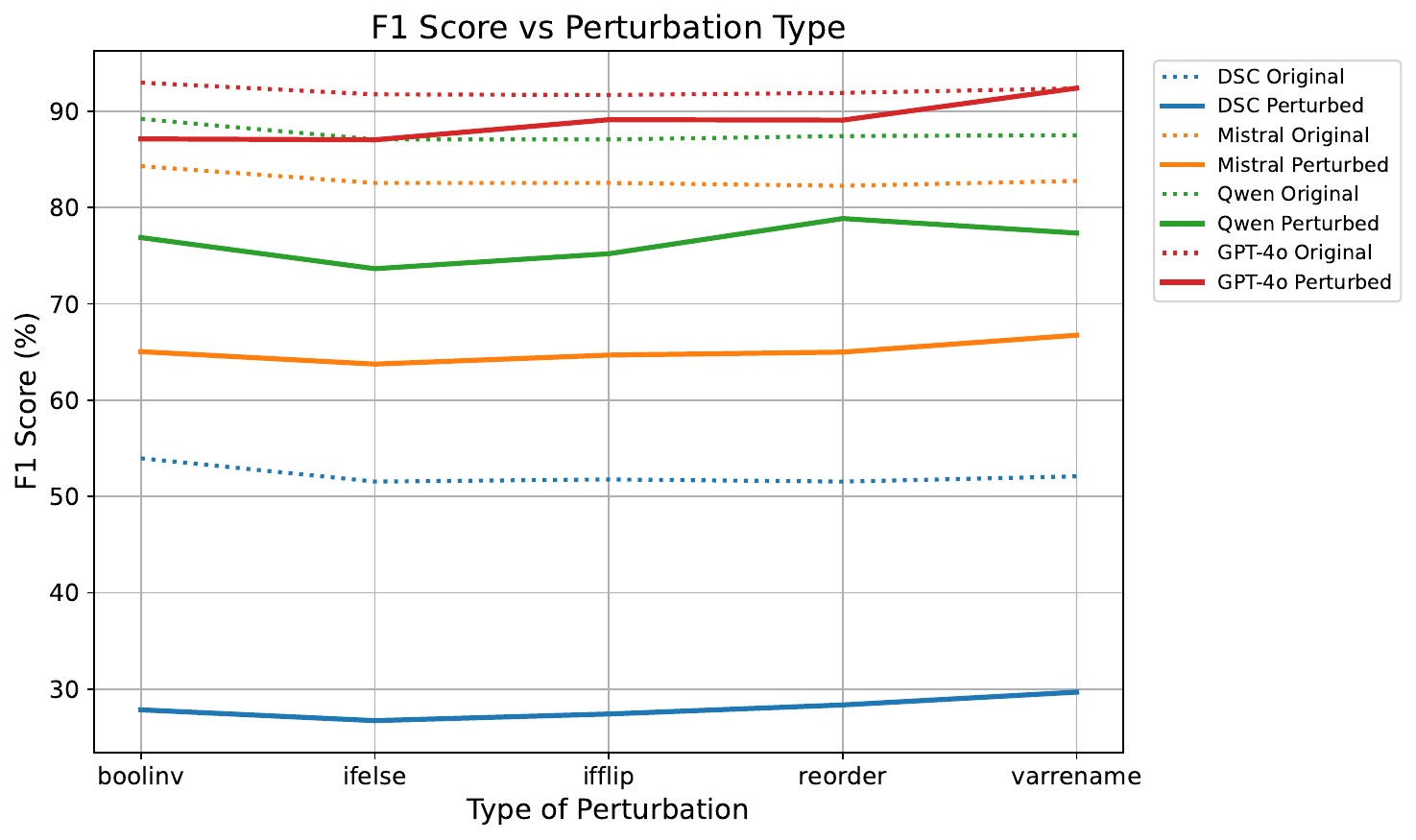}
    \caption{F1 Score vs Perturbation Type}
    \label{fig:f1_vs_perturbation}
\end{subfigure}
\hfill
\begin{subfigure}{.45\textwidth}
    \includegraphics[width=\linewidth]{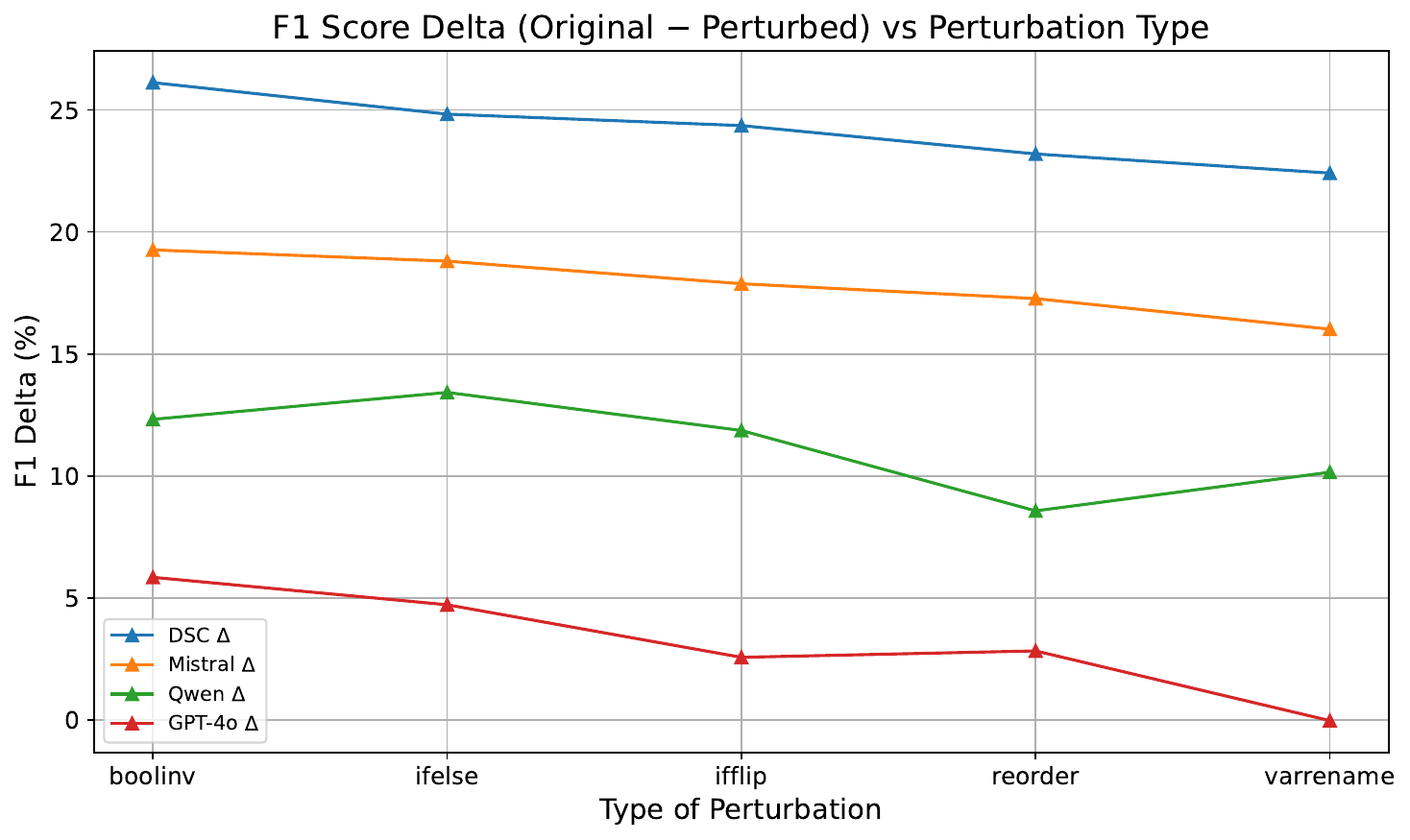}
    \caption{F1 Score Delta vs Perturbation Type}
    \label{fig:f1_vs_perturbation_delta}
    
\end{subfigure}
\caption{Change in F1 score with applying individual Perturbation Types for different Models}
\end{figure*}

Similarly, Fig~\ref{fig:wacc_vs_perturbation} and Fig~\ref{fig:wacc_vs_perturbation_delta} contain these plots for weighted accuracy metric, while Fig~\ref{fig:f1_vs_perturbation} and Fig~\ref{fig:f1_vs_perturbation_delta} contain these plots for F1 metric. 

Tables \ref{tab:ana_diff_gpt-4o} through \ref{tab:ana_per_dsc} contain detailed results of analysis experiments. Table \ref{tab:cpp} contains the results of Python-C++ pairs of cross-lingual programs.

\begin{table*}[t]
    \resizebox{\textwidth}{!}{
    \setlength{\tabcolsep}{1.7mm}

    }
    \captionof{table}{Results comparing performance of various models using self-consistency decoding method.} 
    \label{tab:appendix_self-consistency}
\end{table*}

\begin{table*}[t]
    \resizebox{\textwidth}{!}{
     %
    }
    \captionof{table}{Effect of varying difficulty level on Qwen model.}
    \label{tab:ana_diff_qwen}
\end{table*}

\begin{table*}[t]
    \resizebox{\textwidth}{!}{
     %
    }
    \captionof{table}{Effect of varying difficulty level on Mistral model.}
    \label{tab:ana_diff_mistral}
\end{table*}

\begin{table*}[t]
    \resizebox{\textwidth}{!}{
     %
    }
    \captionof{table}{Effect of varying difficulty level on Deepseek-Coder model.}
    \label{tab:ana_diff_dsc}
\end{table*}

\begin{table*}[t]
    \resizebox{\textwidth}{!}{
     %
    }
    \captionof{table}{Effect of varying difficulty level on GPT-4o model.}
    \label{tab:ana_diff_gpt-4o}
\end{table*}

\begin{table*}[t]
    \resizebox{\textwidth}{!}{
     %
    }
    \captionof{table}{Effect of different perturbations on Qwen model.}
    \label{tab:ana_per_qwen}
\end{table*}

\begin{table*}[t]
    \resizebox{\textwidth}{!}{
     %
    }
    \captionof{table}{Effect of different perturbations on Mistral model.}
    \label{tab:ana_per_mistral}
\end{table*}

\begin{table*}[t]
    \resizebox{\textwidth}{!}{
     %
    }
    \captionof{table}{Effect of different perturbations on Deepseek-Coder model.}
    \label{tab:ana_per_dsc}
\end{table*}

\begin{table*}[t]
    \resizebox{\textwidth}{!}{
     %
    }
    \captionof{table}{Effect of different perturbations on GPT-4o model.}
    \label{tab:ana_per_gpt-4o}
\end{table*}

\begin{table*}
    \centering
    \resizebox{\textwidth}{!}{
    \setlength{\tabcolsep}{2mm}
    %
    }
    \caption{Results comparing various models for C++ and Python pairs of code}
    \label{tab:cpp}
\end{table*}

\begin{table*}[!t]
    \centering
    \resizebox{\textwidth}{!}{
    %
}
    & \makecell{86.7} & \makecell{74.5} & \makecell{92.1}
    & \makecell{74.1} & \makecell{59.2} & \makecell{89.0}
  \\

               \bottomrule
    \end{tabular}
    }
    \caption{Results comparing various models for pairs with one C++ code and one Python code (Both Perturbed) }
    \label{tab:cpp_both}
\end{table*}

\newpage
\subsection{Disclosing the information about the perturbations to models}
We tried to prompt the model to check for equivalence, along with providing information on possible kinds of perturbations applied to the programs. We added the examples of both $np$ and $sp$ types of all 7 types of perturbations. The examples used are the same as Appendix \ref{sec:appendix-exp}. Comparing the results in table \ref{tab:info_prompt} with table \ref{tab:stricter_prompt}, we observe that adding this information hurts the performance of the GPT-4o model. It did not help the Qwen and the Deepseek-Coder models. It improves the performance of Mistral for categories with negative labels, but hurts its performance for categories with positive labels. Similarly, by comparing the results of table \ref{tab:info_cot} to table \ref{tab:s_cot2}, we observe that adding the information of perturbation to the Chain-of-Thought prompt hurts the performance of the Mistral and the Deepseek-Coder models. It doesn't improve GPT-4o's performance. It improves Qwen's performance for some of the negative categories, but the decrease in the positive categories and the $di$ category hurts overall performance. This shows that overall, the models struggle on the perturbed set even after having additional information about the perturbations disclosed to them.

\begin{figure*}[htbp]  
\centering
\begin{minipage}{0.95\textwidth}
\begin{violetlisting}[title={Prompt with Perturbation Information}]
Given two programs, you must check if they solve the same problem - that is, for any input, they must produce identical outputs. \\
Information regarding the programs: One or both of the programs may be modified from their original form using 7 different categories of perturbations. \\
Each of the 7 categories has 2 subtypes of perturbations: semantic preserving or not preserving. Given below are the examples for each of them: \\
\\
1. If-else swapping examples:\\
\# Original\\
if $a < b$:\\
    print("A")\\
else:\\
    print("B")\\
\\
\# Semantic-preserving \\
if not ($a >= b$):\\
    print("B")\\
else:\\
    print("A")\\
\\
\# Semantic-non-preserving \\
if not ($a < b$):\\
    print("A")\\
else:\\
    print("B")\\
\\
2. ... \\
...\\
...\\
\\
The original pair of programs before perturbation may or may not be solving the same problem. Given all this information, now check if the final given pair solves the same problem or not. \\
Your task is to answer only with one word: Yes or No. \\
No explanation, no extra words, no formatting.\\
\\
Program 1: \{prog1\} \\
Program 2: \{prog2\}
\end{violetlisting}

\end{minipage}
\end{figure*}

\begin{table*}[t]
    \resizebox{\textwidth}{!}{
    \setlength{\tabcolsep}{1.7mm}
     \begin{tabular}{ccccccccccc}
        \toprule
       \textbf{Model} & \textbf{id} & \textbf{fe} & \textbf{ne} & \textbf{di} & \textbf{Avg}  &\textbf{W-Avg} & \textbf{Mac-F1} & \textbf{Mic-F1} & \textbf{PosF1} & \textbf{NegF1}  \\ 
       & \makecell{\begin{tabular}{@{}r  r@{}} 
                    \textbf{np} & \textbf{sp}
                \end{tabular}}
        & \makecell{\begin{tabular}{@{}r r@{}} 
                    \textbf{np} & \textbf{sp}
                \end{tabular}}
        & \makecell{\begin{tabular}{@{}r r@{}} 
                    \textbf{np} & \textbf{sp} 
                \end{tabular}}
        & \makecell{\begin{tabular}{@{}r r@{}} 
                    \textbf{np} & \textbf{sp}
                \end{tabular}} & & & & & & \\ \midrule

\rowcolor{rowgray}          \multirow{3}{*}{Qwen}  

    & \makecell{\begin{tabular}{@{}r r@{}}  100.0 & 99.0  \end{tabular}} 
    & \makecell{\begin{tabular}{@{}r r@{}}   78.8 & 78.4 \end{tabular} }
    & \makecell{ \begin{tabular}{@{}r r@{}} 44.0 & 41.5  \end{tabular} }
    & \makecell{ \begin{tabular}{@{}r r@{}} 100.0 & 100.0 \end{tabular} } 
    & \makecell{80.2} & \makecell{80.2} & \makecell{86.9}
    & \makecell{69.6} & \makecell{71.4} & \makecell{67.7}\\
\cline{2-11} \noalign{\vskip 2pt} 
    & \makecell{ \begin{tabular}{@{}r r@{}} 13.5 & 99.0  \end{tabular}}
    & \makecell{ \begin{tabular}{@{}r r@{}}  38.1 & 74.4  \end{tabular}}
    & \makecell{ \begin{tabular}{@{}r r@{}}  60.0 & 47.8  \end{tabular}} 
    & \makecell{ \begin{tabular}{@{}r r@{}} 100.0 & 100.0 \end{tabular}}
    & \makecell{66.6} & \makecell{73.3} & \makecell{75.4}
    & \makecell{59.4} & \makecell{48.8} & \makecell{69.9}
  \\ \hline \noalign{\vskip 3pt}       

\rowcolor{rowgray}          \multirow{3}{*}{Mistral} 

    & \makecell{\begin{tabular}{@{}r r@{}}   100.0 &  99.0  \end{tabular}} 
    & \makecell{\begin{tabular}{@{}r r@{}}   68.4 & 70.4 \end{tabular} }
    & \makecell{ \begin{tabular}{@{}r r@{}} 50.7 & 47.9   \end{tabular} }
    & \makecell{ \begin{tabular}{@{}r r@{}}  90.0 & 93.5 \end{tabular} } 
    & \makecell{77.5} & \makecell{77.5} & \makecell{85.8}
    & \makecell{67.4} & \makecell{67.8} & \makecell{67.0}\\
\cline{2-11} \noalign{\vskip 2pt} 
    & \makecell{ \begin{tabular}{@{}r r@{}} 15.5 & 86.5   \end{tabular}}
    & \makecell{ \begin{tabular}{@{}r r@{}}  46.5 & 58.8   \end{tabular}}
    & \makecell{ \begin{tabular}{@{}r r@{}}  65.6 & 60.7  \end{tabular}} 
    & \makecell{ \begin{tabular}{@{}r r@{}} 95.5 & 98.5 \end{tabular}}
    & \makecell{66.0} & \makecell{68.1} & \makecell{76.1}
    & \makecell{58.6} & \makecell{44.4} & \makecell{72.7}
  \\ \hline \noalign{\vskip 3pt}        

\rowcolor{rowgray}         \multirow{3}{*}{DSC} 

    & \makecell{\begin{tabular}{@{}r r@{}}  100.0 & 98.5  \end{tabular}} 
    & \makecell{\begin{tabular}{@{}r r@{}}  99.8 & 99.6 \end{tabular} }
    & \makecell{ \begin{tabular}{@{}r r@{}} 0.4 & 0.3   \end{tabular} }
    & \makecell{ \begin{tabular}{@{}r r@{}} 0.5 & 0.5  \end{tabular} } 
    & \makecell{49.9} & \makecell{50.0} & \makecell{50.3}
    & \makecell{31.9} & \makecell{63.1} & \makecell{0.8}\\
\cline{2-11} \noalign{\vskip 2pt} 
    & \makecell{ \begin{tabular}{@{}r r@{}} 0.0 & 99.5   \end{tabular}}
    & \makecell{ \begin{tabular}{@{}r r@{}}  0.2 & 99.8    \end{tabular}}
    & \makecell{ \begin{tabular}{@{}r r@{}}  0.1 & 0.3  \end{tabular}} 
    & \makecell{ \begin{tabular}{@{}r r@{}} 0.0 & 0.5 \end{tabular}}
    & \makecell{25.0} & \makecell{49.9} & \makecell{25.2}
    & \makecell{18.9} & \makecell{37.5} & \makecell{0.4}
  \\ \hline \noalign{\vskip 3pt}
\rowcolor{rowgray}        \multirow{3}{*}{GPT-4o} 

    & \makecell{\begin{tabular}{@{}r r@{}} 100.0 & 100.0  \end{tabular}} 
    & \makecell{\begin{tabular}{@{}r r@{}}  79.9 & 79.3  \end{tabular} }
    & \makecell{ \begin{tabular}{@{}r r@{}} 54.3 & 56.0  \end{tabular} }
    & \makecell{ \begin{tabular}{@{}r r@{}} 100.0 & 100.0 \end{tabular} } 
    & \makecell{83.7} & \makecell{83.7} & \makecell{89.9}
    & \makecell{74.9} & \makecell{75.3} & \makecell{74.5}\\
\cline{2-11} \noalign{\vskip 2pt} 
    & \makecell{ \begin{tabular}{@{}r r@{}} 58.0 & 98.0  \end{tabular}}
    & \makecell{ \begin{tabular}{@{}r r@{}}  62.4 & 73.8  \end{tabular}}
    & \makecell{ \begin{tabular}{@{}r r@{}}  82.5 & 61.4  \end{tabular}} 
    & \makecell{ \begin{tabular}{@{}r r@{}} 100.0 & 100.0 \end{tabular}}
    & \makecell{79.5} & \makecell{81.6} & \makecell{87.6}
    & \makecell{70.7} & \makecell{59.2} & \makecell{82.2} 
  \\

               \bottomrule
    \end{tabular}
    }
    \captionof{table}{Results of performance after disclosing information about perturbation to the prompts.}
    \label{tab:info_prompt}
\end{table*}

\section{Performance of Newer Models}
\label{sec:newer-models}
Here we present the results of GPT-5-nano and GPT-5.1 on the remaining experiments. Table~\ref{tab:appendix_cot} shows performance using the Chain of Thought prompt. Table~\ref{tab:appendix_except_var} shows the performance on the dataset with perturbations except variable renaming. And Table~\ref{tab:java} shows performance on the cross-lingual task. Similar to our observation for GPT-4o in Sections \ref{subsec:cot}, \ref{subsec:failure-mode} and \ref{sec:cross-lingual}, we observe a drop in weighted average accuracy and positive-F1 score. Overall, we find these models to be biased towards labelling pairs with differing syntax as negative, as observed by low performance for the \textit{fe} and $<fe, k, sp>$ categories (even lower than open models for many cases).

\begin{table*}
    \centering
    \resizebox{\textwidth}{!}{
    \setlength{\tabcolsep}{1.7mm}
    \begin{tabular}{lcccccccccc}
        \toprule
       \textbf{Model} & \textbf{id} & \textbf{fe} & \textbf{ne} & \textbf{di} & \textbf{Avg} &\textbf{W-Avg} & \textbf{Mac-F1}  & \textbf{Mic-F1} & \textbf{PosF1} & \textbf{NegF1}  \\ 
       & \makecell{\begin{tabular}{@{}r  r@{}} 
                    \textbf{np} & \textbf{sp}
                \end{tabular}}
        & \makecell{\begin{tabular}{@{}r r@{}} 
                    \textbf{np} & \textbf{sp}
                \end{tabular}}
        & \makecell{\begin{tabular}{@{}r r@{}} 
                    \textbf{np} & \textbf{sp} 
                \end{tabular}}
        & \makecell{\begin{tabular}{@{}r r@{}} 
                    \textbf{np} & \textbf{sp}
                \end{tabular}} & & & & & & \\ \midrule

     \rowcolor{rowgray}          \multirow{3}{*}{GPT-5-nano} 
     
    & \makecell{\begin{tabular}{@{}r r@{}} 100.0 & 100.0 \end{tabular}} 
    & \makecell{\begin{tabular}{@{}r r@{}} 58.3 & 64.7 \end{tabular} }
    & \makecell{ \begin{tabular}{@{}r r@{}} 91.1 & 90.5 \end{tabular} }
    & \makecell{ \begin{tabular}{@{}r r@{}} 100.0 & 100.0 \end{tabular} } 
    & \makecell{88.1} & \makecell{88.1} & \makecell{92.8}
    & \makecell{81.0} & \makecell{77.4} & \makecell{84.6}\\
\cline{2-11} \noalign{\vskip 2pt} 
    & \makecell{ \begin{tabular}{@{}r r@{}} 90.0 & 98.5 \end{tabular}}
    & \makecell{ \begin{tabular}{@{}r r@{}} 85.9 & 62.3 \end{tabular}}
    & \makecell{ \begin{tabular}{@{}r r@{}} 97.0 & 90.5 \end{tabular}} 
    & \makecell{ \begin{tabular}{@{}r r@{}} 100.0 & 100.0 \end{tabular}}
    & \makecell{90.5} & \makecell{87.1} & \makecell{94.6}
    & \makecell{81.4} & \makecell{71.1} & \makecell{91.8}
  \\   \hline \noalign{\vskip 3pt}

  \rowcolor{rowgray}          \multirow{3}{*}{GPT-5.1} 
     
    & \makecell{\begin{tabular}{@{}r r@{}} 100.0 & 100.0 \end{tabular}} 
    & \makecell{\begin{tabular}{@{}r r@{}} 61.2 & 63.9 \end{tabular} }
    & \makecell{ \begin{tabular}{@{}r r@{}} 93.8 & 93.7 \end{tabular} }
    & \makecell{ \begin{tabular}{@{}r r@{}} 100.0 & 100.0 \end{tabular} } 
    & \makecell{89.1} & \makecell{89.1} & \makecell{93.4}
    & \makecell{82.5} & \makecell{79.1} & \makecell{86.0}\\
\cline{2-11} \noalign{\vskip 2pt} 
    & \makecell{ \begin{tabular}{@{}r r@{}} 93.0 & 99.5 \end{tabular}}
    & \makecell{ \begin{tabular}{@{}r r@{}} 91.8 & 62.5 \end{tabular}}
    & \makecell{ \begin{tabular}{@{}r r@{}} 98.8 & 93.8 \end{tabular}} 
    & \makecell{ \begin{tabular}{@{}r r@{}} 100.0 & 100.0 \end{tabular}}
    & \makecell{92.4} & \makecell{88.6} & \makecell{95.6}
    & \makecell{84.2} & \makecell{75.1} & \makecell{93.3}
  \\

               \bottomrule
    \end{tabular}
    }
    \caption{Results comparing GPT-5 models using COT prompt}
    \label{tab:appendix_cot}
\end{table*}

\begin{table*}[t]
    \centering
    \resizebox{\textwidth}{!}{
    \setlength{\tabcolsep}{1.7mm}
    \begin{tabular}{lcccccccccc}
        \toprule
       \textbf{Model} & \textbf{id} & \textbf{fe} & \textbf{ne} & \textbf{di} & \textbf{Avg} &\textbf{W-Avg} & \textbf{Mac-F1}  & \textbf{Mic-F1} & \textbf{PosF1} & \textbf{NegF1}  \\ 
       & \makecell{\begin{tabular}{@{}r  r@{}} 
                    \textbf{np} & \textbf{sp}
                \end{tabular}}
        & \makecell{\begin{tabular}{@{}r r@{}} 
                    \textbf{np} & \textbf{sp}
                \end{tabular}}
        & \makecell{\begin{tabular}{@{}r r@{}} 
                    \textbf{np} & \textbf{sp} 
                \end{tabular}}
        & \makecell{\begin{tabular}{@{}r r@{}} 
                    \textbf{np} & \textbf{sp}
                \end{tabular}} & & & & & & \\ \midrule
\rowcolor{rowgray}          \multirow{3}{*}{GPT-5-nano} 
    
    & \makecell{\begin{tabular}{@{}r r@{}} 99.0 & 100.0 \end{tabular}} 
    & \makecell{\begin{tabular}{@{}r r@{}} 75.0 & 74.5 \end{tabular} }
    & \makecell{ \begin{tabular}{@{}r r@{}} 94.0 & 93.5 \end{tabular} }
    & \makecell{ \begin{tabular}{@{}r r@{}} 100.0 & 100.0 \end{tabular} } 
    & \makecell{92.0} & \makecell{92.0} & \makecell{95.5}
    & \makecell{93.2} & \makecell{91.6} & \makecell{94.8}\\
\cline{2-11} \noalign{\vskip 3pt} 
    & \makecell{ \begin{tabular}{@{}r r@{}} 89.5 & 98.5 \end{tabular}}
    & \makecell{ \begin{tabular}{@{}r r@{}} 85.0 & 69.5 \end{tabular}}
    & \makecell{ \begin{tabular}{@{}r r@{}} 97.5 & 95.0 \end{tabular}} 
    & \makecell{ \begin{tabular}{@{}r r@{}} 100.0 & 100.0 \end{tabular}}
    & \makecell{91.9} & \makecell{89.2} & \makecell{95.5}
    & \makecell{90.0} & \makecell{83.8} & \makecell{95.9}
  \\  \hline \noalign{\vskip 1pt}
\rowcolor{rowgray}          \multirow{3}{*}{GPT-5.1} 
    
    & \makecell{\begin{tabular}{@{}r r@{}} 100.0 & 99.5 \end{tabular}} 
    & \makecell{\begin{tabular}{@{}r r@{}} 62.5 & 54.0 \end{tabular} }
    & \makecell{ \begin{tabular}{@{}r r@{}} 92.0 & 89.5 \end{tabular} }
    & \makecell{ \begin{tabular}{@{}r r@{}} 100.0 & 100.0 \end{tabular} } 
    & \makecell{87.2} & \makecell{87.2} & \makecell{92.2}
    & \makecell{89.0} & \makecell{86.0} & \makecell{91.9}\\
\cline{2-11} \noalign{\vskip 3pt} 
    & \makecell{ \begin{tabular}{@{}r r@{}} 90.5 & 91.5 \end{tabular}}
    & \makecell{ \begin{tabular}{@{}r r@{}} 87.0 & 53.0 \end{tabular}}
    & \makecell{ \begin{tabular}{@{}r r@{}} 99.5 & 92.5 \end{tabular}} 
    & \makecell{ \begin{tabular}{@{}r r@{}} 100.0 & 100.0 \end{tabular}}
    & \makecell{89.2} & \makecell{83.6} & \makecell{93.6}
    & \makecell{85.9} & \makecell{77.1} & \makecell{94.7}
  \\
               \bottomrule
    \end{tabular}
    }
    \caption{Results of GPT-5 models on combinations of all other perturbations except variable-renaming}
    \label{tab:appendix_except_var}
\end{table*}

\begin{table*}[!t]
    \centering
    \resizebox{\textwidth}{!}{
    \begin{tabular}{lcccccccccc}
        \toprule
       \textbf{Model} &  \textbf{fe} & \textbf{ne} & \textbf{di} & \textbf{Avg} &\textbf{W-Avg} & \textbf{Mac-F1}  & \textbf{Mic-F1} & \textbf{PosF1} & \textbf{NegF1}  \\ 
       & \makecell{\begin{tabular}{@{}r  r@{}} 
                    \textbf{np} & \textbf{sp}
                \end{tabular}}
        & \makecell{\begin{tabular}{@{}r r@{}} 
                    \textbf{np} & \textbf{sp}
                \end{tabular}}
        & \makecell{\begin{tabular}{@{}r r@{}} 
                    \textbf{np} & \textbf{sp}
                \end{tabular}} & & & & & & & \\ \midrule

\rowcolor{rowgray}          \multirow{3}{*}{GPT-5-nano} 
    
    & \makecell{\begin{tabular}{@{}r r@{}} 73.8 & 74.2 \end{tabular} }
    & \makecell{ \begin{tabular}{@{}r r@{}} 89.5 & 90.7 \end{tabular} }
    & \makecell{ \begin{tabular}{@{}r r@{}} 100.0 & 100.0 \end{tabular} } 
    & \makecell{88.0} & \makecell{84.5} & \makecell{93.3}
    & \makecell{83.8} & \makecell{81.2} & \makecell{86.4}\\
\cline{2-11} \noalign{\vskip 2pt}
   
    & \makecell{ \begin{tabular}{@{}r r@{}} 79.4 & 72.2 \end{tabular}}
    & \makecell{ \begin{tabular}{@{}r r@{}} 96.1 & 91.1 \end{tabular}} 
    & \makecell{ \begin{tabular}{@{}r r@{}} 100.0 & 100.0 \end{tabular}}
    & \makecell{89.8} & \makecell{82.8} & \makecell{94.3}
    & \makecell{81.1} & \makecell{71.0} & \makecell{91.1}
  \\ \hline \noalign{\vskip 3pt}

  \rowcolor{rowgray}          \multirow{3}{*}{GPT-5.1} 
    
    & \makecell{\begin{tabular}{@{}r r@{}} 51.1 & 49.7 \end{tabular} }
    & \makecell{ \begin{tabular}{@{}r r@{}} 91.4 & 92.6 \end{tabular} }
    & \makecell{ \begin{tabular}{@{}r r@{}} 100.0 & 100.0 \end{tabular} } 
    & \makecell{80.8} & \makecell{73.2} & \makecell{87.6}
    & \makecell{72.0} & \makecell{64.1} & \makecell{79.6}\\
\cline{2-11} \noalign{\vskip 2pt}
   
    & \makecell{ \begin{tabular}{@{}r r@{}} 93.2 & 49.0 \end{tabular}}
    & \makecell{ \begin{tabular}{@{}r r@{}} 98.7 & 92.4 \end{tabular}} 
    & \makecell{ \begin{tabular}{@{}r r@{}} 100.0 & 100.0 \end{tabular}}
    & \makecell{89.0} & \makecell{72.9} & \makecell{92.9}
    & \makecell{75.4} & \makecell{60.0} & \makecell{90.7}
  \\

               \bottomrule
    \end{tabular}
    }
    \caption{Results comparing GPT-5 models for pairs with one Java code and one Python code }
    \label{tab:appendix_java}
\end{table*}

\section{Different Configurations of Training Data for Finetuning}
\label{sec:FT-ratios}
\subsection{Training on only unperturbed pairs}

As observed in Table~\ref{tab:appendix_orionly}, clean performance is maintained, but robustness gains on perturbed inputs are modest when finetuning only on clean or unperturbed training data.

\begin{table*}[t]
    \resizebox{\textwidth}{!}{
    \setlength{\tabcolsep}{1.7mm}
     \begin{tabular}{lcccccccccc}
     
        \toprule
       \textbf{Model} & \textbf{id} & \textbf{fe} & \textbf{ne} & \textbf{di} & \textbf{Avg}  &\textbf{W-Avg} & \textbf{Mac-F1} & \textbf{Mic-F1} & \textbf{PosF1} & \textbf{NegF1}  \\ 
       
       & \makecell{\begin{tabular}{@{}r  r@{}} 
                    \textbf{np} & \textbf{sp}
                \end{tabular}}
        & \makecell{\begin{tabular}{@{}r r@{}} 
                    \textbf{np} & \textbf{sp}
                \end{tabular}}
        & \makecell{\begin{tabular}{@{}r r@{}} 
                    \textbf{np} & \textbf{sp} 
                \end{tabular}}
        & \makecell{\begin{tabular}{@{}r r@{}} 
                    \textbf{np} & \textbf{sp}
                \end{tabular}} & & & & & & \\  \midrule

\rowcolor{rowgray}          \multirow{3}{*}{Qwen} 
    
    & \makecell{\begin{tabular}{@{}r r@{}}  100.0 & 100.0  \end{tabular} }
    & \makecell{ \begin{tabular}{@{}r r@{}} 80.5 & 82.2  \end{tabular} }
    & \makecell{ \begin{tabular}{@{}r r@{}} 45.7 & 44.8 \end{tabular} } 
     & \makecell{ \begin{tabular}{@{}r r@{}} 100.0 & 99.0 \end{tabular} } 
    & \makecell{81.5} & \makecell{81.5} & \makecell{87.9}
    & \makecell{71.6} & \makecell{73.4} & \makecell{69.8}\\
\cline{2-11} \noalign{\vskip 2pt}
    & \makecell{ \begin{tabular}{@{}r r@{}} 76.0 & 96.5 \end{tabular}}
    & \makecell{ \begin{tabular}{@{}r r@{}} 68.8 & 80.6  \end{tabular}} 
    & \makecell{ \begin{tabular}{@{}r r@{}} 57.2 & 45.6 \end{tabular}}
    & \makecell{ \begin{tabular}{@{}r r@{}} 99.5 & 99.5 \end{tabular}}
    & \makecell{78.0} & \makecell{78.0} & \makecell{86.3}
    & \makecell{70.4} & \makecell{70.5} & \makecell{70.4}
  \\ \hline \noalign{\vskip 3pt}

\rowcolor{rowgray}          \multirow{3}{*}{Mistral} 
    
    & \makecell{\begin{tabular}{@{}r r@{}}  100.0 & 98.5  \end{tabular} }
    & \makecell{ \begin{tabular}{@{}r r@{}} 56.2 & 57.1  \end{tabular} }
    & \makecell{ \begin{tabular}{@{}r r@{}} 56.2 & 55.0 \end{tabular} } 
     & \makecell{ \begin{tabular}{@{}r r@{}} 94.0 & 94.0 \end{tabular} } 
    & \makecell{76.4} & \makecell{76.4} & \makecell{85.1}
    & \makecell{65.1} & \makecell{62.9} & \makecell{67.4}\\
\cline{2-11} \noalign{\vskip 2pt}
    & \makecell{ \begin{tabular}{@{}r r@{}} 76.5 & 81.5 \end{tabular}}
    & \makecell{ \begin{tabular}{@{}r r@{}} 51.4 & 51.6  \end{tabular}} 
    & \makecell{ \begin{tabular}{@{}r r@{}} 60.1 & 61.3 \end{tabular}}
    & \makecell{ \begin{tabular}{@{}r r@{}} 94.0 & 94.5 \end{tabular}}
    & \makecell{71.4} & \makecell{71.4} & \makecell{82.2}
    & \makecell{63.2} & \makecell{58.8} & \makecell{67.6}
  \\  \hline \noalign{\vskip 3pt}
       
\rowcolor{rowgray}          \multirow{3}{*}{DSC} 
    
    & \makecell{\begin{tabular}{@{}r r@{}} 100.0 & 99.0 \end{tabular} }
    & \makecell{ \begin{tabular}{@{}r r@{}} 99.9 & 99.9 \end{tabular} }
    & \makecell{ \begin{tabular}{@{}r r@{}} 1.4 & 2.7 \end{tabular} } 
    & \makecell{ \begin{tabular}{@{}r r@{}} 96.5 & 93.0 \end{tabular} }
    & \makecell{74.0} & \makecell{74.0} & \makecell{75.2}
    & \makecell{57.4} & \makecell{70.4} & \makecell{44.4}\\
\cline{2-11} \noalign{\vskip 2pt}
    
    & \makecell{ \begin{tabular}{@{}r r@{}} 0.0 & 99.0 \end{tabular}}
    & \makecell{ \begin{tabular}{@{}r r@{}} 0.1 & 99.9 \end{tabular}} 
    & \makecell{ \begin{tabular}{@{}r r@{}} 1.8 & 2.6 \end{tabular}}
    & \makecell{ \begin{tabular}{@{}r r@{}} 97.0 & 92.5 \end{tabular}}
    & \makecell{49.1} & \makecell{65.9} & \makecell{50.4}
    & \makecell{38.1} & \makecell{42.8} & \makecell{33.5}
  \\

               \bottomrule
    \end{tabular}
    }
    \captionof{table}{Results comparing performance of models trained only using unperturbed dataset.} 
    \label{tab:appendix_orionly}
\end{table*}

\subsection{Training on only perturbed pairs}

As observed in Table~\ref{tab:appendix_peronly}, training exclusively on perturbed data improves robustness but leads to noticeable degradation on unperturbed inputs, particularly for Mistral.

\begin{table*}[t]
    \resizebox{\textwidth}{!}{
    \setlength{\tabcolsep}{1.7mm}
     \begin{tabular}{lcccccccccc}
     
        \toprule
       \textbf{Model} & \textbf{id} & \textbf{fe} & \textbf{ne} & \textbf{di} & \textbf{Avg}  &\textbf{W-Avg} & \textbf{Mac-F1} & \textbf{Mic-F1} & \textbf{PosF1} & \textbf{NegF1}  \\ 
       
       & \makecell{\begin{tabular}{@{}r  r@{}} 
                    \textbf{np} & \textbf{sp}
                \end{tabular}}
        & \makecell{\begin{tabular}{@{}r r@{}} 
                    \textbf{np} & \textbf{sp}
                \end{tabular}}
        & \makecell{\begin{tabular}{@{}r r@{}} 
                    \textbf{np} & \textbf{sp} 
                \end{tabular}}
        & \makecell{\begin{tabular}{@{}r r@{}} 
                    \textbf{np} & \textbf{sp}
                \end{tabular}} & & & & & & \\  \midrule

\rowcolor{rowgray}          \multirow{3}{*}{Qwen} 
    
    & \makecell{\begin{tabular}{@{}r r@{}}  96.5 & 97.5  \end{tabular} }
    & \makecell{ \begin{tabular}{@{}r r@{}} 77.1 & 77.2  \end{tabular} }
    & \makecell{ \begin{tabular}{@{}r r@{}} 47.7 & 46.6 \end{tabular} } 
     & \makecell{ \begin{tabular}{@{}r r@{}} 100.0 & 99.5 \end{tabular} } 
    & \makecell{80.3} & \makecell{80.3} & \makecell{87.4}
    & \makecell{70.6} & \makecell{71.6} & \makecell{69.5}\\
\cline{2-11} \noalign{\vskip 2pt}
    & \makecell{ \begin{tabular}{@{}r r@{}} 87.0 & 93.0 \end{tabular}}
    & \makecell{ \begin{tabular}{@{}r r@{}} 80.1 & 84.2  \end{tabular}} 
    & \makecell{ \begin{tabular}{@{}r r@{}} 85.8 & 41.5 \end{tabular}}
    & \makecell{ \begin{tabular}{@{}r r@{}} 100.0 & 98.0 \end{tabular}}
    & \makecell{83.7} & \makecell{85.3} & \makecell{90.0}
    & \makecell{74.4} & \makecell{64.5} & \makecell{84.3}
  \\ \hline \noalign{\vskip 3pt}

\rowcolor{rowgray}          \multirow{3}{*}{Mistral} 
    
    & \makecell{\begin{tabular}{@{}r r@{}}  44.5 & 42.0  \end{tabular} }
    & \makecell{ \begin{tabular}{@{}r r@{}} 66.9 & 68.7  \end{tabular} }
    & \makecell{ \begin{tabular}{@{}r r@{}} 44.0 & 42.8 \end{tabular} } 
     & \makecell{ \begin{tabular}{@{}r r@{}} 86.5 & 87.0 \end{tabular} } 
    & \makecell{60.3} & \makecell{60.3} & \makecell{73.6}
    & \makecell{59.4} & \makecell{59.2} & \makecell{59.7}\\
\cline{2-11} \noalign{\vskip 2pt}
    & \makecell{ \begin{tabular}{@{}r r@{}} 76.0 & 70.5 \end{tabular}}
    & \makecell{ \begin{tabular}{@{}r r@{}} 60.0 & 73.2  \end{tabular}} 
    & \makecell{ \begin{tabular}{@{}r r@{}} 65.1 & 35.4 \end{tabular}}
    & \makecell{ \begin{tabular}{@{}r r@{}} 93.5 & 80.0 \end{tabular}}
    & \makecell{69.2} & \makecell{70.1} & \makecell{80.7}
    & \makecell{60.3} & \makecell{48.2} & \makecell{72.3}
  \\ \hline \noalign{\vskip 3pt}
       
\rowcolor{rowgray}          \multirow{3}{*}{DSC} 
    
    & \makecell{\begin{tabular}{@{}r r@{}} 89.5 & 82.6 \end{tabular} }
    & \makecell{ \begin{tabular}{@{}r r@{}} 85.8 & 85.5 \end{tabular} }
    & \makecell{ \begin{tabular}{@{}r r@{}} 35.2 & 34.8 \end{tabular} } 
    & \makecell{ \begin{tabular}{@{}r r@{}} 97.5 & 98.0 \end{tabular} }
    & \makecell{76.1} & \makecell{76.1} & \makecell{83.9}
    & \makecell{67.6} & \makecell{71.1} & \makecell{64.1}\\
\cline{2-11} \noalign{\vskip 2pt}
    
    & \makecell{ \begin{tabular}{@{}r r@{}} 91.1 & 83.2 \end{tabular}}
    & \makecell{ \begin{tabular}{@{}r r@{}} 81.2 & 88.3 \end{tabular}} 
    & \makecell{ \begin{tabular}{@{}r r@{}} 85.9 & 34.4 \end{tabular}}
    & \makecell{ \begin{tabular}{@{}r r@{}} 100.0 & 97.5 \end{tabular}}
    & \makecell{82.7} & \makecell{83.7} & \makecell{89.0}
    & \makecell{74.0} & \makecell{64.3} & \makecell{83.7}
  \\

               \bottomrule
    \end{tabular}
    }
    \captionof{table}{Results comparing performance of models trained only using perturbed dataset.} 
  
    \label{tab:appendix_peronly}
\end{table*}

\section{Extended Comparision with existing benchmarks}
\label{sec:appendix-sota-cmp}
For comparison with existing benchmarks other than EquiBench~\cite{wei2025equibench} and SeqCoBench~\cite{maveli-etal-2025-large}, we consider CodeNet~\cite{Puri2021CodeNetAL} and CLCDSA~\cite{Nafi2019CLCDSACL}. Due to the absence of executable test cases in these datasets, we could not validate perturbed code. Instead, we evaluated the models on unperturbed samples corresponding to our $id$, $fe$ and $di$ categories (using the same sample counts as CETBench). Results for these are presented in Tables \ref{tab:CLCDSA-A}, \ref{tab:CLCDSA-D} and \ref{tab:CodeNet}. We observe that unperturbed pairs of other datasets are equally or less challenging than the unperturbed CodeContests pairs currently used in CETBench (and less challenging compared to perturbed pairs of CETBench). This highlights that CETBench exposes specific brittleness to simple transformations that other benchmarks do not capture. Also, DeepSeek-Coder (DSC) exhibits a strong bias towards predicting positive labels across all datasets (high accuracy on $id$ or $fe$, near 0\% on $ne$ or $di$), confirming the trend observed in our main paper.

\begin{table*}
    \centering
    \setlength{\tabcolsep}{1.7mm}
    \begin{tabular}{p{2.2cm}lccc}
        \toprule
       \textbf{Model} & \textbf{id} & \textbf{fe} & \textbf{di}   \\ \midrule
        Qwen  & 100.0 & 83.0 & 99.8 \\ \hline
        Mistral & 100.0 & 87.8 & 96.0   \\ \hline
        DSC & 100.0& 100.0  & 2.0  \\
        \bottomrule
       \end{tabular}
    \caption{Results on CLCDSA – A (Easy level problems) dataset}
    \label{tab:CLCDSA-A}
\end{table*}

\begin{table*}
    \centering
    \setlength{\tabcolsep}{1.7mm}
    \begin{tabular}{p{2.2cm}lccc}
        \toprule
       \textbf{Model} & \textbf{id} & \textbf{fe} & \textbf{di}   \\ \midrule
    
        Qwen  & 100.0 & 83.4   & 99.0 \\ \hline
        Mistral & 100.0 & 92.3 & 69.0    \\ \hline
        DSC & 99.0& 99.2 & 8.0      \\
        \bottomrule
       \end{tabular}
    \caption{Results on CLCDSA – D (Difficult level problems) dataset}
    \label{tab:CLCDSA-D}
\end{table*}

\begin{table*}
    \centering
    \setlength{\tabcolsep}{1.7mm}
    \begin{tabular}{p{2.2cm}lccc}
        \toprule
       \textbf{Model} & \textbf{id} & \textbf{fe} & \textbf{di}   \\ \midrule
        Qwen  & 100.0 & 81.2   & 100.0 \\ \hline
        Mistral & 100.0 & 90.8 & 83.8   \\ \hline
        DSC & 100.0 & 100.0  & 7.5  \\
        \bottomrule
       \end{tabular}
    \caption{Results on CodeNet dataset}
    \label{tab:CodeNet}
\end{table*}

\begin{table*}
    \centering
    \resizebox{\textwidth}{!}{
    \setlength{\tabcolsep}{1.7mm}
    \begin{tabular}{ccccccccccc}
        \toprule
       \textbf{Model} & \textbf{id} & \textbf{fe} & \textbf{ne} & \textbf{di} & \textbf{Avg} &\textbf{W-Avg} & \textbf{Mac-F1}  & \textbf{Mic-F1} & \textbf{PosF1} & \textbf{NegF1}  \\ 
       & \makecell{\begin{tabular}{@{}r  r@{}} 
                    \textbf{np} & \textbf{sp}
                \end{tabular}}
        & \makecell{\begin{tabular}{@{}r r@{}} 
                    \textbf{np} & \textbf{sp}
                \end{tabular}}
        & \makecell{\begin{tabular}{@{}r r@{}} 
                    \textbf{np} & \textbf{sp} 
                \end{tabular}}
        & \makecell{\begin{tabular}{@{}r r@{}} 
                    \textbf{np} & \textbf{sp}
                \end{tabular}} & & & & & & \\ \midrule

Qwen
    & \makecell{ \begin{tabular}{@{}r r@{}} 33.5 & 79.0 \end{tabular}}
    & \makecell{ \begin{tabular}{@{}r r@{}} 56.6 & 37.8 \end{tabular}}
    & \makecell{ \begin{tabular}{@{}r r@{}} 68.7 & 68.1 \end{tabular}} 
    & \makecell{ \begin{tabular}{@{}r r@{}}  85.8 & 91.5  \end{tabular}}
    & \makecell{65.1} & \makecell{62.9} & \makecell{77.0}
    & \makecell{54.0} & \makecell{34.4} & \makecell{73.7}
  \\ \hline \noalign{\vskip 3pt}

Mistral
    & \makecell{ \begin{tabular}{@{}r r@{}} 5.5 & 90.5   \end{tabular}}
    & \makecell{ \begin{tabular}{@{}r r@{}} 14.7 & 85.4 \end{tabular}}
    & \makecell{ \begin{tabular}{@{}r r@{}} 28.7 & 24.9\end{tabular}} 
    & \makecell{ \begin{tabular}{@{}r r@{}}  76.0 & 76.0 \end{tabular}}
    & \makecell{50.2} & \makecell{63.0} & \makecell{60.0}
    & \makecell{44.8} & \makecell{42.0} & \makecell{47.6}
  \\ \hline \noalign{\vskip 3pt}

DSC
    & \makecell{ \begin{tabular}{@{}r r@{}} 25.5 & 44.0 \end{tabular}}
    & \makecell{ \begin{tabular}{@{}r r@{}}  14.3 & 36.9 \end{tabular}}
    & \makecell{ \begin{tabular}{@{}r r@{}} 18.9 & 17.5 \end{tabular}} 
    & \makecell{ \begin{tabular}{@{}r r@{}} 33.5 & 43.5 \end{tabular}}
    & \makecell{29.3} & \makecell{33.0} & \makecell{44.1}
    & \makecell{25.0} & \makecell{19.1} & \makecell{30.9}
  \\ \hline \noalign{\vskip 3pt} 
 GPT-4o   & \makecell{ \begin{tabular}{@{}r r@{}} 66.0 & 96.5 \end{tabular}}
    & \makecell{ \begin{tabular}{@{}r r@{}} 50.3 & 74.8 \end{tabular}}
    & \makecell{ \begin{tabular}{@{}r r@{}} 72.2 & 63.1 \end{tabular}} 
    & \makecell{ \begin{tabular}{@{}r r@{}} 100.0 & 100.0 \end{tabular}}
    & \makecell{77.9} & \makecell{80.4} & \makecell{86.4}
    & \makecell{67.6} & \makecell{56.1} & \makecell{79.1}
  \\

               \bottomrule
    \end{tabular}
    }
    \caption{Results on perturbed set after disclosing information about perturbation to the COT prompt}
    \label{tab:info_cot}
\end{table*}

\begin{table*}
    \centering
    \resizebox{\textwidth}{!}{
    \setlength{\tabcolsep}{1.7mm}
    \begin{tabular}{ccccccccccc}
        \toprule
       \textbf{Model} & \textbf{id} & \textbf{fe} & \textbf{ne} & \textbf{di} & \textbf{Avg} &\textbf{W-Avg} & \textbf{Mac-F1}  & \textbf{Mic-F1} & \textbf{PosF1} & \textbf{NegF1}  \\ 
       & \makecell{\begin{tabular}{@{}r  r@{}} 
                    \textbf{np} & \textbf{sp}
                \end{tabular}}
        & \makecell{\begin{tabular}{@{}r r@{}} 
                    \textbf{np} & \textbf{sp}
                \end{tabular}}
        & \makecell{\begin{tabular}{@{}r r@{}} 
                    \textbf{np} & \textbf{sp} 
                \end{tabular}}
        & \makecell{\begin{tabular}{@{}r r@{}} 
                    \textbf{np} & \textbf{sp}
                \end{tabular}} & & & & & & \\ \midrule

\rowcolor{rowgray}          \multirow{3}{*}{Qwen}

    & \makecell{\begin{tabular}{@{}r r@{}} 100.0 & 98.5  \end{tabular}} 
    & \makecell{\begin{tabular}{@{}r r@{}} 100.0 & 100.0  \end{tabular} }
    & \makecell{ \begin{tabular}{@{}r r@{}} 0.0 & 0.1 \end{tabular} }
    & \makecell{ \begin{tabular}{@{}r r@{}} 0.0 & 0.5 \end{tabular} } 
    & \makecell{49.9} & \makecell{49.9} & \makecell{50.0}
    & \makecell{31.7} & \makecell{63.1} & \makecell{0.2}\\
\cline{2-11} \noalign{\vskip 2pt} 

    & \makecell{ \begin{tabular}{@{}r r@{}} 8.5 & 98.5 \end{tabular}}
    & \makecell{ \begin{tabular}{@{}r r@{}} 6.7 & 100.0 \end{tabular}}
    & \makecell{ \begin{tabular}{@{}r r@{}} 8.0 & 0.1 \end{tabular}} 
    & \makecell{ \begin{tabular}{@{}r r@{}}  6.0 & 0.5  \end{tabular}}
    & \makecell{28.5} & \makecell{52.1} & \makecell{31.8}
    & \makecell{23.8} & \makecell{38.6} & \makecell{9.1}
  \\ \hline \noalign{\vskip 3pt}

\rowcolor{rowgray}          \multirow{3}{*}{Mistral}

    & \makecell{\begin{tabular}{@{}r r@{}} 99.5 & 99.0  \end{tabular}} 
    & \makecell{\begin{tabular}{@{}r r@{}} 99.4 & 99.3 \end{tabular} }
    & \makecell{ \begin{tabular}{@{}r r@{}} 0.3 & 1.0 \end{tabular} }
    & \makecell{ \begin{tabular}{@{}r r@{}} 0.0 & 0.5 \end{tabular} } 
    & \makecell{49.9} & \makecell{49.9} & \makecell{50.3}
    & \makecell{32.0} & \makecell{63.0} & \makecell{1.1}\\

\cline{2-11} \noalign{\vskip 2pt} 
    & \makecell{ \begin{tabular}{@{}r r@{}} 0.0 & 99.5   \end{tabular}}
    & \makecell{ \begin{tabular}{@{}r r@{}} 0.2 & 99.3 \end{tabular}}
    & \makecell{ \begin{tabular}{@{}r r@{}} 0.5 & 1.1 \end{tabular}} 
    & \makecell{ \begin{tabular}{@{}r r@{}}  1.0 & 1.0 \end{tabular}}
    & \makecell{25.3} & \makecell{50.0} & \makecell{25.9}
    & \makecell{19.4} & \makecell{37.4} & \makecell{1.3}
  \\ \hline \noalign{\vskip 3pt}
       
\rowcolor{rowgray}          \multirow{3}{*}{DSC} 

    & \makecell{\begin{tabular}{@{}r r@{}} 99.5 & 98.0 \end{tabular}} 
    & \makecell{\begin{tabular}{@{}r r@{}} 99.7 & 99.9 \end{tabular} }
    & \makecell{ \begin{tabular}{@{}r r@{}} 0.1 & 0.4 \end{tabular} }
    & \makecell{ \begin{tabular}{@{}r r@{}} 0.0 & 0.5 \end{tabular} } 
    & \makecell{49.8} & \makecell{49.8} & \makecell{50.1}
    & \makecell{31.7} & \makecell{62.9} & \makecell{0.5}\\

\cline{2-11} \noalign{\vskip 2pt} 
    & \makecell{ \begin{tabular}{@{}r r@{}} 0.5 & 98.0 \end{tabular}}
    & \makecell{ \begin{tabular}{@{}r r@{}}  0.3 & 100.0 \end{tabular}}
    & \makecell{ \begin{tabular}{@{}r r@{}} 0.1 & 0.0 \end{tabular}} 
    & \makecell{ \begin{tabular}{@{}r r@{}} 0.0 & 0.5 \end{tabular}}
    & \makecell{24.9} & \makecell{49.6} & \makecell{25.2}
    & \makecell{18.9} & \makecell{37.4} & \makecell{0.3}
  \\

               \bottomrule
    \end{tabular}
    }
    \caption{Few shot prompt}
    \label{tab:stricter_prompt_fs}
\end{table*}

\newpage

\section{Variance bounds}
\label{sec:appendix-var-bounds}

To analyse the variance in performance of models, we performed the following two experiments:

\subsection{Variance on different original pairs, for the same set of programming problems}
For this experiment, we generated four parallel sets of original and perturbed datasets. Here, we sampled the original pairs from the solution set of the same programming problems. We calculated the standard deviation in accuracies of the Qwen model on sets of original pairs, and as observed in table \ref{tab:var-ori}, we see negligible variance.

\subsection{Variance on different perturbed pairs for the same original pairs}
Here, we generated four parallel perturbed datasets by applying steps 2-17 of algorithm  \ref{alg:perturb} four times on a sampled pair. We calculated the standard deviation in the accuracies of Qwen. We previously observed higher variance for pairs perturbed from $fe$ and $ne$ categories. Due to this, we increased their number of samples in the test set from 400 to 2000 in each category. The table \ref{tab:var-per} shows that the final variance is negligible compared to the drop in performance from the original pairs. 

\begin{table*}[htbp]
    \centering

    \begin{tabular}{p{2.2cm}ccccc}
        \toprule
       \textbf{Model} & \textbf{id} & \textbf{fe} & \textbf{ne} & \textbf{di}   \\ \midrule
        Unperturbed  & 0  & 4.77e-3 & 4.92e-2 & 0 \\
        NP Perturbed & 4.02e-2 & 9.32e-2& 5.66e-2 & 3.54e-3 \\
        SP Perturbed & 6.86e-3 & 5.02e-2& 6.78e-2 & 0\\
        \bottomrule
       \end{tabular}

        \caption{Variance on changing original pairs}
        \label{tab:var-ori}
\end{table*}

\begin{table*}[htbp]
    \centering

    \begin{tabular}{p{2.2cm}ccccc}
        \toprule
       \textbf{Model} & \textbf{id} & \textbf{fe} & \textbf{ne} & \textbf{di}   \\ \midrule
        NP Perturbed  & 8.92e-4 & 8.33e-4 & 4.35e-4 & 9.17e-5 \\
        SP Perturbed  & 3.22e-4 & 1.22e-4 & 1.21e-3 & 4.92e-4\\
        \bottomrule
       \end{tabular}

        \caption{Variance on changing perturbed pairs}
        \label{tab:var-per}
\end{table*}

\section{Error bounds}
\label{sec:appendix-error-bounds}
For calculating the error bounds for each type of perturbation, we calculated the percentage of erroneous samples generated over a total number of perturbation attempts. The results for the same are in Table \ref{tab:error-bounds}. The low values of error percent denote the efficiency of our perturbation techniques.

\begin{table*}[hbt!]
    \centering
    
    \begin{tabular}{lcc}
        \toprule
       \textbf{Perturbation type} & \textbf{error \% in sp} & \textbf{error \% in np}   \\ \midrule
        If-else swapping & 3.86 & 11.68 \\ 
        For-while swapping & 4.64 & 5.529\\ 
        If condition flipping & 0.12 &  4.37\\
        Variable renaming & 5.35 & 9.86\\ 
        Boolean variable inversion & 2.10 & 0.55 \\
        Statement reordering & 1.93 &  8.99\\
        Expression reformatting & 2.88 & 3.33\\

        \bottomrule
       \end{tabular}
        
        \caption{Error bounds in dataset generation}
        \label{tab:error-bounds}
\end{table*}

\section{Error Analysis}
\label{sec:appendix-error-ana}
\begin{figure*}[htbp]
  \centering
  \begin{subfigure}{0.32\textwidth}
    \centering
    \includegraphics[width=\linewidth]{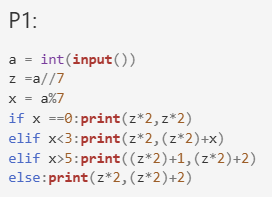}
    \caption{Original Pair program -1}
  \end{subfigure}
  \hfill
  \begin{subfigure}{0.32\textwidth}
    \centering
    \includegraphics[width=\linewidth]{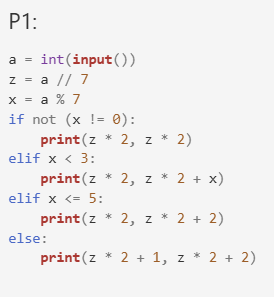}
    \caption{SP Perturbed Pair program -1}
  \end{subfigure}
  \hfill
  \begin{subfigure}{0.32\textwidth}
    \centering
    \includegraphics[width=\linewidth]{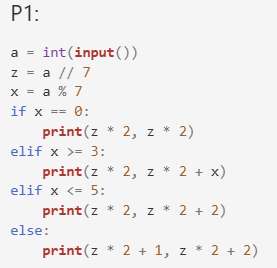}
    \caption{NP Perturbed Pair program -1}
  \end{subfigure}
  \caption{Sample pairs (Program -1 ). }
  \label{fig:prog-ex-p1}
\end{figure*}

\begin{figure*}[htbp]
  \centering
  \begin{subfigure}{0.32\textwidth}
    \centering
    \includegraphics[width=\linewidth]{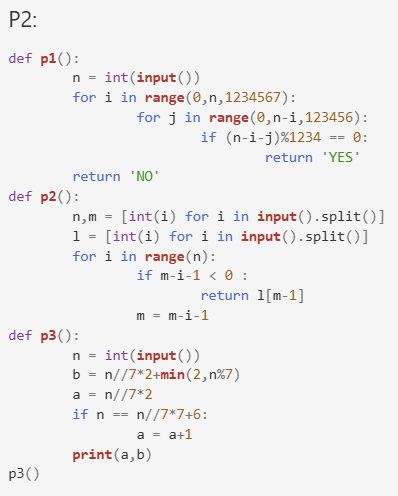}
    \caption{Original Pair program -2}
  \end{subfigure}
  \hfill
  \begin{subfigure}{0.32\textwidth}
    \centering
    \includegraphics[width=\linewidth]{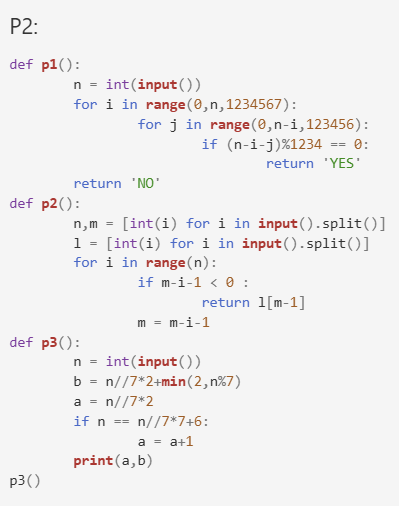}
    \caption{SP Perturbed Pair program -2}
  \end{subfigure}
  \hfill
  \begin{subfigure}{0.32\textwidth}
    \centering
    \includegraphics[width=\linewidth]{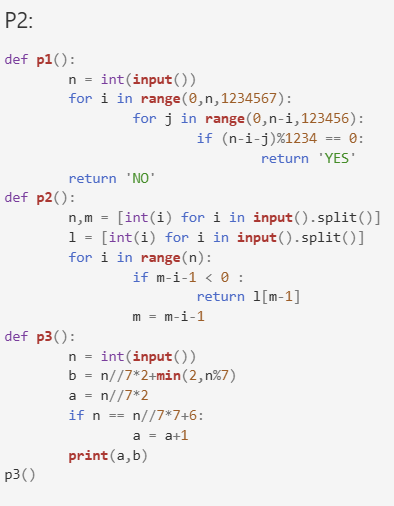}
    \caption{NP Perturbed Pair program -2}
  \end{subfigure}
  \caption{Sample pairs (Program -2 ). }
  \label{fig:prog-ex-p2}
\end{figure*}

\begin{figure*}[htbp]
  \centering
  
    \centering
    \includegraphics[width=\linewidth]{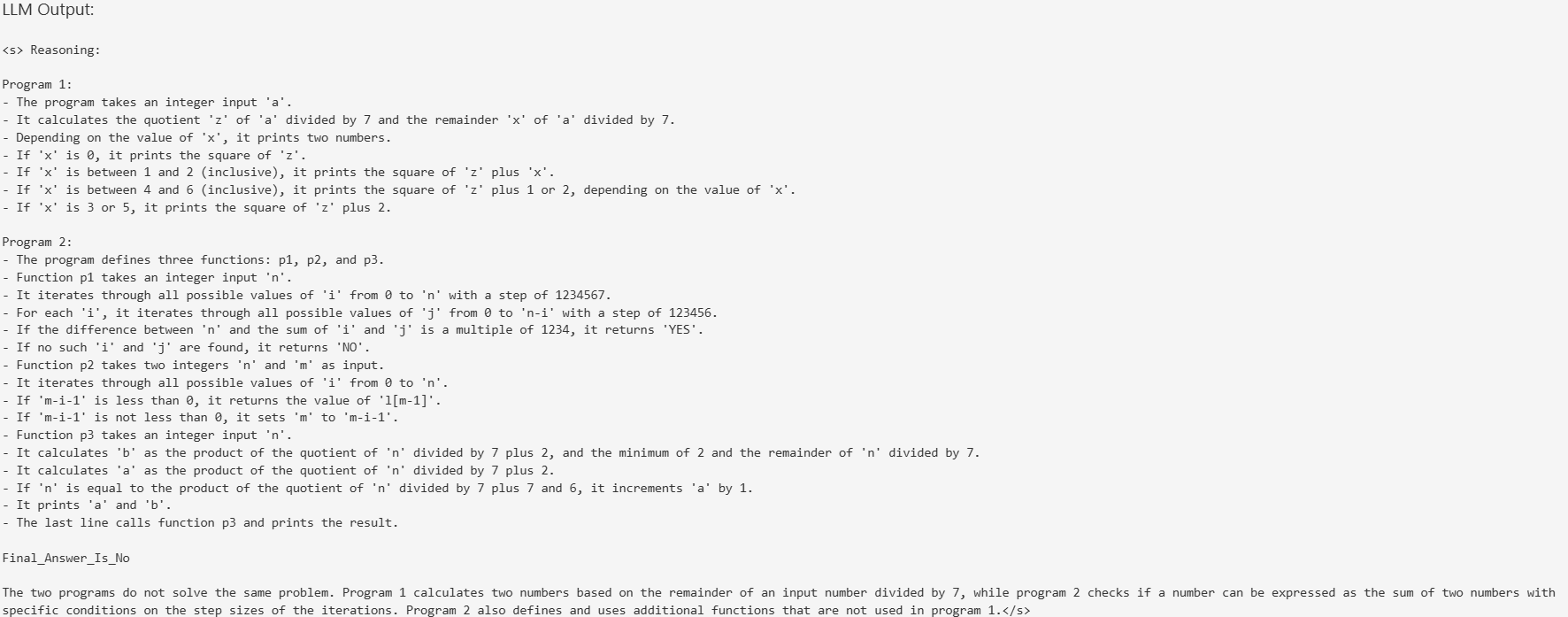}
  \caption{Sample generation of Mistral on original pair}
  \label{fig:gen-sample-prog}
\end{figure*}

\begin{figure*}[htbp]
  \centering
  
    \centering
    \includegraphics[width=\linewidth]{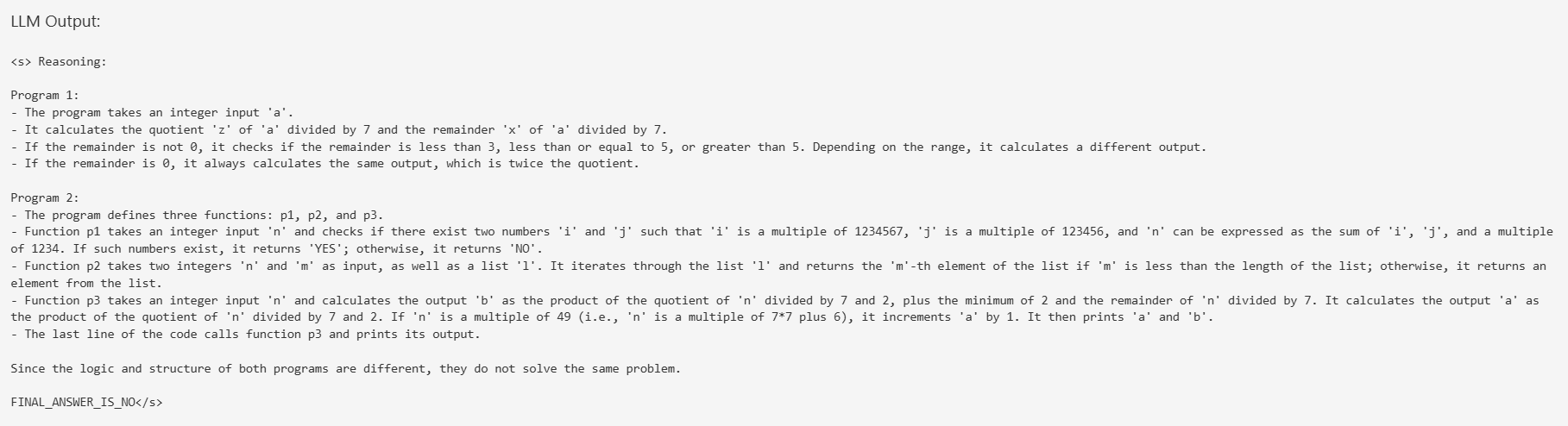}
  \caption{Sample generation of Mistral on SP perturbed pair}
  \label{fig:gen-sample-sp-prog}
\end{figure*}

\begin{figure*}[htbp]
  \centering
  
    \centering
    \includegraphics[width=\linewidth]{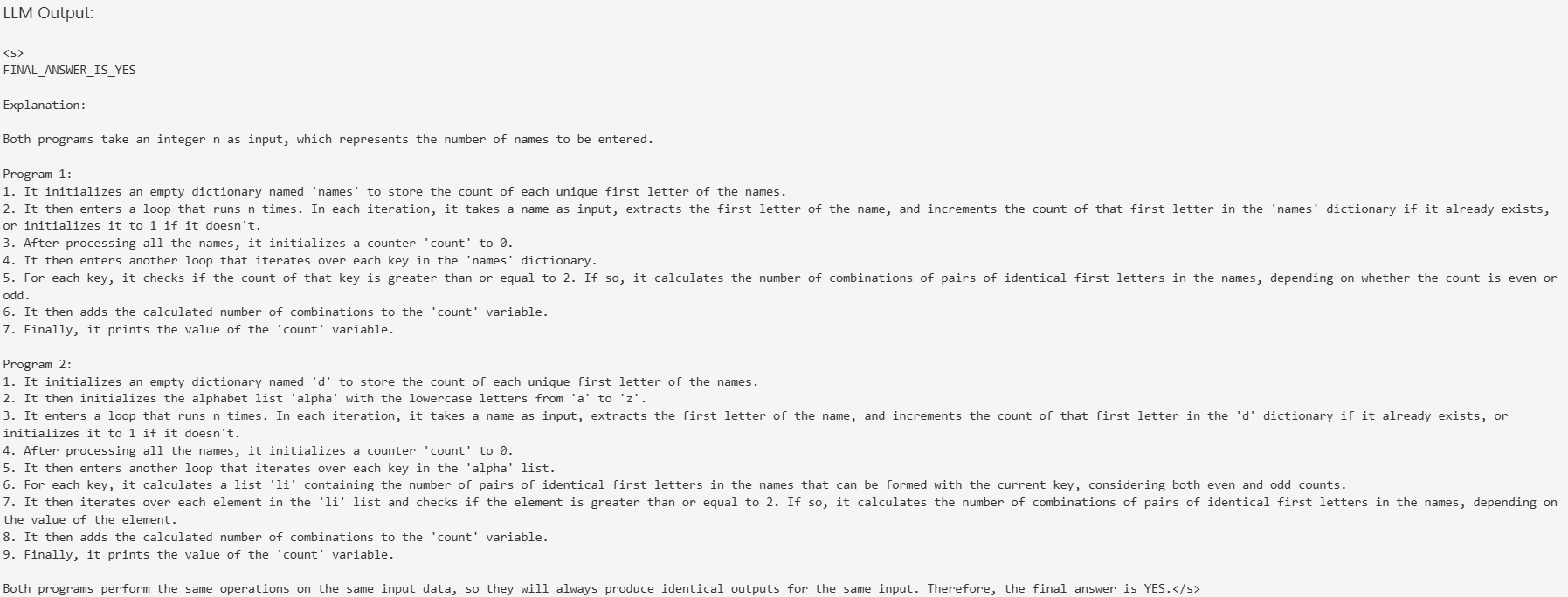}
  \caption{Sample generation of Mistral on NP perturbed pair}
  \label{fig:gen-sample-np-prog}
\end{figure*}

Sample programs from the $fe$ category are shown in Figures~\ref{fig:prog-ex-p1} and~\ref{fig:prog-ex-p2}. The original pair correctly solves the same problem (“670-A Holidays” from Codeforces\footnote{\url{https://codeforces.com/problemset/problem/670/A}}). However, as illustrated in Figure~\ref{fig:gen-sample-prog}, the Chain-of-Thought (CoT) reasoning generated by Mistral fails to recognize that the two programs are functionally equivalent, despite correctly identifying the underlying algorithms. This issue persists under both semantic-preserving (SP) and semantic-non-preserving (NP) perturbations. Figure~\ref{fig:gen-sample-sp-prog} shows that the model continues to misclassify equivalent programs after SP perturbations. Figure~\ref{fig:gen-sample-np-prog} demonstrates that the model fails to detect inequivalence after NP perturbations; although it correctly understands the original algorithms, it overlooks the semantic impact of the perturbation.

Tables~\ref{tab:error_gpt4}, \ref{tab:error_qwen}, \ref{tab:error_mistral}, and~\ref{tab:error_dsc} report how often model predictions in Chain-of-Thought (CoT) prompts change after perturbation. Each entry corresponds to the percentage of instances where the model’s final answer remains unchanged or flips when moving from the original to the perturbed pair. Highlighted values indicate cases where perturbations adversely affect model predictions. For example, values such as 18.5\% (for “No” remaining unchanged in $\langle fe, k, np \rangle$ in Table~\ref{tab:error_gpt4}) suggest that part of the measured performance may stem from incorrect reasoning on the original pairs themselves. Additionally, Table~\ref{tab:error_dsc} reports the percentage of CoT generations from DeepSeek-Coder that omit reasoning and directly produce an answer. This behavior was not observed for other models.

We further analyze the errors through the following questions:

\textbf{Why does CoT hurt performance for GPT-4o?}

\begin{figure*}[htbp]
  \centering
  \begin{subfigure}{0.48\textwidth}
    \centering
    \includegraphics[width=\linewidth]{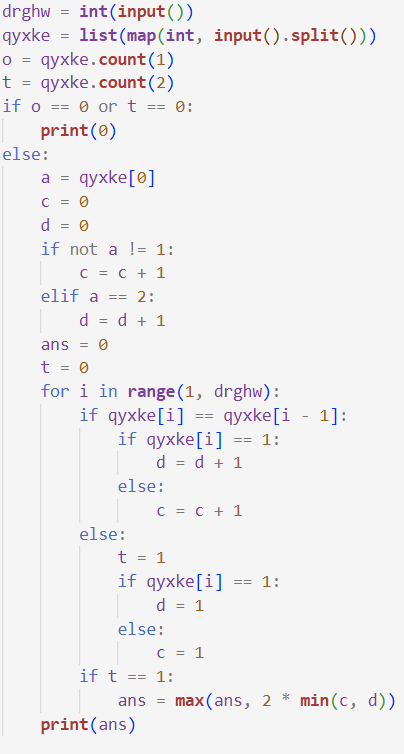}
    \caption{NP Perturbed Pair program -1}
  \end{subfigure}
  \hfill
  \begin{subfigure}{0.48\textwidth}
    \centering
    \includegraphics[width=\linewidth]{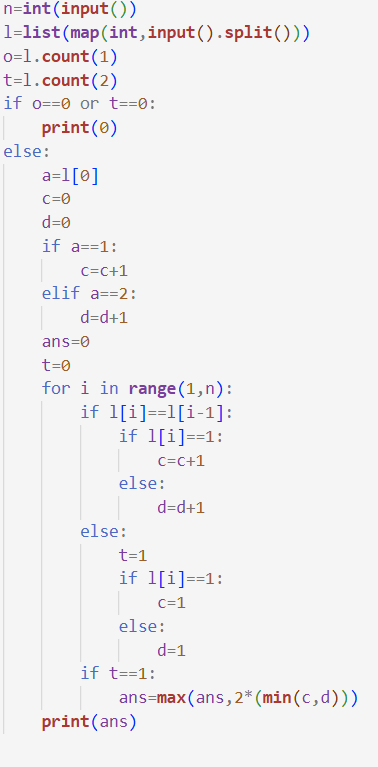}
    \caption{NP Perturbed Pair program -2}
  \end{subfigure}
  \hfill
  
  \caption{Error Analysis of GPT-4o's CoT: < id, k, np > pair - second if statement has the counters of c and d swapped }
  \label{fig:gpt4o-error-1}
\end{figure*}

\begin{figure*}[htbp]
  \centering
  
    \centering
    \includegraphics[width=\linewidth]{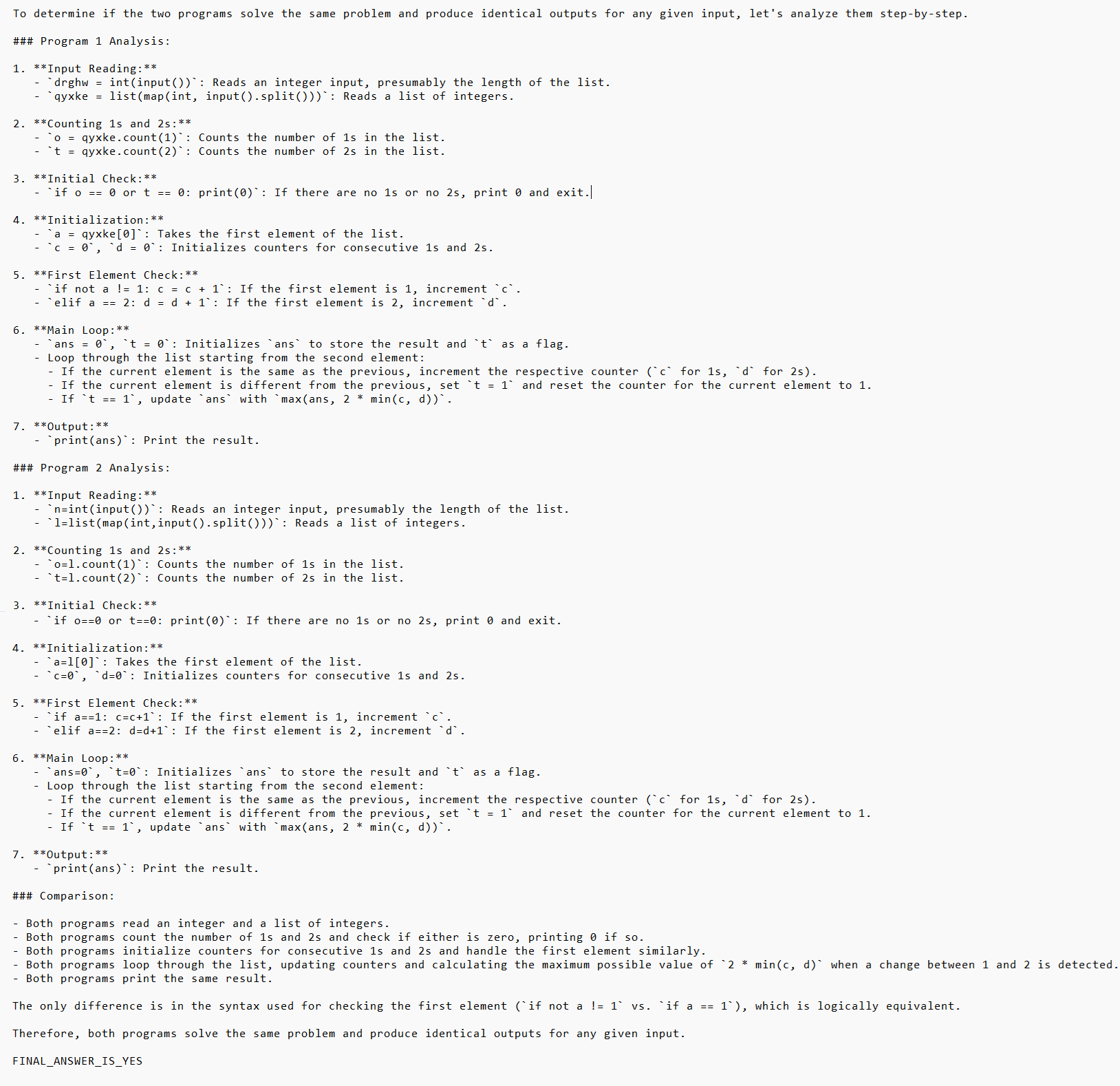}
  \caption{Error Analysis of GPT-4o's CoT: < id, k, np > pair - The response misses the perturbation.}
  \label{fig:gpt4o-error-1-desc}
\end{figure*}

\begin{figure*}[htbp]
  \centering
  \begin{subfigure}{0.35\textwidth}
    \centering
    \includegraphics[width=\linewidth]{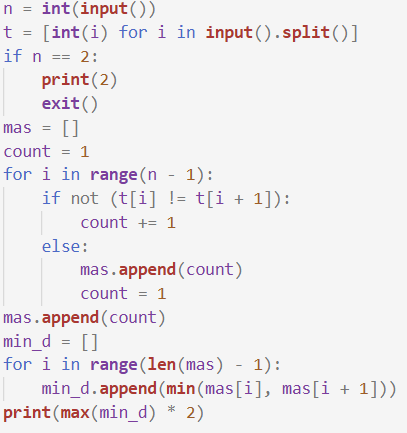}
    \caption{NP Perturbed Pair program -1}
  \end{subfigure}
  \hfill
  \begin{subfigure}{0.62\textwidth}
    \centering
    \includegraphics[width=\linewidth]{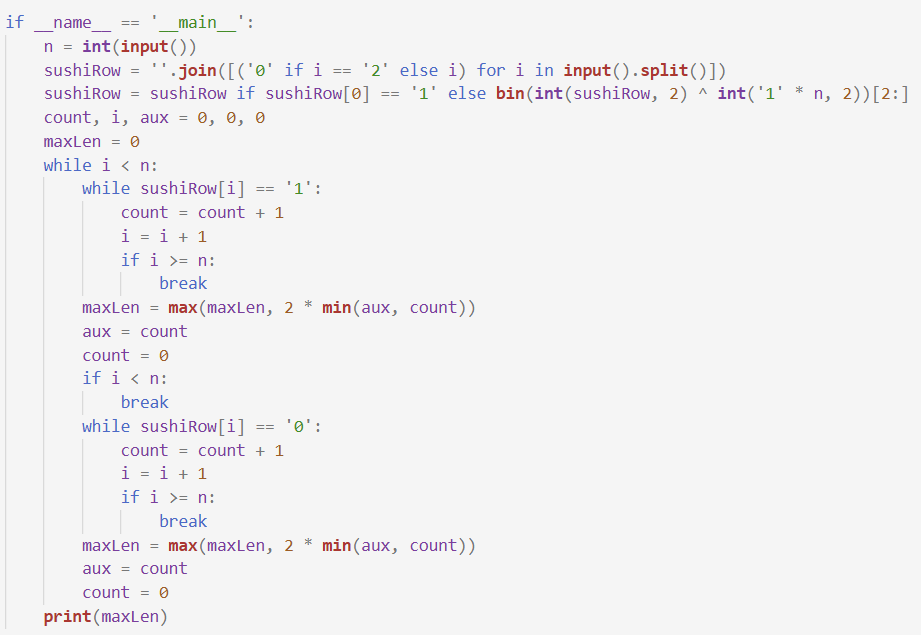}
    \caption{NP Perturbed Pair program -2}
  \end{subfigure}
  \hfill
  
  \caption{Error Analysis of GPT-4o's CoT: < fe, k, np > pair - the break statements in while loops in Program 2 make it exit after processing just 2 rows. }
  \label{fig:gpt4o-error-4}
\end{figure*}

\begin{figure*}[htbp]
  \centering
  
    \centering
    \includegraphics[width=\linewidth]{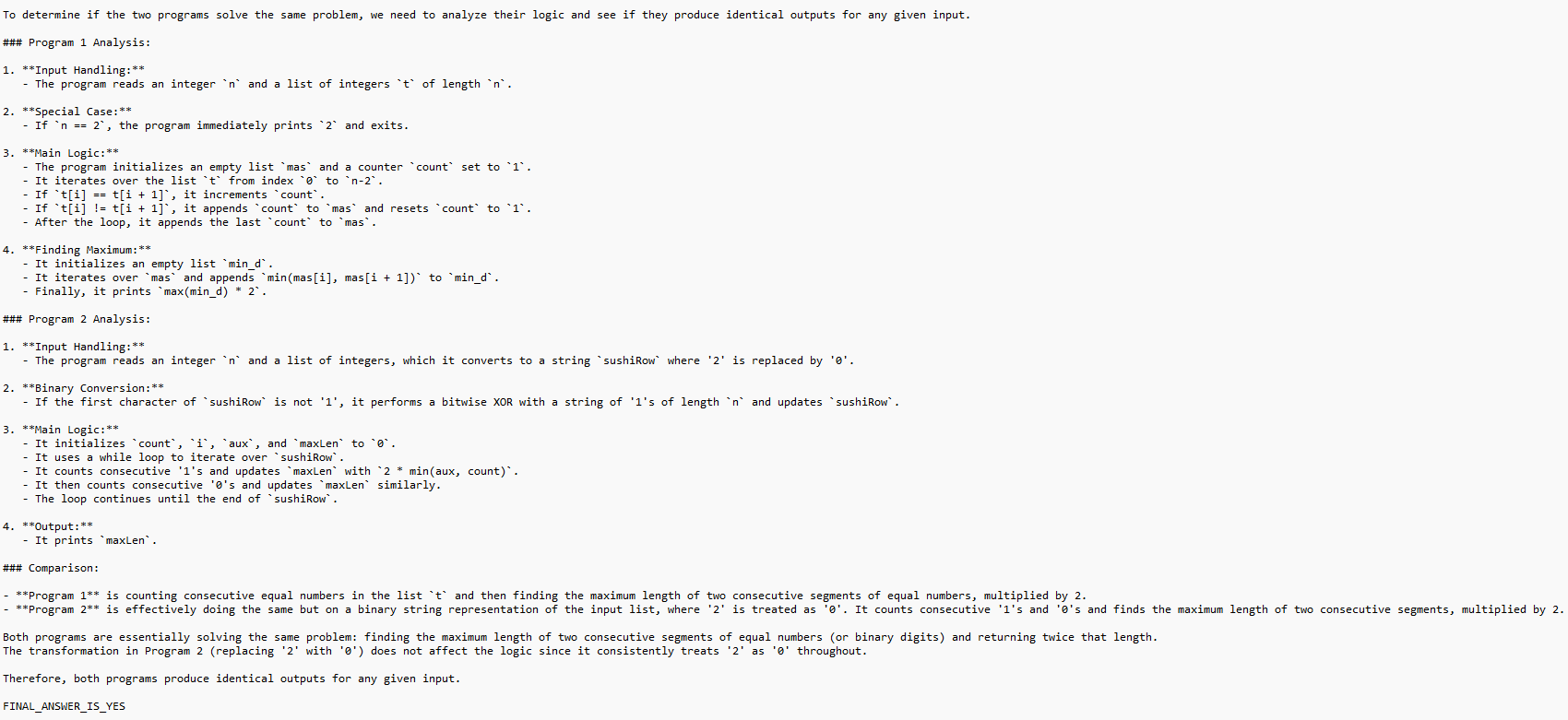}
  \caption{Error Analysis of GPT-4o's CoT: < fe, k, np > pair - The response incorrectly says that the loop continues untill the end of `sushiRow'.}
  \label{fig:gpt4o-error-4-desc}
\end{figure*}

\begin{figure*}[htbp]
  \centering
  \begin{subfigure}{0.48\textwidth}
    \centering
    \includegraphics[width=\linewidth]{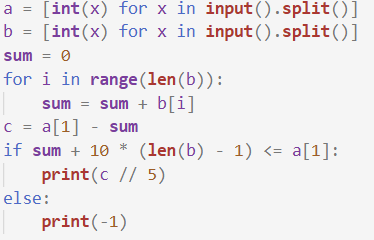}
    \caption{SP Perturbed Pair program -1}
  \end{subfigure}
  \hfill
  \begin{subfigure}{0.48\textwidth}
    \centering
    \includegraphics[width=\linewidth]{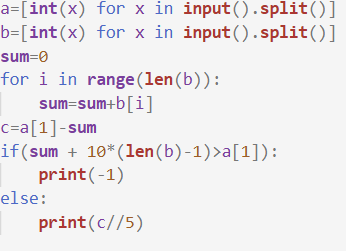}
    \caption{SP Perturbed Pair program -2}
  \end{subfigure}
  \hfill
  
  \caption{Error Analysis of GPT-4o's CoT: < id, k, sp > pair - Semantically equivalent if statements }
  \label{fig:gpt4o-error-2}
\end{figure*}

\begin{figure*}[htbp]
  \centering
  
    \centering
    \includegraphics[width=\linewidth]{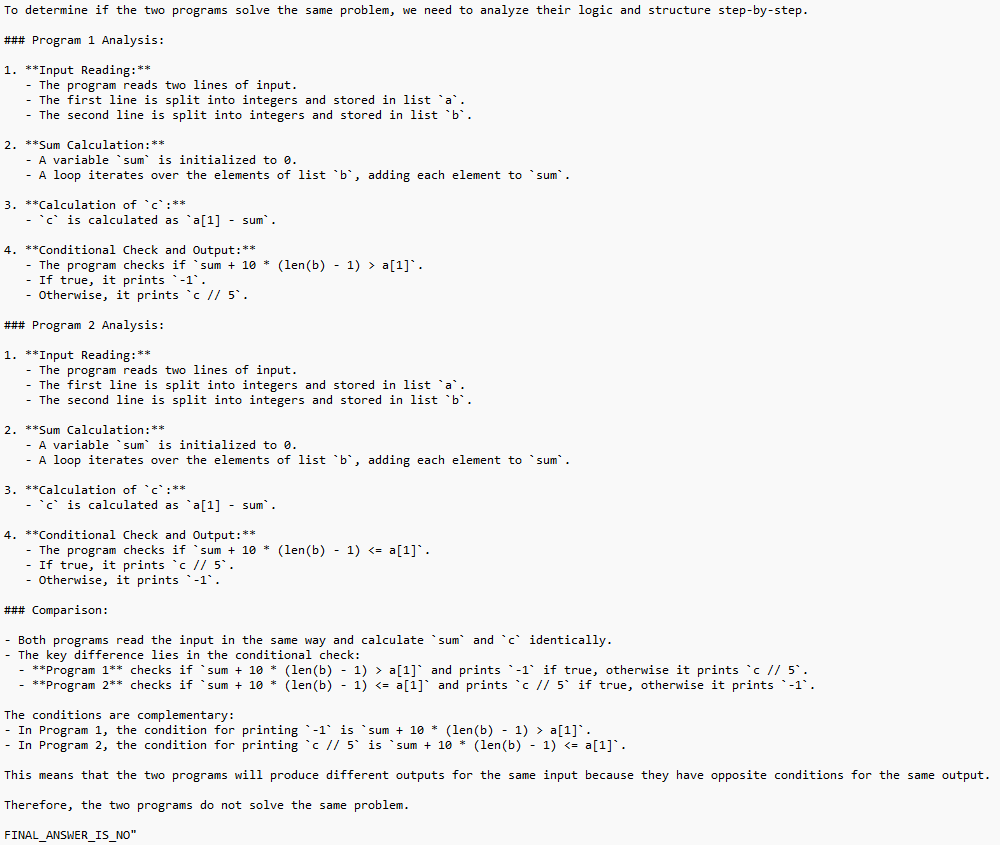}
  \caption{Error Analysis of GPT-4o's CoT: < id, k, sp > pair - The response misses the semantic equivalence of lines with different syntax.}
  \label{fig:gpt4o-error-2-desc}
\end{figure*}

\begin{figure*}[htbp]
  \centering
  \begin{subfigure}{0.65\textwidth}
    \centering
    \includegraphics[width=\linewidth]{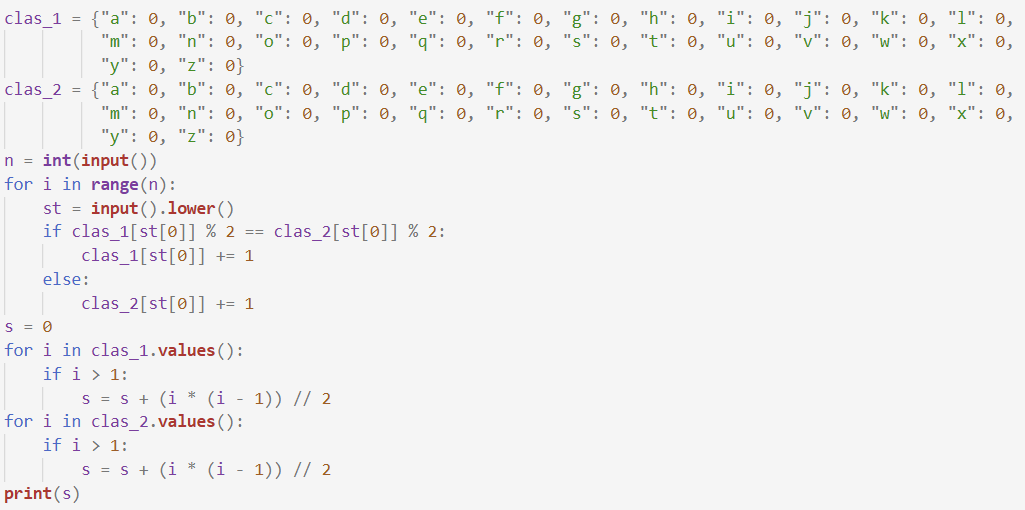}
    \caption{SP Perturbed Pair program -1}
  \end{subfigure}
  \hfill
  \begin{subfigure}{0.33\textwidth}
    \centering
    \includegraphics[width=\linewidth]{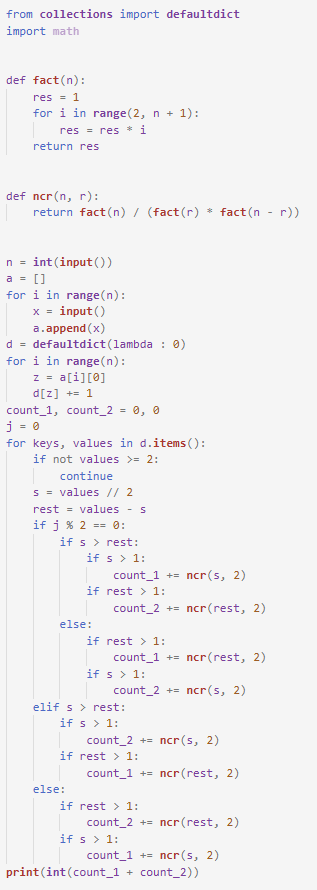}
    \caption{SP Perturbed Pair program -2}
  \end{subfigure}
  \hfill
  
  \caption{Error Analysis of GPT-4o's CoT: < fe, k, sp > pair - Two different programs performing the same operation. }
  \label{fig:gpt4o-error-5}
\end{figure*}

\begin{figure*}[htbp]
  \centering
  
    \centering
    \includegraphics[width=\linewidth]{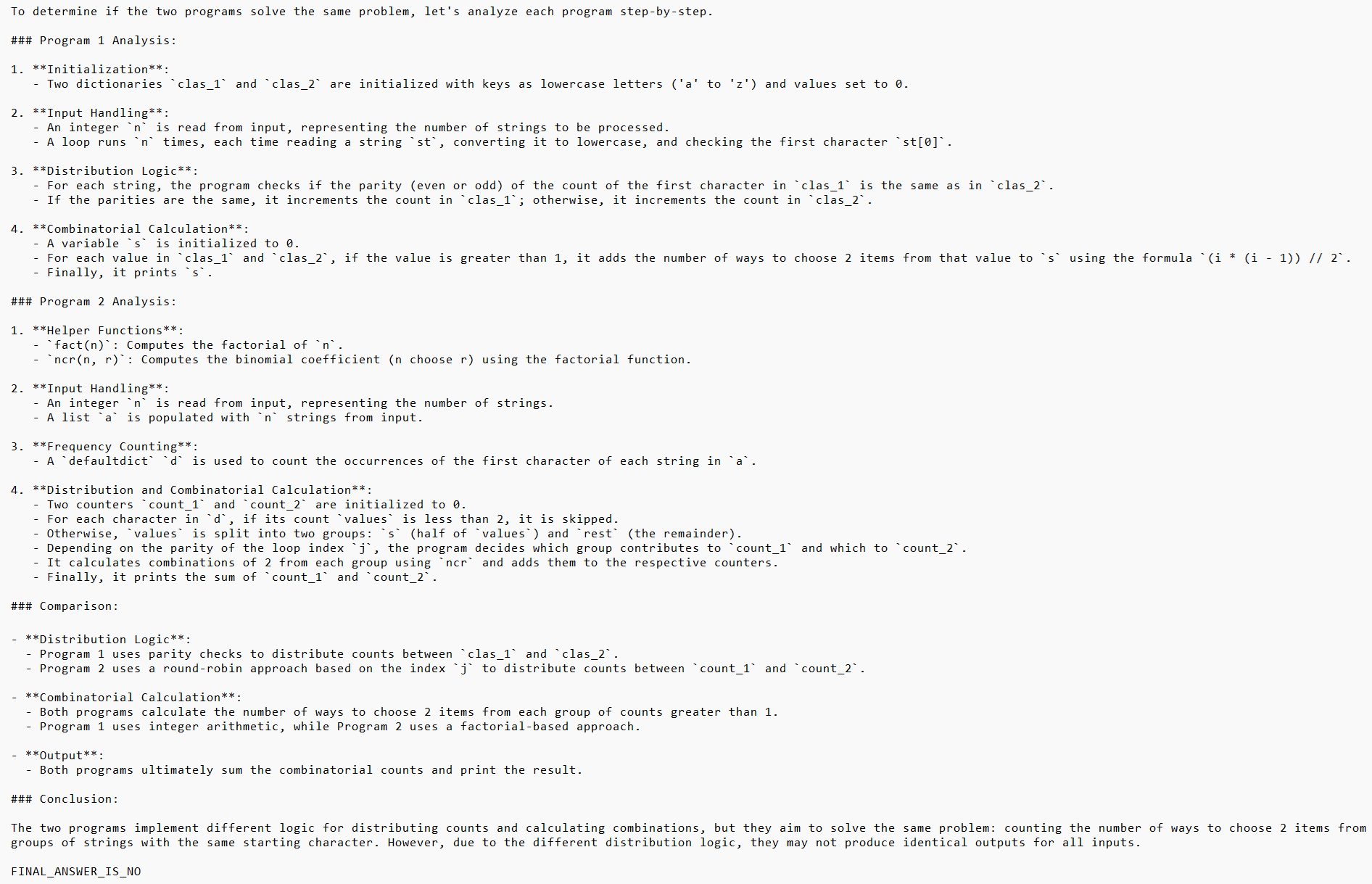}
  \caption{Error Analysis of GPT-4o's CoT: < fe, k, sp > pair.}
  \label{fig:gpt4o-error-5-desc}
\end{figure*}

To investigate this, we qualitatively analyze 50 randomly sampled program pairs where GPT-4o produces correct predictions without CoT but fails with CoT. (Figures \ref{fig:gpt4o-error-1}, \ref{fig:gpt4o-error-2}, \ref{fig:gpt4o-error-4} and \ref{fig:gpt4o-error-5} are few such samples.) We observe that for negative pairs generated via NP perturbations, CoT reasoning often fails to capture subtle semantic differences introduced by the perturbation, leading the model to incorrectly predict equivalence. (As observed in Figures \ref{fig:gpt4o-error-1-desc} and \ref{fig:gpt4o-error-4-desc}.) Conversely, for positive pairs with SP perturbations, the model is misled by local structural changes and incorrectly predicts non-equivalence. (As observed in Figures \ref{fig:gpt4o-error-2-desc} and \ref{fig:gpt4o-error-5-desc}.) These observations suggest that the model struggles to reason about the semantic impact of local perturbations. Additional reasoning steps introduced by CoT appear to amplify this confusion, resulting in degraded performance compared to direct prediction.  Examining this further, and possibly connecting this analysis with how the LLM attends to the reasoning chain during COT, is an interesting direction for future work.

\textbf{Why do some models have low performance before perturbation?}

We analyze model predictions for the $ne$ category and observe that accuracy below 50\% on original pairs indicates a strong bias toward the positive label. In this category, the two programs correspond to different implementations of the same problem, often exhibiting high structural similarity, with one program containing a subtle error. Due to this structural similarity, models tend to predict equivalence based on surface-level patterns rather than correctly identifying semantic differences. This behavior is consistent with our broader findings that LLMs rely more on syntactic cues than deep semantic reasoning. After introducing perturbations, this bias is partially reduced, leading to improved performance in the $ne$ category. This improvement likely arises because perturbations disrupt superficial structural similarity, making it easier for the model to distinguish between equivalent and non-equivalent programs.

\begin{table*}[t]
 \centering
    \resizebox{\textwidth}{!}{
    \setlength{\tabcolsep}{4mm}
     \begin{tabular}{ccccccccccc}
        \toprule
       \textbf{Qwen's Generation} & \textbf{id} & \textbf{fe} & \textbf{ne} & \textbf{di} \\ 
       & \makecell{\begin{tabular}{@{}r  r@{}} 
                    \textbf{np} & \textbf{sp}
                \end{tabular}}
        & \makecell{\begin{tabular}{@{}r r@{}} 
                    \textbf{np} & \textbf{sp}
                \end{tabular}}
        & \makecell{\begin{tabular}{@{}r r@{}} 
                    \textbf{np} & \textbf{sp} 
                \end{tabular}}
        & \makecell{\begin{tabular}{@{}r r@{}} 
                    \textbf{np} & \textbf{sp}
                \end{tabular}}  \\ \midrule
Original: Yes, Perturbed: Yes
    & \makecell{ \begin{tabular}{@{}r r@{}} \textbf{ 73.4} & 82.1  \end{tabular}}
    & \makecell{ \begin{tabular}{@{}r r@{}}  \textbf{ 36.5} & 36.6 \end{tabular}}
    & \makecell{ \begin{tabular}{@{}r r@{}}  15.4 & 16.8  \end{tabular}} 
    & \makecell{ \begin{tabular}{@{}r r@{}} 0.0 & 0.0 \end{tabular}}
    
  \\  \midrule

  Original: Yes, Perturbed: No
    & \makecell{ \begin{tabular}{@{}r r@{}} 26.0 & \textbf{ 14.3}  \end{tabular}}
    & \makecell{ \begin{tabular}{@{}r r@{}}  31.1 & \textbf{ 25.7}  \end{tabular}}
    & \makecell{ \begin{tabular}{@{}r r@{}}  29.2 & 24.4  \end{tabular}} 
    & \makecell{ \begin{tabular}{@{}r r@{}} 0.5 & 1.5 \end{tabular}}
    
  \\  \midrule

  Original: No, Perturbed: Yes
       & \makecell{ \begin{tabular}{@{}r r@{}} 0.4 & 3.6  \end{tabular}}
    & \makecell{ \begin{tabular}{@{}r r@{}} 18.8 & 23.1  \end{tabular}}
    & \makecell{ \begin{tabular}{@{}r r@{}}  \textbf{20.4} & \textbf{23.0} \end{tabular}} 
    & \makecell{ \begin{tabular}{@{}r r@{}} \textbf{0.5} & \textbf{0.5} \end{tabular}}
    
  \\  \midrule

  Original: No, Perturbed: No
    & \makecell{ \begin{tabular}{@{}r r@{}} 0.0 & 0.0  \end{tabular}}
    & \makecell{ \begin{tabular}{@{}r r@{}} 13.6 & 14.6  \end{tabular}}
    & \makecell{ \begin{tabular}{@{}r r@{}}  35.0 & 35.8 \end{tabular}} 
    & \makecell{ \begin{tabular}{@{}r r@{}} 98.9 & 98.0 \end{tabular}}
    
  \\  

               \bottomrule
    \end{tabular}
    }
    \captionof{table}{Analysis of percentage of answers changed or unchanged by Qwen on the program pair before and after perturbation.}
    \label{tab:error_qwen}
\end{table*}

\begin{table*}[t]
 \centering
    \resizebox{\textwidth}{!}{
    \setlength{\tabcolsep}{4mm}
     \begin{tabular}{ccccccccccc}
        \toprule
       \textbf{Mistral's Generation} & \textbf{id} & \textbf{fe} & \textbf{ne} & \textbf{di} \\ 
       & \makecell{\begin{tabular}{@{}r  r@{}} 
                    \textbf{np} & \textbf{sp}
                \end{tabular}}
        & \makecell{\begin{tabular}{@{}r r@{}} 
                    \textbf{np} & \textbf{sp}
                \end{tabular}}
        & \makecell{\begin{tabular}{@{}r r@{}} 
                    \textbf{np} & \textbf{sp} 
                \end{tabular}}
        & \makecell{\begin{tabular}{@{}r r@{}} 
                    \textbf{np} & \textbf{sp}
                \end{tabular}}  \\ \midrule
Original: Yes, Perturbed: Yes
    & \makecell{ \begin{tabular}{@{}r r@{}} \textbf{ 98.5} & 94.5  \end{tabular}}
    & \makecell{ \begin{tabular}{@{}r r@{}}  \textbf{ 38.7} & 42.4  \end{tabular}}
    & \makecell{ \begin{tabular}{@{}r r@{}}  19.2 & 18.7  \end{tabular}} 
    & \makecell{ \begin{tabular}{@{}r r@{}} 0.1 & 0.0 \end{tabular}}
    
  \\  \midrule

  Original: Yes, Perturbed: No
    & \makecell{ \begin{tabular}{@{}r r@{}} 1.5 & \textbf{ 4.5}  \end{tabular}}
    & \makecell{ \begin{tabular}{@{}r r@{}}   18.6 & \textbf{ 16.3}  \end{tabular}}
    & \makecell{ \begin{tabular}{@{}r r@{}}  16.3 & 15.6  \end{tabular}} 
    & \makecell{ \begin{tabular}{@{}r r@{}} 3.0 & 2.6 \end{tabular}}
    
  \\  \midrule

  Original: No, Perturbed: Yes
    & \makecell{ \begin{tabular}{@{}r r@{}} 0.5 & 1.5  \end{tabular}}
    & \makecell{ \begin{tabular}{@{}r r@{}} 19.6 & 19.1  \end{tabular}}
    & \makecell{ \begin{tabular}{@{}r r@{}}  \textbf{20.5} & \textbf{ 20.7}  \end{tabular}} 
    & \makecell{ \begin{tabular}{@{}r r@{}} \textbf{6.0}& \textbf{ 9.9} \end{tabular}}
    
  \\  \midrule

  Original: No, Perturbed: No
    & \makecell{ \begin{tabular}{@{}r r@{}} 0.0 & 0.5  \end{tabular}}
    & \makecell{ \begin{tabular}{@{}r r@{}} 22.9 & 22.2   \end{tabular}}
    & \makecell{ \begin{tabular}{@{}r r@{}}  43.9 & 45.0 \end{tabular}} 
    & \makecell{ \begin{tabular}{@{}r r@{}} 90.0 & 87.4 \end{tabular}}
    
  \\  

               \bottomrule
    \end{tabular}
    }
    \captionof{table}{Analysis of percentage of answers changed or unchanged by Mistral on the program pair before and after perturbation.}
    \label{tab:error_mistral}
\end{table*}

\begin{table*}[t]
 \centering
    \resizebox{\textwidth}{!}{
    \setlength{\tabcolsep}{4mm}
     \begin{tabular}{ccccccccccc}
        \toprule
       \textbf{DSC's Generation} & \textbf{id} & \textbf{fe} & \textbf{ne} & \textbf{di} \\ 
       & \makecell{\begin{tabular}{@{}r  r@{}} 
                    \textbf{np} & \textbf{sp}
                \end{tabular}}
        & \makecell{\begin{tabular}{@{}r r@{}} 
                    \textbf{np} & \textbf{sp}
                \end{tabular}}
        & \makecell{\begin{tabular}{@{}r r@{}} 
                    \textbf{np} & \textbf{sp} 
                \end{tabular}}
        & \makecell{\begin{tabular}{@{}r r@{}} 
                    \textbf{np} & \textbf{sp}
                \end{tabular}}  \\ \midrule
Original: Yes, Perturbed: Yes
    & \makecell{ \begin{tabular}{@{}r r@{}} \textbf{ 83.0} & 92.0  \end{tabular}}
    & \makecell{ \begin{tabular}{@{}r r@{}}  \textbf{ 73.3} & 78.5  \end{tabular}}
    & \makecell{ \begin{tabular}{@{}r r@{}}  59.9 & 63.5  \end{tabular}} 
    & \makecell{ \begin{tabular}{@{}r r@{}} 2.5 & 3.5 \end{tabular}}
    
  \\  \midrule

  Original: Yes, Perturbed: No
    & \makecell{ \begin{tabular}{@{}r r@{}} 16.0 & \textbf{ 6.5}  \end{tabular}}
    & \makecell{ \begin{tabular}{@{}r r@{}}   14.0 & \textbf{ 8.9}  \end{tabular}}
    & \makecell{ \begin{tabular}{@{}r r@{}}  16.3 & 12.5  \end{tabular}} 
    & \makecell{ \begin{tabular}{@{}r r@{}} 8.0 & 10.5 \end{tabular}}
    
  \\  \midrule

  Original: No, Perturbed: Yes
    & \makecell{ \begin{tabular}{@{}r r@{}} 0.5 & 1.5  \end{tabular}}
    & \makecell{ \begin{tabular}{@{}r r@{}} 7.5 & 6.7  \end{tabular}}
    & \makecell{ \begin{tabular}{@{}r r@{}}  \textbf{9.6} & \textbf{ 10.5}  \end{tabular}} 
    & \makecell{ \begin{tabular}{@{}r r@{}} \textbf{ 5.5}& \textbf{ 6.5} \end{tabular}}
    
  \\  \midrule

  Original: No, Perturbed: No
    & \makecell{ \begin{tabular}{@{}r r@{}} 0.5 & 0.0  \end{tabular}}
    & \makecell{ \begin{tabular}{@{}r r@{}} 5.1 & 5.8  \end{tabular}}
    & \makecell{ \begin{tabular}{@{}r r@{}}  14.2 & 13.5 \end{tabular}} 
    & \makecell{ \begin{tabular}{@{}r r@{}} 83.9 & 79.5 \end{tabular}}
    
  \\  \midrule

    Missing Reason
    & \makecell{ \begin{tabular}{@{}r r@{}} 44.5 & 40.5  \end{tabular}}
    & \makecell{ \begin{tabular}{@{}r r@{}} 7.3 & 6.9  \end{tabular}}
    & \makecell{ \begin{tabular}{@{}r r@{}}  5.9 & 5.1 \end{tabular}} 
    & \makecell{ \begin{tabular}{@{}r r@{}} 0.5 & 0.5 \end{tabular}}
    
  \\

               \bottomrule
    \end{tabular}
    }
    \captionof{table}{Analysis of percentage of answers changed or unchanged by Deepseek-Coder on the program pair before and after perturbation.}
    \label{tab:error_dsc}
\end{table*}

\begin{table*}[t]
 \centering
    \resizebox{\textwidth}{!}{
    \setlength{\tabcolsep}{4mm}
     \begin{tabular}{ccccccccccc}
        \toprule
       \textbf{GPT-4o's Generation} & \textbf{id} & \textbf{fe} & \textbf{ne} & \textbf{di} \\ 
       & \makecell{\begin{tabular}{@{}r  r@{}} 
                    \textbf{np} & \textbf{sp}
                \end{tabular}}
        & \makecell{\begin{tabular}{@{}r r@{}} 
                    \textbf{np} & \textbf{sp}
                \end{tabular}}
        & \makecell{\begin{tabular}{@{}r r@{}} 
                    \textbf{np} & \textbf{sp} 
                \end{tabular}}
        & \makecell{\begin{tabular}{@{}r r@{}} 
                    \textbf{np} & \textbf{sp}
                \end{tabular}}  \\ \midrule
Original: Yes, Perturbed: Yes
    & \makecell{ \begin{tabular}{@{}r r@{}} \textbf{ 34.5} & 99.0  \end{tabular}}
    & \makecell{ \begin{tabular}{@{}r r@{}}  \textbf{ 47.4} & 66.4  \end{tabular}}
    & \makecell{ \begin{tabular}{@{}r r@{}}  21.3 & 27.9  \end{tabular}} 
    & \makecell{ \begin{tabular}{@{}r r@{}} 0.0 & 0.0 \end{tabular}}
    
  \\  \midrule

  Original: Yes, Perturbed: No
    & \makecell{ \begin{tabular}{@{}r r@{}} 65.5 & \textbf{ 1.0}  \end{tabular}}
    & \makecell{ \begin{tabular}{@{}r r@{}}   29.0 & \textbf{ 1.0}  \end{tabular}}
    & \makecell{ \begin{tabular}{@{}r r@{}}  18.2 & 11.0  \end{tabular}} 
    & \makecell{ \begin{tabular}{@{}r r@{}} 0.0 & 0.0 \end{tabular}}
    
  \\  \midrule

  Original: No, Perturbed: Yes
    & \makecell{ \begin{tabular}{@{}r r@{}} 0.0 & 0.0  \end{tabular}}
    & \makecell{ \begin{tabular}{@{}r r@{}}  5.1 & 8.2  \end{tabular}}
    & \makecell{ \begin{tabular}{@{}r r@{}}  \textbf{ 4.7} & \textbf{ 5.7}  \end{tabular}} 
    & \makecell{ \begin{tabular}{@{}r r@{}} \textcolor{black}{ 0.0}& \textcolor{black}{ 0.0} \end{tabular}}
    
  \\  \midrule

  Original: No, Perturbed: No
    & \makecell{ \begin{tabular}{@{}r r@{}} 0.0 & 0.0  \end{tabular}}
    & \makecell{ \begin{tabular}{@{}r r@{}}  18.5 & 15.5  \end{tabular}}
    & \makecell{ \begin{tabular}{@{}r r@{}}  55.8 & 55.4  \end{tabular}} 
    & \makecell{ \begin{tabular}{@{}r r@{}} 100.0 & 100.0 \end{tabular}}
    
  \\  

               \bottomrule
    \end{tabular}
    }
    \captionof{table}{Analysis of percentage of answers changed or unchanged by GPT-4o on the program pair before and after perturbation.}
    \label{tab:error_gpt4}
\end{table*}

\section{Licence Details and Ethics}

Please refer table \ref{tab:licenses} for details of Licences.
The data used from CodeContests dataset contains code submissions of various users on codeforces platform. The userid or name of the owner of the submissions are not present in the dataset.

\begin{table*}[htbp]
\centering
\begin{tabular}{@{}lll@{}}
\toprule
\multicolumn{1}{l}{\textbf{Model / Tool / Dataset}} & \multicolumn{2}{l}{\textbf{License}}                          \\ \midrule
\textbf{HuggingFace}                        & \multicolumn{2}{l}{Apache License, Version 2.0} \\
\textbf{DeepSeek-Coder}                     & \multicolumn{2}{l}{Code: MIT License; Model: DeepSeek License Agreement} \\
\textbf{Qwen}                               & \multicolumn{2}{l}{Tongyi Qianwen License Agreement} \\
\textbf{GPT-4o}                             & \multicolumn{2}{l}{Proprietary (OpenAI Terms of Use)} \\
\textbf{Mistral}                            & \multicolumn{2}{l}{Apache License, Version 2.0} \\
\textbf{CodeContests Dataset}               & \multicolumn{2}{l}{CC BY 4.0 (Creative Commons Attribution 4.0)} \\ \bottomrule
\end{tabular}
\caption{Licenses of the different models, tools, and datasets used in our experiments.}
\label{tab:licenses}
\end{table*}

\end{document}